\documentclass{article}
\usepackage{setspace}
\usepackage[english]{babel}

\newtheorem{assumption}{Assumption}

\newtheorem{assumptionalt}{Assumption}[assumption]

\newenvironment{assumptionp}[1]{
  \renewcommand\theassumptionalt{#1}
  \assumptionalt
}{\endassumptionalt}

\usepackage[letterpaper,top=2cm,bottom=2cm,left=3cm,right=3cm,marginparwidth=1.75cm]{geometry}

\usepackage{amsmath}
\usepackage{graphicx}
\usepackage[colorlinks=true, allcolors=blue]{hyperref}
\usepackage{enumerate}
\usepackage{enumitem}
\usepackage{adjustbox}
\usepackage{soul}
\usepackage{multirow}

\usepackage{natbib}

\newtheorem{theorem}{Theorem}[section] 

\title{Automated, efficient and model-free inference for randomized clinical trials
 via data-driven covariate adjustment}
\author{Kelly Van Lancker$^{1,2}$, Iv\'an D\'iaz$^3$, and Stijn Vansteelandt$^1$}

\date{%
    $^1$Department of Mathematics, Computer Science and Statistics, Ghent University, Ghent, Belgium\\%
    $^2$Department of Mathematics and Data Science, Vrije Universiteit Brussel, Brussels, Belgium\\%
    $^3$Division of Biostatistics, Department of Population Health, New York University School of Medicine, NY, United States%
}

\begin{document}
\maketitle

\begin{abstract}
In 2023, the U.S. Food and Drug Administration issued guidance for adjustment of covariates in randomized clinical trials, emphasizing its role in enhancing precision and power through prognostic baseline variables. Despite its potential, many trials underutilize this method partly due to challenges in pre-specifying optimal baseline covariates and their functional forms.

We explore the potential of automated, data-adaptive methods—including stepwise regression, Lasso and flexible machine learning algorithms—for covariate adjustment, addressing the challenge of pre-specification. Our approach ensures valid and interpretable treatment effect estimates and standard errors, even when outcome models are misspecified or biased outcome predictions are used. 
This differs from most competing methods, which assume correctly specified models for consistent standard errors. Our estimators require cross-fitting for reliable standard error estimation, though it can be omitted when variable selection is used, provided the outcome model satisfies an ultra-sparsity assumption. As such, we arrive at simple estimators and standard errors for marginal treatment effects in randomized clinical trials (or similar studies like A/B-testing), exploiting data-adaptive predictions from prognostic baseline covariates, with little (or no) bias in finite samples even when predictions are biased.

Empirical and methodological results demonstrate promise of automated covariate adjustment for improving statistical power of trial analyses.
\end{abstract}

\section{Introduction}

For many years, the debate on whether or not to adjust for baseline covariates in the primary intention-to-treat analysis of randomized clinical trials has been obscured by the lack of focus on a single pre-specified estimand. Unadjusted comparisons deliver marginal measures of treatment effect, but tend to be inefficient as a result of ignoring potential imbalances in prognostic baseline covariates. Correction for such covariates happens in adjusted comparisons delivered by standard regression analyses; however, at the risk of bias resulting from misspecification of the regression model that is needed for adjustment (with the exception of targeting marginal effects with linear regression). Moreover, the potential efficiency gain delivered by those analyses is difficult to quantify \citep{robinson1991some} as they, conversely, deliver conditional measures of treatment effect, whose meaning and magnitude may depend on precisely what covariates have been adjusted for \citep{daniel2021making}.  \cite{tsiatis2008covariate} brought much clarity to the debate. They focused on estimation of the marginal treatment effect and found (semi-parametric) efficient estimators to exploit covariate adjustment for baseline covariates in order to boost precision, whilst being insulated - in sufficiently large samples - against the bias that might otherwise result from model misspecification. 

Despite these advances, practical implementation of (semi-parametric) efficient estimators of the marginal treatment effect in randomized clinical trials has been hindered by the need to pre-specify the baseline covariates for adjustment, as mandated by regulatory agencies such as the U.S. Food and Drug Administration (FDA) and the European Medicine Agency (EMA) \citep{FDA1998, EMA2015}. This poses challenges in determining the appropriate covariates and their functional forms (e.g., which interactions between covariates will be included in the model for the outcome) prior to seeing the data, often leading to difficult or even infeasible tasks in many trials. Indeed, while correct model specification is not strictly required to mitigate bias in treatment effect estimates, it remains indispensable for ensuring valid statistical inference. To address this, there seems promise in the use of automated data-adaptive methods, which `adapt' to the data by making use of automated variable selection strategies (e.g., stepwise AIC-based selection, the lasso, ...) or even machine learning methods (e.g., random forest regression, gradient boosting, neural networks, ...). Such algorithms allow for objective, guided selection of variables and construction of statistical models, and can be pre-specified \citep{tsiatis2008covariate,van2011targeted,wager2016high,wu2018loop,vansteelandt2021statistical}.

Nonetheless, the application of data-adaptive methods also introduces potential concerns. These methods are generally optimized for predictive accuracy, and may be sub-optimal for the estimation of treatment effects. In particular, they may deliver finite-sample biases that are non-negligible to the extent that they exceed the magnitude of the standard error at all sample sizes, thereby distorting subsequent inferences \citep{chernozhukov2018double}. In the literature on variable and model selection, this is often referred to as the problem of post-selection inference \citep{leeb2006can,berk2013valid,belloni2014inference,belloni2017program}. It stems from the estimator's distribution being different depending on the selected model, with standard confidence intervals disregarding the resulting uncertainty in model selection. 

Debiasing procedures have been developed to remedy this, thereby ensuring more robust treatment effect estimates. Targeted learning procedures \citep{moore2009covariate,van2011targeted} work by cleverly tuning the predictions delivered by data-adaptive algorithms in a way that removes bias in the resulting treatment effect estimator. Debiased machine learning methods \citep{chernozhukov2018double} instead make use of default predictions, but subsequently debias the obtained treatment effect estimate. Interestingly, the (semi-parametric) efficient estimators of \cite{tsiatis2008covariate} incorporate such debiasing by default. Unlike many other covariate-adjusted estimators of treatment effect, they can thus readily be used in conjunction with data-adaptive procedures, such as automated variable selection or machine learning. They thereby deliver treatment effect estimates that invoke covariate adjustment in a manner that can be pre-specified - as it can be automated - and have low bias even when the employed data-adaptive predictions are themselves biased (e.g., as a result of model misspecification or the use of poorly fitting/converging machine learning predictions). \cite{wager2016high} show the resulting estimators to be exactly unbiased, even in finite samples, provided linear models are used and a specific sample splitting procedure (known as cross-fitting) is employed, whereby the model building process runs on a different part of the data than the part on which the estimator is calculated. Importantly, this finite-sample unbiasedness does not rely on model correctness or consistency assumptions, but follows from randomization and cross-fitting.

Unfortunately, however, this robustness to model misspecification or poor convergence of machine learning predictions does not generally extend to standard errors, whether obtained using sandwich estimators or the bootstrap. In general, the standard errors delivered by 
the proposal of \cite{tsiatis2008covariate} or by more general targeted learning procedures \citep{van2011targeted} may be biased when based on biased data-adaptive predictions. 
This disconnect between consistent estimation and exact asymptotic coverage (i.e., as $n$ goes to infinity, the confidence interval has coverage exactly equal to the nominal level, even if the outcome working models are misspecified) has likely constrained the practical adoption of covariate adjustment—particularly in regulatory settings where pre-specification and transparency are paramount.

\paragraph{Our Contributions.}
We study covariate-adjusted estimators of the average treatment effect in randomized trials, with a focus on understanding standard error estimation under outcome model misspecification and data-adaptive modeling. We provide new theoretical insights into the conditions under which standard plug-in standard error estimators --based on the efficient influence function --yield exact asymptotic coverage, and clarify when sample splitting is required or advantageous. Our main contributions are as follows:
\begin{enumerate}
    \item Our results demonstrate that simple plug-in standard error estimators derived from the efficient influence function can yield exact asymptotic coverage for covariate-adjusted estimators with canonical GLMs --constructed via data-adaptive methods like LASSO or stepwise regression-- even under model misspecification.
    \item We show that, while the usual (efficient influence function-based) standard error estimator of the augmented inverse probability weighted estimator (AIPW) is biased when inconsistent data-adaptive outcome predictions are used, it remains applicable in targeted maximum likelihood estimation (TMLE) when combined with cross-fitting. While our results apply to flexible machine learning predictions, our primary focus is on parametric predictions (e.g., based on stepwise regression or Lasso), where efficiency gains are nearly maximal and randomness in predictions is typically lower than with more complex machine learning methods.
    \item We show that the above results support parametric estimation of the propensity score (e.g., using a logistic regression model) to improve precision (see Section \ref{sec:par_ps}). Specifically, we show that exact asymptotic coverage can be obtained in the non-trivial case where the propensity score may become a complicated expression of the estimated parameters, even when the outcome model is misspecified or biased outcome predictions are used. 
    \item We theoretically demonstrate that AIPW estimators of the marginal treatment effect remain exactly unbiased when employing data-adaptive methods --including LASSO, stepwise regression, or machine learning --so long as cross-fitting is applied and the true propensity score is used. This no longer holds for TMLE.
    \item We establish that blinded estimation of the outcome models eliminates the need for sample splitting, even with machine learning, though possibly at the expense of some efficiency loss. 
\end{enumerate}

As such, we will arrive at simple estimators and standard errors for the marginal treatment effect in randomized clinical trials (or other randomized studies such as A/B-testing), which exploit data-adaptive outcome predictions based on prognostic baseline covariates, and have low (or no) bias even when those predictions are themselves biased. 
These features are appealing as they allow for the utilization of exceedingly complex models, such as those involving splines or machine learning algorithms, without distorting the interpretation or validity of the treatment effect estimate. This flexibility guarantees that the analysis remains solely evidence-based and  leverages covariate information to maximize statistical power \citep{tsiatis2008covariate}.

\paragraph{Organization of the Paper.}
The remainder of this paper is organized as follows. 
In Section \ref{sec:background}, we review related literature, introduce notation and the estimand of interest, and present a simple covariate-adjusted estimator. Section \ref{sec:data-adaptive} extends this approach by allowing data-adaptive estimation of outcome predictions within a parametric framework. Section \ref{sec:samplesplitting} introduces a more general framework for incorporating data-adaptive predictions through cross-fitting.
Section \ref{sec:sim} presents the results of our simulation study, which mimics key features of a real dataset. Section \ref{sec:par_ps} touches on the results when estimating the propensity score parametrically. Finally, Section \ref{sec:discussion} concludes with a summary of our findings and potential future research directions.

\section{Background}\label{sec:background}
\subsection{Relation to Prior Literature}
This paper contributes to a growing literature on covariate adjustment in randomized experiments, particularly at the intersection of data-adaptive methods—such as stepwise regression and Lasso, and more generally machine learning—and valid inference.

A central modern reference point is \citet{lin2013agnostic}, who showed that ordinary least squares regression with treatment–covariate interactions yields valid and efficient inference under random assignment, even if the regression model is misspecified, for the sample average treatment effect (see also \citet{berk2013covariance}). However, Lin's approach assumes a pre-specified model and does not accommodate variable selection. 

More recently, several authors have extended covariate adjustment methods to high-dimensional or nonlinear models. \citet{wu2018loop} introduced the leave-one-out potential outcomes (LOOP) estimator, which leverages flexible regression estimators while preserving unbiasedness for the sample average treatment effect. However, this approach lacks a simple standard error formula and is more difficult to implement. \citet{guo2023generalized} extended covariate adjustment for the sample average treatment effect beyond linear models via the generalized Oaxaca–Blinder estimator, which uses nonlinear regressions to produce covariate-adjusted confidence intervals valid under randomization alone, without requiring model correctness or assuming a constant treatment effect. Their variance estimates are conservative under finite population assumptions but can be adapted for superpopulation settings. 
\citet{cohen2024no} proposed a calibration method that ensures data-adaptive covariate-adjusted estimators never perform worse in terms of asymptotic efficiency than the unadjusted difference-in-means estimator (the ``no-harm'' property) in randomized trials, generalizing the classical OLS efficiency guarantee to nonlinear models. They use sample splitting for finite-sample unbiasedness and bridge finite population and superpopulation frameworks. \citet{bannick2023general} extended these results to covariate-adaptive randomization in a superpopulation framework, proposing estimators with asymptotically exact standard errors that can incorporate machine learning. Our current work was developed concurrently \citep[see e.g.,][page 38]{vanlancker2021phd}.

Closely related lines of work are the Double/Debiased Machine Learning (DML) framework developed by \citet{chernozhukov2018double} and the targeted learning approaches developed by Mark van der Laan and colleagues \citep{moore2009covariate,van2011targeted, williams2021, balzer2024adaptive}. Like our approach, DML and TMLE use cross-fitting and data-adaptive estimators to adjust for high-dimensional or nonparametric nuisance functions. However, a key distinction with our work is that DML requires the nuisance estimators to be consistent—typically at sufficiently fast rates—to ensure valid inference. 
In contrast, we allow for misspecification of the outcome model, such as when performing variable selection using LASSO under a nonlinear or misspecified structure. \citet{moore2009covariate} and \citet{van2011targeted} showed that TMLE yields conservative inference in randomized settings if models are misspecified. In this article, we show that exact asymptotic coverage is still attainable in such cases, either by leveraging an ultra-sparsity condition or via cross-fitting.

In the context of doubly robust estimation, earlier work has identified the challenge of obtaining valid standard errors under inconsistent or slowly converging nuisance estimators. \citet{vermeulen2015bias} proposed bias correction methods for parametric models, later extended to variable selection \citep{avagyan2022high}. This has been independently developed for arbitrary learners by \cite{benkeser2017doubly}, though at the expense of possibly non-regular behavior and instability under certain data-generating laws \citep{dukes2021doubly}. More recently, \citet{van2024doubly} addressed this issue through calibrated debiased machine learning, in which cross-fitted nuisance estimates are calibrated—via, for example, isotonic regression—to achieve doubly robust asymptotic normality for linear functionals. Their approach guarantees asymptotic normality if either the outcome regression or the (conditional) treatment probability is estimated sufficiently well, allowing the other to converge arbitrarily slowly or even inconsistently, and can be implemented as a lightweight augmentation of existing DML pipelines. In contrast, in our setting the treatment is randomized and the propensity score is known, which ensures regularity and stability even when using highly data-adaptive nuisance estimators; we therefore do not need the use of the more general, but more technically demanding strategies listed above

A related strand of work is \citet{rothe2018flexible}, who shows that average treatment effect estimators expressed as sample averages of an empirical analogue of the uncentered efficient influence function maintain valid inference despite model misspecification and can even achieve unbiasedness in finite samples when combined with leave-one-out cross-fitting. His estimators resemble doubly robust approaches but plug in the known treatment assignment probability (rather than an estimated propensity score), simplifying the problem. Our results generalize this insight by allowing for estimated propensity scores, both constant and covariate-dependent, and by showing that similar asymptotic properties—consistency, unbiasedness, and asymptotic normality—hold for AIPW, TMLE, and standardization estimators, even under flexible, nonparametric nuisance estimation, provided cross-fitting is employed (or under an ultra-sparsity condition). Crucially, exact asymptotic coverage in this setting requires adjusting standard errors to account for the extra variability from estimating the propensity score.

Our work is most closely related to \citet{williams2021}, who demonstrated for TMLE that exact asymptotic coverage is possible under model misspecification, so long as outcome regressions are cross-fitted and the propensity score is known or estimated independently of baseline covariates. Concurrently with our current work on simple randomization \citep[see e.g.,][page 38]{vanlancker2021phd}, \citet{bannick2023general} developed related results, showing that asymptotically exact standard errors can also be obtained under covariate-adaptive randomization with flexible machine learning. We generalize the results to a broader class of estimators—including AIPW, TMLE, and G-computation—under outcome model misspecification, for different types of endpoints and estimated propensity scores (e.g., using a logistic regression model). Although dependent randomization is less common, estimating the propensity score can still improve efficiency in simple randomization, where the true propensity score is constant and independent of baseline covariates, yielding simple, robust inference procedures valid even under imperfect or data-adaptive outcome models.


\subsection{Notation and Estimand}\label{sec:notation}
We adopt a super-population framework in which the observed data consist of $n$ independent and identically distributed (i.i.d.) units. For participant $i$ ($i = 1, \dots, n$), let $X_i$ denote a vector of baseline covariates measured prior to treatment assignment, and let $Z_i \in \{0,1\}$ denote the treatment assignment indicator, where $Z_i = 1$ indicates assignment to the experimental treatment arm and $Z_i = 0$ indicates assignment to the control arm. We assume that treatment is assigned according to a Bernoulli randomization scheme, in which $Z_i \sim \mathrm{Bernoulli}(\pi)$ independently across $i$, where $\pi$ is a known probability (often 0.5), possibly depending on baseline covariates $X_i$. 
Let $Y_i$ denote the observed primary outcome of interest for participant $i$, which may be continuous, binary, or ordinal. When all variables are fully observed, participant $i$ contributes the observed data tuple $\mathbf{O}_i = (X_i, Z_i, Y_i)$. We assume that the collection $\{\mathbf{O}_i\}_{i=1}^n$ comprises i.i.d.\ realizations from a joint distribution $P$ over the space of $(X, Z, Y)$.

We are interested in making inferences about marginal / unconditional intention-to-treat effects, which we will denote by $\theta$ and refer to as the \textit{policy estimand} of interest. 
Examples include the difference in means of the primary outcome between study arms for continuous and binary outcomes, relative risk and odds ratio for binary outcomes, and Mann-Whitney estimand for ordinal outcomes. 
In what follows, without loss of generality, we focus on estimation and inference for $$\theta=E\left(Y|Z=1\right)-E\left(Y|Z=0\right).$$

The next section outlines a regular covariate-adjusted estimator for $\theta$, which sets the stage for the data-adaptive covariate adjustment methods developed in the subsequent sections.

\subsection{Regular Covariate-Adjusted Estimation}\label{sec:regular}
Unadjusted estimators are readily obtained by replacing population means with sample means in the estimand definition. For example, the unadjusted estimator of the average treatment effect is the difference in sample means.
To take full advantage of prognostic baseline covariates, we focus on a generalization of the covariate adjusted estimator suggested in the recent U.S. Food and Drug Administration guidance on covariate adjustment \citep{FDA2021}. 
To focus ideas, we consider a model-based covariate-adjusted estimator of $\mu_1=E(Y|Z=z)$ (with $z\in\{0, 1\}$):
\begin{enumerate}
    \item[\textbf{Step 1:}] \textbf{Model fitting:}\\ Fit a generalized linear model (GLM) with a canonical link and intercept using maximum likelihood to regress $Y$ on covariates $X$ using only participants with $Z = z$.
    \item[\textbf{Step 2:}] \textbf{Prediction:} Use the fitted model to predict outcomes $\hat{Y}_{z,i}$ for all participants $i = 1, \ldots, n$.
    \item[\textbf{Step 3:}] \textbf{Averaging:} Compute $\hat{\mu}_z = n^{-1} \sum_{i=1}^n \hat{Y}_{z,i}$.
\end{enumerate}
A covariate-adjusted estimator for the estimand of interest $\theta$ can then be obtained as 
$$\hat\theta = \hat{\mu}_1-\hat{\mu}_0 = \frac{1}{n} \sum_{i=1}^n \left(\hat{Y}_{1,i}- \hat{Y}_{0,i}\right).$$ 
It is known as a standardization, G-computation or imputation estimator \citep{tsiatis2008covariate}. 

This estimator has several desirable properties in randomized controlled trials (see \cite{van2024covariate, negi2025robust}). When the GLMs are fitted using maximum likelihood, the predictions satisfy the following prediction-unbiasedness condition:
\begin{equation}
\sum_{i: Z_i = z} \hat{Y}_{z,i} = \sum_{i: Z_i = z} Y_i \text{\quad for }z\in\{0,1\}, \label{eq:predictionUnbiasedness}
\end{equation}
which guarantees that the estimator is unbiased in large samples even if the working model is misspecified. For other types of models, additional debiasing is needed to restore this robustness by using augmented inverse probability weighting \citep{robins1994estimation}. The latter estimator typically improves efficiency compared to the unadjusted estimator, leading to narrower confidence intervals and increased statistical power \citep{tsiatis2008covariate, benkeser2020improving}.

\section{Data-Adaptive Estimation via Canonical GLMs}\label{sec:data-adaptive}
Pre-defining a data-adaptive covariate-selection rule offers greater flexibility and efficiency than fixing covariates and model form a priori \citep{tsiatis2008covariate}. We show that, even when the outcome model is misspecified, the estimator in Section \ref{sec:regular} remains valid and permits consistent standard error estimation (leading to exact asymptotic coverage) when the candidate set and selection algorithm are pre-specified—for instance, by inserting a model-selection step (Step 1a) before model fitting (Step 1b) \citep{tsiatis2008covariate}:
\begin{itemize}[left=2em]
    \item[\textbf{Step 1a:}]\textbf{Model selection} Fit a generalized linear regression model with a canonical link (e.g., logistic, linear, \dots) that regresses the outcome $Y$ on pre-specified baseline covariates $X$ among the treated participants (i.e., with $Z=1$), using a data-adaptive method (e.g., stepwise regression or Lasso). Select the covariates — which can for instance include interaction terms or higher order terms — with non-zero estimated coefficients.
	\item[\textbf{Step 1b:}]\textbf{Model fitting} Fit a generalized linear regression model with a canonical link (e.g., logistic, linear, \dots) and intercept via maximum likelihood that regresses the outcome $Y$ on all covariates selected in Step 1a among the treated participants. This regression may also include additional variables (such as stratification factors) that were not selected in the previous steps, but that were identified \textit{a priori} as being important. 
\end{itemize}
The estimated treatment effect is denoted by 
\[
\hat\theta_{\mathrm{DA}} = \hat\mu_{1,\mathrm{DA}} - \hat\mu_{0,\mathrm{DA}} = \frac{1}{n} \sum_{i=1}^n \big(\hat{Y}_{1,i} - \hat{Y}_{0,i}\big),
\] 
where `DA' stands for `data-adaptive', and $\hat{Y}_{z,i}$ are predictions obtained from a data-adaptive procedure using a canonical GLM.

\subsection{Inference and Theoretical Properties}\label{sec:Characteristics}
The estimator $\hat\theta_{\mathrm{DA}}$ inherits key robustness properties from the fixed-model estimator in Section~\ref{sec:regular}, provided that canonical generalized linear models are fitted via maximum likelihood. This ensures the prediction-unbiasedness condition~\eqref{eq:predictionUnbiasedness}, so that outcome model misspecification or imperfect covariate selection does not introduce bias in large samples. 
Surprisingly, under randomized treatment assignment independent of baseline covariates, this robustness extends to the estimation of standard errors as well. They are asymptotically unbiased even when the set of candidate covariates is high-dimensional, provided the \textit{ultra-sparsity} condition holds—that is, the number of covariates with non-zero coefficients remains small relative to the sample size (see Appendix~A.1). This requirement is relatively mild, as users retain full flexibility in choosing the predictors and model size; however, we recommend keeping the number of predictors below $\sqrt{n}$. In typical applications, this essentially ensures that the theoretical condition of ultra-sparsity (see Appendix~A.1) is satisfied. The following theorem formalizes this result:

\begin{theorem}[Asymptotic validity of $\hat\theta_{DA}$ and its variance estimator under misspecification]\label{thm:asymptotic_se}
Assume treatment is assigned completely at random and that the outcome regression models used in Step~1b of the data-adaptive procedure are canonical generalized linear models fitted by maximum likelihood, which we assume to converge to some (possibly misspecified) limit. We moreover assume ultra-sparsity on the candidate covariate set (see Appendix~A.1). Then, even if the working models are misspecified and covariates are selected via a data-adaptive rule from a high-dimensional set, $\sqrt{n}\left(\hat\theta_{DA} - \theta\right)$ is asymptotically normal with variance consistently estimated by the sample variance of the values
\begin{align}\label{eq:var}
    \frac{Z_i}{\hat\pi}(Y_i-\hat{Y}_{1,i})+\hat{Y}_{1,i} - \left[\frac{1-Z_i}{1-\hat\pi}(Y_i-\hat{Y}_{0,i})+\hat{Y}_{0,i}\right],
\end{align}
where $\hat{\pi} = n^{-1} \sum_{i=1}^n Z_i$ and $\hat{Y}_{z,i}$ are the predicted values obtained from the data-adaptive procedure in Section~\ref{sec:data-adaptive}. 
\end{theorem}

To clarify the origin of the variance expression in \eqref{eq:var} and its connection to more general estimators, note that the quantity inside the sample variance corresponds to the empirical analogue of the efficient influence function for the AIPW estimator of $\theta$, which is given by
\[
\hat\theta_{\mathrm{AIPW}}
=
\frac{1}{n}\sum_{i=1}^n
\left[
\frac{Z_i}{\hat\pi}(Y_i-\hat{Y}_{1,i})
-
\frac{1-Z_i}{1-\hat\pi}(Y_i-\hat{Y}_{0,i})
+
\hat{Y}_{1,i}-\hat{Y}_{0,i}
\right].
\]
When $\hat{Y}_{z,i}$ are arbitrary outcome predictions, this estimator explicitly corrects for potential bias in the working outcome model through the augmentation terms $\frac{Z_i}{\hat\pi}(Y_i-\hat{Y}_{1,i})$ and $\frac{1-Z_i}{1-\hat\pi}(Y_i-\hat{Y}_{0,i})$. The variance expression in \eqref{eq:var} is precisely the sample variance of the corresponding estimated efficient influence function.

When canonical generalized linear models are fitted via maximum likelihood as in Section~\ref{sec:data-adaptive}, the prediction-unbiasedness condition \eqref{eq:predictionUnbiasedness} holds. In that case, the augmentation terms average to zero within each treatment arm, and the AIPW estimator reduces exactly—also in finite samples—to the standardization estimator $\hat\theta_{\mathrm{DA}}$. Thus, in the canonical GLM setting, no explicit augmentation step is required: the debiasing is implicitly enforced by maximum likelihood estimation.

Importantly, the asymptotic normality and variance consistency result stated in Theorem~\ref{thm:asymptotic_se} extends more generally to the AIPW estimator, even when the outcome regressions are non-canonical parametric models for which prediction-unbiasedness does not hold. For example, if one uses a negative binomial regression model for count outcomes, or a quasi-likelihood model with a non-canonical link, the standardization estimator alone is no longer robust to misspecification. However, the corresponding AIPW estimator retains the same asymptotic validity under analogous sparsity conditions. Canonical GLMs therefore constitute a special subclass of working models for which standardization and AIPW coincide exactly, explaining why the influence-function-based variance formula in \eqref{eq:var} applies directly in that setting.

The result Theorem~\ref{thm:asymptotic_se} implies that standard Wald-type confidence intervals for $\hat\theta_{\mathrm{DA}}$ (and $\hat\theta_{\mathrm{AIPW}}$) attain the correct asymptotic coverage even when the outcome model is misspecified or when variable selection is imperfect. As long as ultra-sparsity holds—i.e., only a small number of covariates among the candidate set are truly prognostic—valid inference proceeds as if the selected covariates had been fixed in advance. A full proof is provided in Appendix~A.3.

This is a strong result as it implies that, besides the aforementioned sparsity conditions, only weak assumptions (see Appendix A.1) are needed to ensure that the data-adaptive approach in Section \ref{sec:data-adaptive} is insensitive to selection mistakes in large samples (i.e., preserves nominal Type I error). We conjecture that ultra-sparsity is sufficient but not necessary—further research is needed to clarify this point. A proof is given in Appendix A.3.

\paragraph{Efficiency considerations.} 
When the outcome model is correctly specified and the variable selection procedure is consistent (e.g., stepwise regression or Lasso), $\hat\theta_{\mathrm{DA}}$ is asymptotically efficient in the subclass of estimators that are unbiased as soon as $Z$ is independent of $X$ \citep{yang2001efficiency, tsiatis2006semiparametric}. Even under misspecification, efficiency gains are common in practice \citep{colantuoni2015leveraging}.

Furthermore, following the literature on empirical efficiency maximization \citep{rubin2008empirical, cao2009improving, gruber2012targeted}, one can optimize the outcome predictions to minimize mean squared error (MSE), yielding an estimator of the average treatment effect that is at least as efficient as the unadjusted estimator—even under model misspecification (see Appendix A~6). This, however, requires fitting the model via least squares, which breaks the prediction-unbiasedness property and thus forfeits robustness. An alternative that retains all desirable properties is discussed in Section~\ref{sec:samplesplitting}, based on cross-fitting.\\

\begin{table}[h!]
	\caption{\label{tab:characteristics} Overview of important characteristics of the different covariate adjustment methods. Cross-fitting: whether cross-fitting is considered; Algorithm: the prediction algorithms that are considered [Can. GLMs: only allowing for automated variable selection approaches (e.g., lasso, stepwise AIC-based selection, \dots) for canonical generalized linear models fitted via maximum likelihood; ML: also allowing for more general machine learning algorithms; GLMs$^{*}$: automated variable selection approach for parametric models, for which the prediction unbiasedness condition \eqref{eq:predictionUnbiasedness} does not necessarily hold]; SP efficient: whether the estimators are semi-parametric efficient under the model that assumes treatment to be independent of covariates; Sample bounded: whether estimates of $\mu_1$ and $\mu_0$ are both contained in the sample outcome range; Robust estimation: whether estimators are unbiased in large samples when the outcome working models are misspecified [Asympt.: estimator is (only) asymptotically unbiased, when models are misspecified; Exact$^{**}$: estimator is exactly unbiased - also in finite samples - as long as true $\pi$ has been used, even when models are misspecified]; SN w/o SS: can one avoid sample splitting when testing strong null (without any additional assumptions)? [$\dag$: We refer to Section \ref{subsubsec:strongNull} as it requires an adaptation of the regular and data-adaptive covariate-adjusted estimators with canonical GLMs.] 
Note: The two TMLE rows reflect TMLE without cross-fitting (low-dimensional parametric models) and cross-validated TMLE (CV-TMLE) for flexible machine learning; see Appendix A.4 for full formulas and implementation details.
}
\centering
\begin{tabular}{ l cc|cccc}
  \hline
Method  & Cross- & Algorithm &SP &Sample & Robust & SN  \\
 & fitting & &efficient  &bounded & estimation& w/o SS  \\
\hline
Unadjusted  & No &/ &/&$\surd$ &  Exact$^{**}$ & $\surd$ \\ 
G-computation ($\hat\theta$) & No & Can. GLMs & $\surd$ & $\surd$  & Asympt. & $\surd^\dag$ \\
Data-adapt. G-comp. ($\hat\theta_{DA}$) & No & Can. GLMs & $\surd$ & $\surd$  & Asympt. & $\surd^\dag$ \\
Data-adapt. AIPW ($\hat\theta_{AIPW}$) & No & GLMs$^{*}$ & $\surd$ & $\surd$  & Asympt. & $\surd$ \\
ML-based cov. adj. ($\hat\theta_{CF}$) & Yes & ML & $\surd$ &  / & Exact$^{**}$ & $\surd$\\
TMLE   & No & GLMs$^{*}$ & $\surd$ &$\surd$   & Asympt. & /  \\
CV-TMLE & Yes  & ML & $\surd$ &$\surd$ & Asympt. & / \\
   \hline
\end{tabular}\\
	 {\raggedright  \par}
\end{table}

All these appealing characteristics -- which are summarized in Table \ref{tab:characteristics} -- also hold when the association between $Z$ and $X$ is parametrically estimated (e.g., using a logistic regression model), as demonstrated in Appendix A.5.2. This extension can be accomplished by fitting a regression model in Step 1b via weighted maximum likelihood with weights equal to the inverse of the probability of receiving the assigned treatment conditional on baseline covariates. Further elaboration on this is provided in Section \ref{sec:par_ps}.

\section{Data-Adaptive Estimation Using Cross-Fitting}\label{sec:samplesplitting}

The sparsity assumptions required for the method(s) proposed in the previous section(s) justify the use of certain data-adaptive approaches such as the (post-) Lasso or stepwise selection. However, these assumptions are no longer necessary when the predictions $\hat{Y}_{1,i}$ and $\hat{Y}_{0,i}$ are estimated from a separate sample. This can be achieved via sample splitting and cross-fitting (see next paragraph; \cite{zheng2011cross, chernozhukov2018double}). Their use is appealing as they enable the use of more complex data-adaptive methods, such as machine learning algorithms. 

Cross-fitting is performed as follows. First, we split the data into $K$ non-overlapping folds. 
Let $\mathcal{V}_1, \dots, \mathcal{V}_K$ denote the partition of the index set $\{1, \dots, n\}$ into $K$ non-overlapping subsets. 
Next, for the participants belonging to a certain fold $k=1, \dots, K$, the prediction models (under both treatments) are estimated based on the $K-1$ remaining folds, that is, $\{1, \dots, n\}\backslash \mathcal{V}_k$ (Step 1). 
This might involve more complex data-adaptive methods such as machine learning algorithms as well as variable selection approaches for parametric models.
Then, for each participant $i$ in fold $k$, predictions $\hat{Y}_{1,k,i}$ (under $Z=1$) and $\hat{Y}_{0,k,i}$ (under $Z=0$) are obtained based on these models (Step 2). 
Define $\hat \pi_k=\frac{1}{|\mathcal{V}_k |}\sum_{i\in\mathcal{V}_k}Z_i$ as an estimate of $\pi=P(Z=1)$ based on fold $k$, with $|\mathcal{V}_k |$ the number of participants in fold $k$ (i.e., participants with index in $\mathcal{V}_k$). As the prediction unbiasedness condition \eqref{eq:predictionUnbiasedness} no longer holds under sample splitting, the estimator $\hat{\theta}_{CF}$ of $\theta$, is defined as \citep{robins1994estimation}
\begin{align}
\hat\theta_{CF}
=
\frac{1}{n}
\sum_{i=1}^n
\left[
\frac{Z_i}{\hat{\pi}_{k(i)}}\{Y_i - \hat{Y}_{1,k(i),i}\}
-
\frac{1-Z_i}{1-\hat{\pi}_{k(i)}}\{Y_i - \hat{Y}_{0,k(i),i}\}
+
\hat{Y}_{1,k(i),i} - \hat{Y}_{0,k(i),i}
\right],\label{eq:IF_aipw_samplesplitting}
\end{align}
where $k(i) \in \{1,\dots,K\}$ denotes the fold such that $i \in \mathcal{V}_{k(i)}$.

\subsection{Inference and Theoretical Properties}\label{sec:CF_Characteristics}
This estimator as well as the variant where  the association between $Z$ and $X$ is parametrically estimated (see Section \ref{sec:par_ps}) inherits the appealing properties discussed in the previous sections. 
It should be noted, however, that the variance cannot simply be derived as $1/n$ times the sample variance of the values from \eqref{eq:IF_aipw_samplesplitting}, due to the need for extra error terms to account for the uncertainty associated with the (estimated) randomization probabilities:

\begin{theorem}[Asymptotic validity of $\hat\theta_{CF}$ and its variance estimator under misspecification]\label{thm:asymptotic_seCF}
Assume treatment is assigned completely at random and that the outcome predictions converge to some (possibly misspecified) limit. Then, even if covariates are selected via a data-adaptive rule from a high-dimensional set, $\sqrt{n}\left(\hat\theta_{\mathrm{CF}}-\theta\right)$ is asymptotically normal with variance consistently estimated by the sample variance of the values
\begin{align*}
    &\left\{\left(\frac{Z_i}{\hat\pi_{k(i)}}\left\{Y_i-\hat{Y}_{1,k(i),i}\right\}+\hat{Y}_{1,k(i),i}-\frac{1}{|\mathcal{V}_{k(i)}|}\sum_{j\in \mathcal{V}_{k(i)}}\left[\frac{Z_j}{\hat\pi_{k(i)}^2}\left\{Y_j-\hat{Y}_{1,k(i),j}\right\}\right](Z_i - \hat\pi_{k(i)})\right)\right.\\
    &-\left.\left(\frac{1-Z_i}{1-\hat\pi_{k(i)}}\left\{Y_i-\hat{Y}_{0,k(i),i}\right\}+\hat{Y}_{0,k(i),i}+\frac{1}{|\mathcal{V}_{k(i)}|}\sum_{j\in \mathcal{V}_{k(i)}}\left[\frac{1-Z_j}{(1-\hat\pi_{k(i)})^2}\left\{Y_j-\hat{Y}_{0,k(i),j}\right\}\right](Z_i - \hat\pi_{k(i)})\right)\right\},
\end{align*}
where $k(i)$ indicates to which fold participant $i$ belongs.
\end{theorem}
A proof is provided in Appendix A.2. 
While related results exist for AIPW estimators with pre-specified models \citep{tsiatis2006semiparametric, vermeulen2015bias}, our result extends these to settings where models are selected in a data-adaptive manner. This illustrates the necessity of accounting for additional variability when the prediction-unbiasedness condition is violated and highlights the simplicity of the variance estimator in Theorem \ref{thm:asymptotic_se} when applied to canonical GLMs.

\paragraph{Efficiency considerations}
As mentioned in Section \ref{sec:Characteristics}, some form of efficiency is attainable by fitting the outcome model in a specific manner even when the outcome model is misspecified. 
Concretely, the estimator based on Equation \eqref{eq:IF_aipw_samplesplitting} with minimal variance (over a broad class of estimators, including the unadjusted estimator; see Appendix A.6)
can be obtained by selecting and fitting the outcome model to have minimal MSE over that class (see Appendix A.6). This is especially appealing as the approach in this section allows the use of machine learning algorithms, many of which allow to choose the MSE as the loss function of interest. A similar approach was proposed by  \cite{cohen2024no} and \cite{bannick2023general}.

\paragraph{Finite-sample unbiasedness}
Suprisingly, when the known randomization probability $\pi$ (e.g., 0.5 for a 1:1 randomized trial) replaces the estimated propensity scores $\hat{\pi}_k$, the estimator $\hat{\theta}_{CF}$ is exactly unbiased, even in finite samples, regardless of model misspecification:
\begin{theorem}[Finite sample unbiasedness]\label{thm:finiteSample_unbiasedness}
Assume treatment is assigned completely at random and that the outcome predictions converge to some (possibly misspecified) limit. Then, even if the working models are misspecified and covariates are selected adaptively from a high-dimensional set, the estimator $\hat{\theta}_{CF}$ remains unbiased in finite samples when $\hat{\pi}_k$ is replaced by the known randomization probability $\pi$.
\end{theorem}
A proof is given in Appendix A.2.2. While the core insight aligns with results such as the finite sample unbiasedness of the LOOP estimator in \citet{wu2018loop}, the efficient influence function–based estimator in \citet{rothe2018flexible}, and the calibration estimator in \citet{cohen2024no}, our proof takes a different route to be more accessible to applied researchers.

This also means that, when using estimated propensity scores $\hat\pi_k$, any potential bias is solely attributed to the estimation of these randomization probabilities, which we can reasonably expect to be small in magnitude. Indeed, note that unadjusted estimators (e.g., mean differences) likewise exhibit finite sample bias due to the denominator (representing number of people in each arm) being subject to randomness. The weak assumptions needed to obtain exact asymptotic coverage with this approach are described in Appendix A.1. 
A possible issue of sample splitting is that it may introduce additional variation, which is a potential concern in general \citep[see e.g.,][]{guo2023rank} although likely to a lesser extent in RCTs.

\subsection{Avoiding cross-fitting when testing the strong null -- without additional assumptions}\label{subsubsec:strongNull}
Interestingly, sample splitting can be avoided — even when using machine learning predictions — if interest lies solely in testing the (strong) null hypothesis that $Z$ has no effect on $Y$ (i.e., $Z$ is independent of $Y$). 
In particular, one may then estimate $E(Y|Z=1, X)$ and $E(Y|Z=0, X)$ under the null hypothesis that $Z$ is jointly independent of $Y$ and $X$. Specifically, under the null we can learn $E(Y|Z=1, X)=E(Y|Z=0, X)=E(Y|X)$ by fitting one model in all participants only including the covariates $X$ (and not the study arm assignment indicator $Z$). The prediction for participant $i$ ($i=1,\dots,n$) under this model is denoted by $\hat{Y}_i$.
Under the assumption that $Z$ is jointly independent of $Y$ and $X$, $\mu_1=E(Y|Z=1)$ can be estimated as $\hat\mu_{1, SN}=\frac{1}{n}\sum_{i=1}^{n}\frac{Z_i}{\hat\pi}\left\{Y_i-\hat{Y}_i\right\}+\hat{Y}_i$. Here, `SN' stands for `strong null'. Note again that there are no restrictions on how the predictions are obtained, meaning that they are potentially obtained by machine learning algorithms. Similarly, $\hat\mu_{0, SN}=\frac{1}{n}\sum_{i=1}^{n}\frac{1-Z_i}{1-\hat\pi}\left\{Y_i-\hat{Y}_i\right\}+\hat{Y}_i$, which uses the same predictions $\hat{Y}_i$ as $\hat\mu_{1, SN}$.

In Appendix A.2.1, we give a variance estimator and prove the validity of this approach. 
The resulting estimator, $\hat\theta_{SN}=\hat\mu_{1, SN}-\hat\mu_{0, SN}$, also has the advantage of being exactly unbiased, even in finite samples as long as the estimated probability $\hat\pi$ is replaced with the known randomization probability $\pi$ (e.g., 0.5 for a 1:1 randomized trial). We note, however, that under the weaker null hypothesis that only requires $E(Y|Z=1)=E(Y|Z=0)$, but does not assume statistical independence between $Z$ and $Y$, the estimator may be anti-conservative \citep{cohen2022gaussian}.

\section{Simulation Study}\label{sec:sim}
In this section, we examine whether the data-adaptive covariate adjustment approaches control the Type I error and achieve desirable power, and compare their performance (in terms of Type I error and power) to that of an unadjusted estimator and a regular standardization estimator (see Section \ref{sec:regular}). Our simulation studies mimic features of the data generating distributions from the MISTIE III trial \citep{hanley2019efficacy}.

The MISTIE III trial was an open-label, blinded endpoint, Phase III clinical trial of minimally invasive surgery with thrombolysis in intracerebral haemorrhage evacuation. The aim was to assess whether minimally invasive catheter evacuation followed by thrombolysis, to decrease clot size to 15 mL or less, improves functional outcome in patients with intracerebral haemorrhage (a severe form of stroke). A good functional outcome was defined as having a modified Rankin Scale (mRS) score of 0-3 measured 365 days from enrollment  (defined as a ``success"). The estimand of interest was defined as the (absolute) risk difference between the treatment and control (i.e., standard of care using medical management) arm.
Although the trial used covariate-adaptive randomization, we ignore that for simplicity and use simple randomization in our simulation study.

\subsection{Data-Generating Mechanisms}\label{subsec:sim_datagen}
For every simulation setting, we simulate a trial of $n$ (with $n$ equal to 200, 500 and 1000) participants using the original dataset as follows. We resample with replacement from the original (modified intention-to-treat) study population of 499 patients and only extract the baseline covariate information from the subjects in the resampled population. The baseline covariates $X$ are stability intracerebral haemorrhage clot size (\textit{ich volume}; continuous in mL), age at randomization (\textit{age}; continuous in years), severity of impairment at randomization as measured by GCS (\textit{severity}; categorical with levels: 3-8, severe; 9-12, moderate; 13-15, mild), stability intraventricular haemorrhage size (\textit{ivh volume}; continuous in mL), intracerebral haemorrhage clot location (\textit{ich location}; binary with levels lobar and deep), gender (\textit{gender}; binary with levels male and female), hyperlipidaemia medication compliant history (\textit{hyperlipidemia}; binary with levels yes and no), hormone replacement therapy (\textit{hormone}; binary with levels yes and no), smoker (\textit{smoker}; binary with levels yes and no),
cocaine usage (\textit{cocaine}; binary with levels yes and no),
alcohol abuse (\textit{alcohol}; binary with levels yes and no), 
race (\textit{race}; categorical with levels: African or American black; white; other), and
diabetes (\textit{diabetes}; binary with levels yes and no).

Treatment and control are assigned with probability 0.5 to each hypothetical participant. Then, the primary endpoint data, measured 365 days after randomization, are generated under different scenarios depending on the simulation experiment. 
We consider two scenarios; in the first one there is a positive average treatment effect (i.e., the alternative hypothesis), while in the second one there is a zero average treatment effect (i.e., the null hypothesis). 
For the first scenario, the following data-generating mechanisms are employed: $Y_i(0)|X_i \stackrel{d}{=} Ber\{m_0(X_i)\}$ and $Y_i(1)|X_i \stackrel{d}{=} Ber\{m_1(X_i)\}$ $(i=1, \dots, n)$. Here, $m_0(X)=logit^{-1}(7.84-0.05\cdot  \textit{\text{ich volume}}-2.28\cdot \textit{\text{ich location}}-0.08\cdot \textit{\text{age}})
$ and 
$m_1(X)=logit^{-1}(8.25-0.04\cdot  \textit{\text{ich volume}}-2.40\cdot \textit{\text{ich location}}-0.08\cdot \textit{\text{age}})
$, which were obtained by fitting a model for the outcome given \textit{ich volume}, \textit{ich location} and \textit{age} in respectively the control and treatment arm. For the second scenario, we first generate data as in scenario 1. Then, for each simulated participant with initial values $Y(1)=1$, we randomly replaced $Y(1)$ by an independent Bernoulli draw with probability 0.40/0.45 of being 1. This adjustment is based on the outcome probabilities under scenario 1: the probability of $Y(0)=1$ in the control arm is 0.40, and the probability of $Y(1)=1$ in the treatment arm is 0.45. To simulate the null hypothesis of no treatment effect, we resample participants with $Y(1)=1$ using a Bernoulli draw with probability $0.40/0.45$, so that the expected mean outcome in the treatment arm matches that of the control arm.

\subsection{Methods of Analysis}\label{subsec:sim_methods}
Each simulated dataset is analyzed at a one-sided 2.5\% significance level  using Wald test statistics based on the following estimators:
\begin{itemize}
	\item \textbf{Unadjusted} estimator of  difference in means.
	\item \textbf{General covariate adjustment (i.e., standardization)} estimator using a logistic regression model including (1) \textit{ich volume}, (2) \textit{ich location}, (3) \textit{age}, (4) main effects of \textit{ich volume}, \textit{ich location} and \textit{age} (i.e., true model), (5) main effects of \textit{ich volume}, \textit{ich location}, \textit{age}, \textit{ivh volume}, \textit{gender} and \textit{severity}, (6) main effects of all baseline covariates, (7) main effects of log(\textit{ich volume}), \textit{ich location} and 1/log(\textit{age}), (8) main effects of log(\textit{ich volume}), \textit{ich location}, 1/log(\textit{age}), \textit{ivh volume}, \textit{gender} and \textit{severity}, and (9) main effects of all baseline covariates but by replacing \textit{ich volume} and \textit{age} by respectively log(\textit{ich volume}) and 1/log(\textit{age}).
	\item \textbf{Data-adaptive  covariate adjustment} estimator (see Section \ref{sec:data-adaptive}) using backward stepwise regression based on AIC starting from a logistic regression model defined by settings (4) to (9) above. 
 We consider these estimation methods with and without cross-fitting (see Section \ref{sec:samplesplitting}). For the settings with cross-fitting, 5 folds are used.
	\end{itemize}

A common concern with large sample methods, such as standardization, is their tendency to overlook the small sample corrections typically applied in exact methods, which can result in underestimated standard errors \citep{tackney2023comparison}. To address this, it may be particularly useful to multiply the variance estimator in Equation \eqref{eq:var} by the following small sample correction factor:
$$\frac{(n_0-p_0-1)^{-1}+(n_1-p_1-1)^{-1}}{(n_0-1)^{-1}+(n_1-1)^{-1}},$$
where $n_j$ ($j=0,1$) are the number of participants used to fit the outcome working model in treatment arm $j$ and $p_j$ the numbers of parameters fitted in these models, exclusive of intercepts \citep{tsiatis2008covariate}. As this small sample correction has shown promising results in earlier simulation studies \citep{van2022combining}, we made use of this in our main simulation study for the general and data-adaptive covariate estimator without cross-fitting (see Table \ref{tab:MISTIE_Type1_Power}). For data-adaptive methods (without cross-fitting), $p_j$ represents the number of parameters in the final model. 
We did not apply a small sample size correction when cross-fitting was used, as cross-fitting inherently mitigates the underestimation of standard errors. This is because cross-fitting prevents own-observation bias by ensuring that each prediction is made using a model trained on data separate from the observation being predicted, which avoids overly optimistic estimates and the resulting variance underestimation.
Results without this correction (for the general and data-adaptive covariate adjustment without cross-fitting) can be found in Table 3 in Appendix B.
In Table 4 in Appendix B, we show results with a (too) stringent correction (for all estimators) with $p_j$ the number of parameters in the upper model (i.e., largest possible model allowed in the selection procedure). This leads to a conservative approach.

\subsection{Results}

\begin{table}[h!]
\caption{\label{tab:MISTIE_Type1_Power}Results for Type I error and Power. A small sample correction was used for the general covariate adjustment approaches and data-adaptive covariate adjustment approaches without cross-fitting. $\#$ parameters depicts the (maximum) number of parameters (not including the intercept) in the (final) outcome working models. Results based on 100,000 simulations.
	}
 \centering
\begin{tabular}{ l  c ccc c ccc c ccc}
  \hline
  &&\multicolumn{3}{c}{\multirow{2}{*}{General}}&&\multicolumn{3}{c}{Data-adaptive}&&\multicolumn{3}{c}{Data-adaptive}\\
  &&\multicolumn{3}{c}{}&&\multicolumn{3}{c}{No cross-fitting}&&\multicolumn{3}{c}{Cross-fitting}\\
  \cline{3-5} \cline{7-9} \cline{11-13}\\
Setting & $\#$ parameters & 200 & 500 & 1000 & & 200 & 500 & 1000 & & 200 & 500 & 1000\\
  \hline  \\
  \multicolumn{13}{c}{\textbf{Type I error}}\\\\
  Unadjusted &0&2.6\%&2.5\%&2.4\%&&&&&&&&\\
  Cov. adj. (1) &1&2.6\% & 2.5\% & 2.4\%&&&&&&&&\\
  Cov. adj. (2) &1&2.6\% & 2.6\% & 2.4\%&&&&&&&&\\
  Cov. adj. (3) &1&2.6\% & 2.5\% & 2.4\%&&&&&&&&\\
  Cov. adj. (4) &$\leq 3$&2.6\%&2.5\%&2.5\%&&2.6\%&2.5\%&2.5\%&&2.9\%&2.5\%&2.5\%\\
  Cov. adj. (5) &$\leq 7$&2.7\%&2.4\%&2.5\%&&2.7\%&2.5\%&2.5\%&&2.8\%&2.5\%&2.5\%\\
  Cov. adj. (6) &$\leq 16$&2.7\%&2.5\%&2.5\%&&3.0\%&2.6\%&2.5\%&&2.8\%&2.6\%&2.5\%\\
  Cov. adj. (7) &$\leq 3$&2.6\%&2.5\%&2.5\%&&2.6\%&2.5\%&2.5\%&&2.9\%&2.5\%&2.5\%\\
  Cov. adj. (8) &$\leq 7$&2.6\%&2.4\%&2.5\%&&2.7\%&2.5\%&2.5\%&&2.8\%&2.5\%&2.5\%\\
  Cov. adj. (9) &$\leq 16$&2.6\%&2.5\%&2.5\%&&3.0\%&2.6\%&2.5\%&&2.8\%&2.6\%&2.5\%\\\\
  \multicolumn{13}{c}{\textbf{Power}}\\\\
  Unadjusted &0&11\%&22\%&38\%&&&&&&&&\\
  Cov. adj. (1) &1&12\% & 23\% & 41\%&&&&&&&&\\
  Cov. adj. (2) &1&12\% & 23\% & 40\%&&&&&&&&\\
  Cov. adj. (3) &1&12\% & 23\% & 40\%&&&&&&&&\\
  Cov. adj. (4) &$\leq 3$&14\%&28\%&49\%&&14\%&28\%&49\%&&14\%&28\%&49\%\\
  Cov. adj. (5) &$\leq 7$&14\%&28\%&49\%&&14\%&28\%&49\%&&14\%&28\%&49\%\\
  Cov. adj. (6) &$\leq 16$&13\%&27\%&49\%&&15\%&28\%&49\%&&14\%&28\%&49\%\\
  Cov. adj. (7) &$\leq 3$&14\%&28\%&49\%&&14\%&28\%&49\%&&14\%&28\%&49\%\\
  Cov. adj. (8) &$\leq 7$&14\%&28\%&49\%&&14\%&28\%&49\%&&14\%&28\%&49\%\\
  Cov. adj. (9) &$\leq 16$&13\%&27\%&48\%&&15\%&28\%&49\%&&14\%&28\%&49\%\\\\
   \hline
\end{tabular}\\
	 {\raggedright  \par}
\end{table}

The empirical Type I error and power for the Wald-tests based on the different estimators are summarized in Table \ref{tab:MISTIE_Type1_Power}. At large sample sizes, all estimators yield correct Type I error. For small sample sizes, we observe a Type I error rate that is (slightly) larger than the nominal level. This inflation is most pronounced for estimators (6) and (9), due to a relatively high number of parameters in the working models compared to the number of participants. In cases with small sample sizes, employing a more conservative small sample size correction, defined by $p_j$ as the number of parameters in the upper model, might be worthwhile for data-adaptive covariate adjustment estimators that do not use cross-fitting (see Table 4 in Appendix B). Covariate adjusted estimators -- with and without variable selection -- generally lead to a larger power compared to an unadjusted estimator.

Without a small sample size correction (see Table 3 in Appendix B), similar results are obtained albeit with a higher Type I error (inflation) for small samples (i.e., of 200 participants). At large sample sizes all estimators, except the general covariate adjustment estimators (6) and (9), yield (approximately) correct Type I error. 
These estimators, which use working models including main effects of all baseline covariates (without and with model misspecification), have a Type I error rate that is slightly larger than the nominal level. This is a consequence of a relatively high number of parameters in the working models compared to the number of participants. In such settings, data-adaptive methods such as the one discussed in this paper can be used to select the most prognostic covariates. 
Here, `Data-adaptive covariate adjustment (6)' with and without cross-fitting both start from the same (correctly specified) working models which include main effects of all baseline covariates, but lead to more acceptable Type I error rates at all sample sizes compared to the general covariate adjustment (i.e., standardization) estimator including main effects of all baseline covariates (i.e., `General covariate adjustment estimator (6)'). This is a consequence of the fact that only the (most) prognostic covariates are selected. We see similar results for the data-adaptive covariate adjustment methods that start from a misspecified model (i.e., `Data-adaptive covariate adjustment (9)' with and without cross-fitting'). This also confirms our theoretical results. 
Although cross-fitting is not strictly necessary (see Section \ref{sec:data-adaptive}) because the number of covariates with a non-null effect is low compared to the sample size (i.e., ultra-sparsity holds), it seems to lessen the degree of Type I error inflation when the number of covariates is large compared to the sample size when no small sample size correction is used; see for example data-adaptive covariate adjustment method (6) and (9).

Additional simulation results, investigating the finite sample bias of the different estimators, can be found in Table 5 in Appendix B. As expected, they show that the finite sample bias is negligible for all considered estimators, especially for the data-adaptive covariate adjustment approach using cross-fitting (with and without using the true propensity score).
In Table 6 and Table 7 in Appendix B, we show that the Wald tests based on the covariate adjustment estimates (general and data-adaptive) and the conditional treatment effect estimates based on the underlying generalized linear models (i.e., logistic regression models in our case) both control the Type I error rate and are equally powerful \citep{rosenblum2016matching}. Note that the outcome working models are fitted on all patients (not on treatment arms separately), and include an indicator for treatment in addition to the (selected) baseline covariates. Nevertheless, enhanced covariate adjustment estimators (e.g., by fitting separate outcome working models) have the potential for greater efficiency gains. 
Although randomization often allows valid inference for the treatment effect despite variable selection in the outcome model, this is not universally guaranteed \citep{belloni2012sparse}. 
For example, in trials with time-to-event endpoints, variable selection is not only required for the outcome model but also for the censoring model \citep{van2021principled}.

\section{Enhancing precision: Estimating propensity scores parametrically}\label{sec:par_ps}
Up to this point, we have estimated the propensity score $\pi=P(Z=1|X)$ as the (marginal) proportion of treated individuals, reflecting the assumption of simple randomization. Even so, under misspecification of the outcome model, more efficiency can be attained by estimating the propensity score as a function of baseline covariates — even when it is constant by design. This becomes essential when the randomization scheme itself depends on baseline covariates.
The results from the previous sections remain valid in that case, so long as a parametric propensity score model is used (e.g., logistic regression). 

Consider $X_{PS}\subseteq X$, which is pre-defined at the design stage and does not grow with $n$. Let us define $P(Z=1|X_{PS})=p(X_{PS}; \boldsymbol\beta_0)$, where $p(X_{PS}; \boldsymbol\beta)$ represents a known function of an unknown parameter $\boldsymbol\beta$, the population value of which is denoted as $\boldsymbol\beta_0$. Let $\hat{\boldsymbol\beta}$ be an consistent estimator of $\boldsymbol\beta_0$ under this model.
The data-adaptive covariate-adjustment approach using canonical generalized linear models (see Section \ref{sec:data-adaptive}) can then be extended by retaining all steps (i.e., Step 1a, Step 2 and Step 3), but replacing Step 1b by:
\begin{itemize}[left=2em]
	\item[\textbf{Step 1b:}]\textbf{Model fitting} Fit a generalized linear regression model with a canonical link (e.g., logistic, linear, \dots) and intercept via weighted maximum likelihood that regresses the outcome $Y$ on all covariates selected in Step 1a among the treated participants (i.e., with $Z=1$), with weights $1/p(X_{PS}; \hat{\boldsymbol\beta})$. 
\end{itemize}
A similar approach can be followed to estimate $\mu_0$, but now using weights $1/\{1-p(X_{PS}; \hat{\boldsymbol\beta})\}$ in Step 1b. As we do not allow variable or model selection for the propensity score, we need to pre-specify the covariates that will be adjusted for along with the functional form of the propensity score model. We then have the following result:
\begin{theorem}[Asymptotic results for $\hat\theta_{DA}$ with parametrically estimated $\pi$]\label{thm:asymptotic_se_paramaetricPS}
Assume treatment is assigned completely at random and that the outcome regression models used in Step~1b of the data-adaptive procedure are canonical generalized linear models fitted by weighted maximum likelihood (using inverse probability of treatment weights $1/p(X_{PS}; \hat{\boldsymbol\beta})$ and $1/(1-p(X_{PS}; \hat{\boldsymbol\beta}))$), which we assume to converge to some (possibly misspecified) limit. We moreover assume ultra-sparsity on the candidate covariate set (see Appendix~A.1). Then, even if the working outcome models are misspecified and covariates in these models are selected via a data-adaptive rule from a high-dimensional set, $\sqrt{n}\left(\hat\theta_{DA}-\theta\right)$ is asymptotically normal with variance consistently estimated by the sample variance of the values
\begin{align}\label{eq:var_parPS}
    \frac{Z_i}{p(X_{PS,i}; \hat{\boldsymbol\beta})}(Y_i-\hat{Y}_{1,i})+\hat{Y}_{1,i} - \left[\frac{1-Z_i}{1-p(X_{PS,i}; \hat{\boldsymbol\beta})}(Y_i-\hat{Y}_{0,i})+\hat{Y}_{0,i}\right],
\end{align}
where $\hat{Y}_{z,i}$ are the predicted values obtained from the data-adaptive procedure --using weighted maximum likelihood-- described above. 
\end{theorem}
A proof is provided in Appendix A.5.2. The same result also applies to other estimators such as CV-TMLE. A detailed discussion of the TMLE case is given in in Appendix A.5.3 and the Discussion section.
Accordingly, the augmented inverse probability weighting estimator using cross-fitting in Section \ref{sec:samplesplitting} can be adapted by replacing $\hat\pi_k$ by $p(X_{PS}; \hat{\boldsymbol\beta}_k)$ in Equation \eqref{eq:IF_aipw_samplesplitting}. The corresponding distribution (with a proof) is provided in Appendix A.5.1. As in Theorem \ref{thm:asymptotic_seCF}, it is necessary to account for the extra variability that arises when the (weighted) prediction-unbiasedness condition does not hold.

\section{Discussion}\label{sec:discussion}
It is imperative that the analysis of randomized experiments is based on statistical procedures that are completely pre-specified \citep{FDA2021,EMA2015}. 
This requirement is nonetheless difficult to reconcile with the need of most statistical procedures to be based on correctly specified models. 
This has kept trialists away from considering extensive covariate adjustment beyond simple adjustment for a handful of discrete covariates, in turn leading trial analyses to have suboptimal power. 
Given the rapidly growing interest in the use of data-adaptive methods in statistical inference in general \citep{van2011targeted, belloni2014inference, chernozhukov2018double, VanLancker2021Ensuring, van2021principled}, we wanted to address the conditions under which these methods can be safely used to estimate (marginal) treatment effects in clinical trials. 
In particular, we have shown that (a generalization of) the covariate-adjusted estimator suggested in the recent U.S. Food and Drug Administration guidance on covariate adjustment \citep{FDA2021} can be combined with data-adaptive methods. While it is well established that such estimators yield unbiased treatment effect estimates under randomization, it has been less clear whether they also enable unbiased inference when model selection is data-adaptive—a question we specifically address in this work.  

The independence between treatment and baseline covariates renders these covariate-adjusted estimators immune to model misspecification as well as variable selection uncertainty in randomized clinical trials. It moreover ensures that exact asymptotic coverage can be obtained, 
based on simple analytic expressions,  
despite these complications.
This is attractive because it enables the use of arbitrarily complex models (e.g., including splines) or (machine learning) algorithms to be used without compromising the validity of statistical inference for the treatment effect. This remains true even when the outcome predictions converge arbitrarily slowly or to the wrong limit,  highlighting the adaptability of this framework. This flexibility ensures that the resulting analysis is purely evidence-based and allows covariate information to be optimally used to gain power \citep{tsiatis2008covariate}. Our derivations relied on ultra-sparsity, which we could bypass either by limiting the number of predictors in the model or via cross-fitting (even for parametric models). 

By enabling the use of data-adaptive analysis strategies, while at the same time guaranteeing a valid inference that is insulated against model misspecification, the approaches outlined in this article offer the possibility to incorporate covariate adjustment in an automated manner and on a large scale. They simplify the difficulty of pre-specifying the statistical analysis to pre-specification of the full covariate set, along with the algorithm for variable selection / model building that will be employed. This form of pre-specification is what we should strive for in the future. Note that without such pre-specification of the algorithm, there is a general danger for data fishing, e.g., conducting multiple analyses and reporting only the one with largest estimated effect, which can lead to bias and / or inflated $p$-values. 

Our considered estimators are closely related to the Targeted Maximum Likelihood Estimation (TMLE; see \cite{rosenblum2010simple} and Appendix A.4) framework, which accommodates non-canonical generalized linear models or machine learning in Step 1a of our procedure (see Section~\ref{sec:data-adaptive}). In Step 1b, TMLE implements a targeting step to satisfy the prediction unbiasedness condition \eqref{eq:predictionUnbiasedness}. This is an alternative correction to obtain prediction unbiasedness compared to the AIPW estimator, which explicitly adjusts for potential bias in the outcome model; in large samples, AIPW and TMLE are equivalent, though TMLE can additionally ensure that the final estimate respects natural parameter constraints (e.g., probabilities remain in $[0,1]$).
When machine learning is used, cross-fitting is required to avoid overfitting, leading to the CV-TMLE procedure.
In Appendix A.4, we show that the standard errors of Section \ref{sec:Characteristics} 
are asymptotically unbiased for the TMLE (with parametric models), under the same weak conditions as discussed in that section. This has been conjectured before \citep{moore2009covariate} and recently been shown for CV-TMLE (w.r.t. ordinal and time-to-event outcomes) by \citet{williams2021}, but to the best of our knowledge has not been formally shown in full generality such as for parametrically estimated propensity scores (see Section \ref{sec:par_ps} and Appendix A.5.3). 
Remarkably, CV-TMLE no longer has the attractive finite-sample-unbiasedness property (when using the known randomization probability $\pi$) due to the targeting step (over all folds). Nevertheless, we still expect at most small finite sample bias as any potential bias is solely attributed to the estimation of the randomization probabilities and the targeting step which only involves the estimation of an intercept. Full formulas and implementation details for TMLE and CV-TMLE, including the targeting step and cross-fitting, are provided in Appendix~A.4; TMLE is included primarily to illustrate an alternative correction for prediction unbiasedness with flexible machine learning, while this main manuscript focuses on conceptual properties for readability.

Due to the concern of underestimated standard errors for large-sample methods such as standardization, we suggested multiplying the variance estimators by the small-sample correction factor proposed by \citet{tsiatis2008covariate}. We applied this correction in our simulation study. However, further research is needed to determine how to best account for degrees of freedom when variable selection or machine learning methods are used, as standard approximations may not apply in these settings \citep[see e.g.,][]{zou2007degrees}.

Finally, a drawback of covariate-adjusted methods is that they may be prone to additional missing (baseline covariate) data in the statistical analysis.
However, if covariates have missing values, this does not necessarily pose problems for the analysis as it is justified to adopt missing indicator and mean imputation procedures \citep{zhao2024adjust}.
These imputed covariates, along with the missing indicators, can be considered as a separate set of baseline covariates. As our procedures require treatment assignment to be independent of all baseline covariates, including the imputed ones, it is crucial not to use any treatment or outcome information during the imputation process; in particular, this implies that multiple imputation methods that rely on such information are not appropriate in this context.

\section*{Funding}
Kelly Van Lancker is supported by Bijzonder Onderzoeksfonds UGent Grant BOF.01P08419.

\section*{Data Availability}
The MISTIE III data that support the findings in this paper are available from the National Institute of Neurological Disorders and Stroke (\url{https://www.ninds.nih.gov/}). 
Software in the form of R code is available at The American Statistician website.

\newpage
\appendix
\section{Theoretical derivations}
\subsection{Assumptions}\label{app:assumptions}
We denote the regression coefficients obtained by the data-adaptive procedure in Section 3 as $\tilde{\boldsymbol\gamma}$ for estimation of $\mu_1$, and $\tilde{\boldsymbol\eta}$ for estimation of $\mu_0$. To allow for model misspecification, we introduce the limiting value of the estimated value $\tilde{\boldsymbol\gamma}$ as $\boldsymbol\gamma^*$ and the limiting value of the estimated value $\tilde{\boldsymbol\eta}$ as $\boldsymbol\eta^*$. 
Let $p$ denote the number of considered terms in the model (including possible interactions, transformations, higher order terms, \dots).
Define the sets of terms that have coefficients that are `truly' non-zero (i.e., active set) as $S_\gamma = \text{support}(\boldsymbol\gamma^*)$ and $S_\eta = \text{support}(\boldsymbol\eta^*)$, and define $s_\gamma=|S_\gamma|$ and $s_\eta=|S_\eta|$. 

\begin{assumption}[Ultra-sparsity]
	\label{as:ultraSparsity} 
	$$n^{-1/2}(s_\gamma)\log(p\lor n)=o(1)$$
    $$n^{-1/2}(s_\eta)\log(p\lor n)=o(1)$$
\end{assumption}

This assumption frequently finds application in the literature \citep[see e.g.,][]{farrell2015robust, belloni2014high}. Although in general this might be a strong assumption that may be difficult to validate in practice \citep{bradic2019sparsity}, in our setting this seems rather a weak assumption. The rationale behind this is our allowance for a misspecified outcome working model, enabling us to keep the number of covariates ($p$) limited. 

We will also make use of the following rate conditions: 

\begin{assumption}[Rates on error of estimated coefficients]\label{as:ratescoeff} 
	$$ $$ \vspace{-1cm}
	\begin{enumerate}
		\item $||\boldsymbol\gamma^*-\tilde{\boldsymbol\gamma}||_1=O_{P_n}\left(s_\gamma\sqrt{\log(p\lor n)/n}\right)$
		\item $||\boldsymbol\gamma^*-\tilde{\boldsymbol\gamma}||_2=O_{P_n}\left(\sqrt{s_\gamma\log(p\lor n)/n}\right)$
        \item $||\boldsymbol\eta^*-\tilde{\boldsymbol\eta}||_1=O_{P_n}\left(s_\eta\sqrt{\log(p\lor n)/n}\right)$
		\item $||\boldsymbol\eta^*-\tilde{\boldsymbol\eta}||_2=O_{P_n}\left(\sqrt{s_\eta\log(p\lor n)/n}\right)$
	\end{enumerate}
 where $\{P_n\}$ refers to a sequence of data-generating processes that obey, for each $n$, the assumptions above. 
	We refer the reader to \cite{belloni2013} and \cite{belloni2017program} who show that these rates hold for the Lasso estimator in the context of linear and logistic regression.
\end{assumption}

Besides the aforementioned sparsity conditions, we only need the following weak assumptions to ensure that the data-adaptive approach in Section 3 is asymptotically insensitive to selection mistakes (i.e., preserves Type I error). 
\begin{assumption}[Positivity assumption]
\label{as:positivity}
    $$1>1-\rho\geq P_n(Z=1)\geq\rho>0,$$ 
    with $\rho$ a constant.
\end{assumption}

\begin{assumption}[Convergence in probability of predictions]
\label{as:convergence}
$$ $$ \vspace{-1cm}

The imputations $\hat{Y}_{1,i}=h_1(X_i; \tilde{\boldsymbol\gamma})$ and $\hat{Y}_{0,i}=h_0(X_i, \tilde{\boldsymbol\eta})$ converge in probability to (possibly misspecified) limits $h_1(X_i; \boldsymbol{\gamma}^*)$ and $h_0^*(X_i; \boldsymbol\eta^*)$, respectively, in the sense that the $L_2$-norms
    $$\sqrt{\frac{1}{n}\sum_{i=1}^n\left\{h_1(X_i; \boldsymbol\gamma^*)-h_1(X_i; \tilde{\boldsymbol\gamma})\right\}^2}=o_{P_n}(1),$$ and 
    $$\sqrt{\frac{1}{n}\sum_{i=1}^n\left\{h_0(X_i; \boldsymbol\eta^*)-h_0(X_i; \tilde{\boldsymbol\eta})\right\}^2}=o_{P_n}(1).$$
Here, $o_{P_n}(1)$ denotes a stochastic variable that converges in probability to zero as $n\rightarrow \infty$. 
\end{assumption}

We denote the imputations in Section 4, $\hat{Y}_{1,k,i}$ and $\hat{Y}_{0,k,i}$, which can be based on machine learning algorithms, as $\hat h_{1, k}(X_i)$ and $\hat h_{0, k}(X_i)$ (for $k=1,\dots, K$), respectively. The following assumption focuses on these imputations $\hat h_{1, k}(X_i)$ and $\hat h_{0, k}(X_i)$.
\begin{assumptionp}{\ref*{as:convergence}$'$ (Convergence in probability of predictions for cross-fitting)}\label{as:convergenceSampleSplitting}
$$ $$ \vspace{-1cm}

    The imputations  $\hat h_{1, k}(X_i)$ and $\hat h_{0, k}(X_i)$ (for $k=1,\dots, K$) converge in probability to (possibly misspecified) limits $h_{1}^*(X_i)$ and $h_{0}^*(X_i)$, respectively, in the sense that the $L_2$-norms
    $$\sqrt{\frac{1}{|\mathcal{V}_k |}\sum_{i\in\mathcal{V}_k}\left\{h_{1}^*(X_i)-\hat h_{1,k}(X_i)\right\}^2}$$ and 
    $$\sqrt{\frac{1}{|\mathcal{V}_k |}\sum_{i\in\mathcal{V}_k}\left\{h_{0}^*(X_i)-\hat h_{0,k}(X_i)\right\}^2}$$
    converge in probability to zero.
\end{assumptionp}

\begin{assumption}[Uniform integrability without cross-fitting]
\label{as:UniformlyIntegrable}
$$\frac{1}{n}\sum_{i=1}^{n}\left\{h^*_1(X_i)-\hat{h}_1(X_i)\right\}^2$$ is uniformly integrable.\\ A collection of random variables $\{G_n:n\geq n_0>0\}$ with $G_n=\frac{1}{|\mathcal{V}_k|}\sum_{i=\in \mathcal{V}_k}\left\{h_1^*(X_i)-\hat{h}_{1}(X_i)\right\}^2$ is uniformly integrable if $sup_{n\geq n_0}E\left\{|G_n|I(|G_n|>g)\right\}=0$, as $g\rightarrow \infty$.  Here, $n_0$ can be seen as a minimal (total) sample size. 
\end{assumption}
We make a similar assumption for the imputations $\hat h_{1, k}(X_i)$ and $\hat h_{0, k}(X_i)$:
\begin{assumptionp}{\ref*{as:UniformlyIntegrable}$'$ (Uniform integrability with cross-fitting)}\label{as:UniformlyIntegrableSampleSplitting}
    $$\frac{1}{|\mathcal{V}_k|}\sum_{i\in \mathcal{V}_k}\left\{h_1^*(X_i)-\hat{h}_{1,k}(X_i)\right\}^2$$ is uniformly integrable.
\end{assumptionp}

\begin{assumption}[Concentration of gradients]\label{ass:concGradients}
$$ $$ \vspace{-1cm}

For the weighted data-adaptive covariate-adjustment approach using canonical generalized linear models in Section 6, using post-Lasso or stepwise regression for $h_{1}(X_i;\hat{\boldsymbol\gamma})$ and 
$h_{0}(X_i;\hat{\boldsymbol\eta})$, 

    $$\left|\left|\frac{1}{n}\sum_{i=1}^nZ_i\left\{Y_i-h_{1}(X_i;\boldsymbol\gamma^* )\right\}\left(\begin{array}{c c}
    1&
    X_i'\end{array}\right)\right|\right|_\infty=O_{P_n}\left(\sqrt{\log(p\lor n)/n}\right)$$ and
    $$\left|\left|\frac{1}{n}\sum_{i=1}^n(1-Z_i)\left\{Y_i-h_{0}(X_i;\boldsymbol\eta^* )\right\}
    \left(\begin{array}{c c}
    1&
    X_i'\end{array}\right)
    \right|\right|_\infty=O_{P_n}\left(\sqrt{\log(p\lor n)/n}\right)$$
\end{assumption}
We refer the reader to \citet{Ning2017general}. 

In the next section, we first explain why the data-adaptive approach with cross-fitting/sample splitting in Section 4 gives exact asymptotic coverage (i.e., asymptotically exact inference). Only in Appendix \ref{app:withoutsamplesplitting}, we will give a proof to show that we can avoid cross-fitting when a parametric outcome regression is fitted, possibly data-adaptive (using consistent variable selection, regardless of the loss function). 

\subsection{Theorem 4.1: Data-adaptive covariate adjustment with cross-fitting}\label{app:sampleSplitting}
In this appendix we first show that the estimator in fold $k$ (similar to Equation (3) in the main paper, but only focusing on fold $k$ for simplicity),
$$
\hat\theta_k = \frac{1}{|\mathcal{V}_k|}\sum_{i\in \mathcal{V}_k}\frac{Z_i}{\hat{\pi}_k}\left\{Y_i-\hat h_{1,k}(X_i)\right\}+\hat h_{1,k}(X_i)-\left[\frac{1-Z_i}{1-\hat{\pi}_k}\left\{Y_i-\hat h_{0,k}(X_i)\right\}+\hat h_{0,k}(X_i)\right],
$$
is asymptotically linear, assuming the predictions $\hat{h}_{1,k}(X_i)$ and $\hat{h}_{0,k}(X_i)$ are obtained from a separate sample. This can be accommodated by sample splitting or cross-fitting \citep{zheng2011cross, chernozhukov2018double}. 
In particular, for the participants belonging to a certain fold $k=1, \dots, K$, the predictions $\hat h_{1, k}(X)$ and $\hat h_{0, k}(X)$ are estimated based on the $K-1$ remaining folds, that is, $\{1, \dots, n\}\backslash \mathcal{V}_k$.

For simplicity, we focus on estimation of $\mu_1=E(Y|Z=1)$, which is estimated by $\hat\mu_{1,k} = \frac{1}{|\mathcal{V}_k|}\sum_{i\in \mathcal{V}_k}\frac{Z_i}{\hat{\pi}_k}\left\{Y_i-\hat h_{1,k}(X_i)\right\}+\hat h_{1,k}(X_i)$ in fold $k$.
We then have

\begin{align}
    \frac{1}{\sqrt{|\mathcal{V}_k|}}\sum_{i\in \mathcal{V}_k}\left[\frac{Z_i}{\hat{\pi}_k}\left\{Y_i-\hat h_{1,k}(X_i)\right\}+\hat h_{1,k}(X_i)\right] &= \frac{1}{\sqrt{|\mathcal{V}_k|}}\sum_{i\in \mathcal{V}_k}\left[\frac{Z_i}{\pi}\left\{Y_i-\hat h_{1,k}(X_i)\right\}+\hat h_{1,k}(X_i)\right]\label{eq:term1}\\
&+\frac{1}{|\mathcal{V}_k|}\sum_{i\in \mathcal{V}_k}Z_i\left\{Y_i-\hat h_{1,k}(X_i)\right\}\left(\frac{1}{\hat{\pi}_k}-\frac{1}{\pi}\right)\sqrt{|\mathcal{V}_k|}.\label{eq:term2}
\end{align}

For term \eqref{eq:term2}, we have
\begin{align}
    \frac{1}{|\mathcal{V}_k|}\sum_{i\in \mathcal{V}_k}&Z_i\left\{Y_i-\hat h_{1,k}(X_i)\right\}\left(\frac{1}{\hat{\pi}_k}-\frac{1}{\pi}\right)\sqrt{|\mathcal{V}_k|} \nonumber\\
    &= \frac{1}{|\mathcal{V}_k|}\sum_{i\in \mathcal{V}_k}Z_i\left\{Y_i-\hat h_{1,k}(X_i)+h_{1}^*(X_i)-h_{1}^*(X_i)\right\}\left(\frac{1}{\hat{\pi}_k}-\frac{1}{\pi}\right)\sqrt{|\mathcal{V}_k|}\nonumber\\
    &= \frac{1}{|\mathcal{V}_k|}\sum_{i\in \mathcal{V}_k}Z_i\left[Y_i-h_{1}^*(X_i)+\left\{h_{1}^*(X_i)-\hat h_{1,k}(X_i)\right\}\right]\left(\frac{1}{\hat{\pi}_k}-\frac{1}{\pi}\right)\sqrt{|\mathcal{V}_k|}\nonumber\\
    &= \frac{1}{|\mathcal{V}_k|}\sum_{i\in \mathcal{V}_k}Z_i\left\{Y_i-h_{1}^*(X_i)\right\}\left(\frac{1}{\hat{\pi}_k}-\frac{1}{\pi}\right)\sqrt{|\mathcal{V}_k|}\label{eq:term2.1}\\
    &+ \frac{1}{|\mathcal{V}_k|}\sum_{i\in \mathcal{V}_k}Z_i\left\{h_{1}^*(X_i)-\hat h_{1,k}(X_i)\right\}\left(\frac{1}{\hat{\pi}_k}-\frac{1}{\pi}\right)\sqrt{|\mathcal{V}_k|}.\label{eq:term2.2}
\end{align}

Then, following a Taylor expansion for term \eqref{eq:term2.1} with respect to $\pi$, we have 
\begin{align*}
   \frac{1}{|\mathcal{V}_k|}\sum_{i\in \mathcal{V}_k}Z_i\left\{Y_i-h_{1}^*(X_i)\right\}\left(\frac{1}{\hat{\pi}_k}-\frac{1}{\pi}\right)\sqrt{|\mathcal{V}_k|} = -\frac{1}{\pi^2}(\hat \pi_k - \pi)\frac{1}{|\mathcal{V}_k|}\sum_{i\in \mathcal{V}_k}Z_i\left\{Y_i-h_1^*(X_i)\right\}\sqrt{|\mathcal{V}_k|}+o_{P_n}(1),
\end{align*}
where the $o_{P_n}(1)$ term follows from the fact that $|\hat \pi_k-\pi|^2=O_{P_n}(n^{-1})$. Note that we use ``$n$'' instead of $|\mathcal{V}_k|$, which is justified as both quantities are proportional. By the weak law of large numbers, this equals 
\begin{align}
    \sqrt{|\mathcal{V}_k|}\left\{-\frac{1}{\pi^2}(\hat \pi_k - \pi)\right\}E[Z\left\{Y-h_1^*(X)\right\}]+o_{P_n}(1).\label{eq:term2.1b}
\end{align}
Here, the first term is $O_{P_n}(1)$ when $E[Z\left\{Y-h_1^*(X)\right\}]$ is bounded as $|\hat \pi_k-\pi|=O_{P_n}(n^{-1/2})$. The boundedness of $E[Z\left\{Y-h_1^*(X)\right\}]$ is often generally satisfied; e.g., for learners that provide predictions in range of the data. 

Next, applying a Taylor expansion as for term \eqref{eq:term2.1}, term \eqref{eq:term2.2} yields
\begin{align*}
    \sqrt{|\mathcal{V}_k|}&\left|\frac{1}{|\mathcal{V}_k|}\sum_{i\in \mathcal{V}_k}Z_i\left\{h_1^*(X_i)-\hat h_{1,k}(X_i)\right\}\left(\frac{1}{\hat \pi_k}-\frac{1}{\pi}\right)\right|\\
    &\leq \sqrt{|\mathcal{V}_k|}\rho^{-2}\left|\frac{1}{|\mathcal{V}_k|}\sum_{i\in \mathcal{V}_k}\left[Z_i\left\{h_1^*(X_i)-\hat h_{1,k}(X_i)\right\}\right](\hat \pi_k-\pi)\right|+o_{P_n}(1)\\
    &\leq \sqrt{|\mathcal{V}_k|}\rho^{-2}\sqrt{\frac{1}{|\mathcal{V}_k|}\sum_{i\in \mathcal{V}_k}\left[Z_i\left\{h_1^*(X_i)-\hat h_{1,k}(X_i)\right\}\right]^2}|\hat \pi_k-\pi|+o_{P_n}(1)\\
    &\leq \sqrt{|\mathcal{V}_k|}\rho^{-2}\sqrt{\frac{1}{|\mathcal{V}_k|}\sum_{i\in \mathcal{V}_k}\left\{h_1^*(X_i)-\hat h_{1,k}(X_i)\right\}^2}|\hat \pi_k-\pi|+o_{P_n}(1)\\
    &=\sqrt{|\mathcal{V}_k|}\rho^{-2} o_{P_n}(1)O_{P_n}(n^{-1/2})+o_{P_n}(1)\\
    &=o_{P_n}(1),
\end{align*}
where the second inequality follows from the Cauchy-Schwarz inequality, the third inequality from the fact that $Z$ is bounded (by 1) and the first equality from Assumption \ref{as:convergenceSampleSplitting}. We conjecture that this term can probably be handled under weaker assumptions.

In order to handle term \eqref{eq:term1}, we define $\hat\mu_{1,k}^{(\pi)}=\frac{1}{|\mathcal{V}_k|}\sum_{i\in \mathcal{V}_k}\left[\frac{Z_i}{\pi}\left\{Y_i-\hat h_{1,k}(X_i)\right\}+\hat h_{1,k}(X_i)\right]$.
Then, it holds that
\begin{align}
\sqrt{|\mathcal{V}_k|}(\hat\mu_{1,k}^{(\pi)}-\mu_1)&=\frac{1}{\sqrt{|\mathcal{V}_k|}}\sum_{i\in \mathcal{V}_k}\left[\frac{Z_i}{\pi}\left\{Y_i-h_1^*(X_i)\right\}+h_1^*(X_i)-\mu_1\right]\nonumber\\
&+\frac{1}{\sqrt{|\mathcal{V}_k|}}\sum_{i\in \mathcal{V}_k}\left[\left(\frac{Z_i}{\pi}-1\right)\left\{h_1^*(X_i)-\hat h_{1,k}(X_i)\right\}\right].\label{eq:term1.2}
\end{align}

Assuming the predictions $\hat h_{1, k}(X_i)$ for fold $k$ are obtained from a ``separate'' sample $\mathbf{O}^{\bar k}=\{\mathbf{O}_j: j\in \{1, \dots, n\}\backslash \mathcal{V}_k\}$ (i.e., independent from the sample $\mathbf{O}^{k}=\{\mathbf{O}_j: j\in \mathcal{V}_k\}$ which is used for calculation of $\hat\mu_{1,k}$), we can reason conditional on the sample $\mathbf{O}^{\bar k}$ to handle term \eqref{eq:term1.2}. First, note that 
term \eqref{eq:term1.2} has mean zero as
\begin{align*}
    E\left(\frac{1}{|\mathcal{V}_k|}\sum_{i\in \mathcal{V}_k}\left[\left(\frac{Z_i}{\pi}-1\right)\left\{h_1^*(X_i)-\hat{h}_{1,k}(X_i)\right\}\right]\right)
    &=E\left\{E\left(\frac{1}{|\mathcal{V}_k|}\sum_{i\in \mathcal{V}_k}\left[\left(\frac{Z_i}{\pi}-1\right)\left\{h_1^*(X_i)-\hat{h}_{1,k}(X_i)\right\}\right]\bigg\vert\mathbf{O}^{\bar k}\right)\right\}\\
    &=E\left(E\left[\left(\frac{Z}{\pi}-1\right)\left\{h_1^*(X)-\hat{h}_{1,k}(X)\right\}\bigg\vert\mathbf{O}^{\bar k}\right]\right)\\
    &=E\left(\frac{Z}{\pi}-1\right)E\left[E\left\{h_1^*(X)-\hat{h}_{1,k}(X)\bigg\vert\mathbf{O}^{\bar k}\right\}\right]\\
    &=0,
\end{align*}
where the third equation follows from the randomization (i.e., $Z$ being independent of $X$) and the last equation from $E\left(\frac{Z}{\pi}-1\right)=0$.
By the law of total variance, it follows from the above that
\begin{align*}
Var&\left(\frac{1}{|\mathcal{V}_k|}\sum_{i\in \mathcal{V}_k}\left[\left(\frac{Z_i}{\pi}-1\right)\left\{h_1^*(X_i)-\hat{h}_{1,k}(X_i)\right\}\right]\right)\\
    &=
    E\left\{Var\left(\frac{1}{|\mathcal{V}_k|}\sum_{i\in \mathcal{V}_k}\left[\left(\frac{Z_i}{\pi}-1\right)\left\{h_1^*(X_i)-\hat{h}_{1,k}(X_i)\right\}\right]\bigg\vert\mathbf{O}^{\bar k}\right)\right\}\\
    &+Var\left\{E\left(\frac{1}{|\mathcal{V}_k|}\sum_{i\in \mathcal{V}_k}\left[\left(\frac{Z_i}{\pi}-1\right)\left\{h_1^*(X_i)-\hat{h}_{1,k}(X_i)\right\}\right]\bigg\vert\mathbf{O}^{\bar k}\right)\right\}\\
    &=
    E\left\{Var\left(\frac{1}{|\mathcal{V}_k|}\sum_{i\in \mathcal{V}_k}\left[\left(\frac{Z_i}{\pi}-1\right)\left\{h_1^*(X_i)-\hat{h}_{1,k}(X_i)\right\}\right]\bigg\vert\mathbf{O}^{\bar k}\right)\right\}.
\end{align*}
By randomization, this can be rewritten as
\begin{align*}
    &\frac{1}{|\mathcal{V}_k|^2}E\left(\sum_{i\in \mathcal{V}_k}Var\left[\left(\frac{Z_i}{\pi}-1\right)\left\{h_1^*(X_i)-\hat{h}_{1,k}(X_i)\right\}\bigg\vert\mathbf{O}^{\bar k}\right]\right)\\
    &=\frac{1}{|\mathcal{V}_k|^2}E\left(\sum_{i\in \mathcal{V}_k}E\left[\left(\frac{Z_i}{\pi}-1\right)^2\left\{h_1^*(X_i)-\hat{h}_{1,k}(X_i)\right\}^2\bigg\vert\mathbf{O}^{\bar k}\right]\right)\\
    &=\frac{1}{|\mathcal{V}_k|^2}E\left\{\left(\frac{Z}{\pi}-1\right)^2\right\}E\left(\sum_{i\in \mathcal{V}_k}E\left[\left\{h_1^*(X_i)-\hat{h}_{1,k}(X_i)\right\}^2\bigg\vert\mathbf{O}^{\bar k}\right]\right)\\
    &=\frac{1}{|\mathcal{V}_k|}E\left\{\left(\frac{Z}{\pi}-1\right)^2\right\}E\left[\frac{1}{|\mathcal{V}_k|}\sum_{i\in \mathcal{V}_k}\left\{h_1^*(X_i)-\hat{h}_{1,k}(X_i)\right\}^2\right]\\
\end{align*}
where $E\left\{\left(\frac{Z}{\pi}-1\right)^2\right\}$ is bounded and $E\left[\frac{1}{|\mathcal{V}_k|}\sum_{i\in \mathcal{V}_k}\left\{h_1^*(X_i)-\hat{h}_{1,k}(X_i)\right\}^2\right]$ converges to zero assuming that $\frac{1}{|\mathcal{V}_k|}\sum_{i\in \mathcal{V}_k}\left\{h_1^*(X_i)-\hat{h}_{1,k}(X_i)\right\}^2$ is uniformly integrable (Assumption \ref{as:UniformlyIntegrableSampleSplitting}).
The latter follows from Theorem 3.5 in \cite{Billingsley1999} which states that convergence in distribution (which is implied by convergence in probability) implies convergence of the means under extra regularity conditions, that is, under uniform integrability. Specifically,
assuming that $\frac{1}{|\mathcal{V}_k|}\sum_{i=\in \mathcal{V}_k}\left\{h_1^*(X_i)-\hat{h}_{1,k}(X_i)\right\}^2$ is uniformly integrable, it follows from Theorem 3.5 in \cite{Billingsley1999} that $\sqrt{\frac{1}{|\mathcal{V}_k |}\sum_{i\in\mathcal{V}_k}\left\{h_{1}^*(X_i)-\hat h_{1,k}(X_i)\right\}^2}=o_{P_n}(1)$ (see Assumption \ref{as:convergenceSampleSplitting}) implies that $E\left[\frac{1}{|\mathcal{V}_k|}\sum_{i\in \mathcal{V}_k}\left\{h_1^*(X_i)-\hat{h}_{1,k}(X_i)\right\}^2\right]$ converges to zero. 
Application of Chebyshev's inequality then shows that term \eqref{eq:term1.2} converges in probability to zero:
for any constant $l>0$,
$$P\left(\left|\frac{1}{\sqrt{|\mathcal{V}_k|}}\sum_{i\in \mathcal{V}_k}\left[\left(\frac{Z_i}{\pi}-1\right)\left\{h_1^*(X_i)-\hat h_{1,k}(X_i)\right\}\right]\right|\geq l\right)\leq \frac{E\left\{\left(\frac{Z}{\pi}-1\right)^2\right\}E\left[\frac{1}{|\mathcal{V}_k|}\sum_{i\in \mathcal{V}_k}\left\{h_1^*(X_i)-\hat{h}_{1,k}(X_i)\right\}^2\right]}{l^2},$$
where the right hand side converges to zero.

Consequently, 
\begin{align*}
\sqrt{|\mathcal{V}_k|}(\hat\mu_{1,k}^{(\pi)}-\mu_1)&=\frac{1}{\sqrt{|\mathcal{V}_k|}}\sum_{i\in \mathcal{V}_k}\left[\frac{Z_i}{\pi}\left\{Y_i-h_1^*(X_i)\right\}+h_1^*(X_i)-\mu_1\right]+o_{P_n}(1),
\end{align*}
and thus
\begin{align*}
    \frac{1}{\sqrt{|\mathcal{V}_k|}}\sum_{i\in \mathcal{V}_k}\left[\frac{Z_i}{\pi}\left\{Y_i-\hat h_{1,k}(X_i)\right\}+\hat h_{1,k}(X_i)\right]=\frac{1}{\sqrt{|\mathcal{V}_k|}}\sum_{i\in \mathcal{V}_k}\left[\frac{Z_i}{\pi}\left\{Y_i-h_1^*(X_i)\right\}+h_1^*(X_i)\right]+o_{P_n}(1).
\end{align*}

Finally, we have
\begin{align*}
    \frac{1}{\sqrt{|\mathcal{V}_k|}}&\sum_{i\in \mathcal{V}_k}\left[\frac{Z_i}{\hat \pi_k}\left\{Y_i-\hat h_{1,k}(X_i)\right\}+\hat h_{1,k}(X_i)\right]\\&=
    \frac{1}{\sqrt{|\mathcal{V}_k|}}\sum_{i\in \mathcal{V}_k}\left(\frac{Z_i}{\pi}\left\{Y_i-h_1^*(X_i)\right\}+h_1^*(X_i)-E\left[\frac{Z}{\pi^2}\left\{Y-h_1^*(X)\right\}\right](Z_i - \pi)\right)
    +o_{P_n}(1),
\end{align*}
which leads to
\begin{align}
\label{eq:influenceFunction1}
\begin{split}
    &\sqrt{|\mathcal{V}_k|}(\hat{\mu}_{1,k}-\mu_1)\\
    &=
    \frac{1}{\sqrt{|\mathcal{V}_k|}}\sum_{i\in \mathcal{V}_k}\left(\frac{Z_i}{\pi}\left\{Y_i-h_1^*(X_i)\right\}+h_1^*(X_i)-E\left[\frac{Z}{\pi^2}\left\{Y-h_1^*(X)\right\}\right](Z_i - \pi)-\mu_1\right)+o_{P_n}(1).
    \end{split}
\end{align}
A similar reasoning can be followed to prove that
\begin{align}
\label{eq:influenceFunction0}
\begin{split}
    &\sqrt{|\mathcal{V}_k|}(\hat{\mu}_{0,k}-\mu_0)\\
    &=
    \frac{1}{\sqrt{|\mathcal{V}_k|}}\sum_{i\in \mathcal{V}_k}\left(\frac{1-Z_i}{1-\pi}\left\{Y_i-h_0^*(X_i)\right\}+h_0^*(X_i)+E\left[\frac{1-Z}{(1-\pi)^2}\left\{Y-h_0^*(X)\right\}\right](Z_i - \pi)-\mu_0\right)+o_{P_n}(1),
    \end{split}
\end{align}
where $\hat{\mu}_{0,k} = \frac{1}{|\mathcal{V}_k|}\sum_{i\in \mathcal{V}_k}\frac{1-Z_i}{1-\hat{\pi}_k}\left\{Y_i-\hat h_{0,k}(X_i)\right\}+\hat h_{0,k}(X_i)$.

The final estimate $\hat\mu_{1, CF}$ for $\mu_1$ is then obtained by taking the mean over these $K$ estimates $\hat{\mu}_{1,1}, \dots, \hat{\mu}_{1,K}$. Then,
\begin{align*}
    \sqrt{n}(\hat\mu_{1, CF}-\mu_1)&=\sqrt{n}\left(\frac{1}{K}\sum_{k=1}^K\hat{\mu}_{1,k}-\mu_1\right)\\
    &=\sqrt{n}\frac{1}{K}\sum_{k=1}^K(\hat{\mu}_{1,k}-\mu_1)\\
    &=\sqrt{n}\frac{1}{K}\sum_{k=1}^K\frac{1}{\sqrt{|\mathcal{V}_k|}}\sqrt{|\mathcal{V}_k|}(\hat{\mu}_{1,k}-\mu_1)\\
    &=\sqrt{n}\frac{1}{K}\sum_{k=1}^K\frac{1}{\sqrt{|\mathcal{V}_k|}}\\
    &\hspace{0.25cm}\left\{\frac{1}{\sqrt{|\mathcal{V}_k|}}\sum_{i\in \mathcal{V}_k}\left(\frac{Z_i}{\pi}\left\{Y_i-h_1^*(X_i)\right\}+h_1^*(X_i)-E\left[\frac{Z}{\pi^2}\left\{Y-h_1^*(X)\right\}\right](Z_i - \pi)-\mu_1\right)+o_{P_n}(1)\right\}\\
    &=\sqrt{n}\frac{1}{K}\sum_{k=1}^K\frac{1}{|\mathcal{V}_k|}\sum_{i\in\mathcal{V}_k}\left(\frac{Z_i}{\pi}\left\{Y_i-h_1^*(X_i)\right\}+h_1^*(X_i)-E\left[\frac{Z}{\pi^2}\left\{Y-h_1^*(X)\right\}\right](Z_i - \pi)-\mu_1\right)\\
    &+o_{P_n}(1).
\end{align*}
Assuming that $n$ is a multiple of $K$, this can be simplified as
\begin{align*}
    \sqrt{n}(\hat\mu_{1, CF}-\mu_1)&=\frac{1}{\sqrt{n}}\sum_{k=1}^K\sum_{i\in\mathcal{V}_k}\left(\frac{Z_i}{\pi}\left\{Y_i-h_1^*(X_i)\right\}+h_1^*(X_i)-E\left[\frac{Z}{\pi^2}\left\{Y-h_1^*(X)\right\}\right](Z_i - \pi)-\mu_1\right)+o_{P_n}(1).
\end{align*}
Following a similar reasoning for $\hat\mu_{0, CF}$, we have (for the general case)
\begin{align*}
    \sqrt{n}(\hat\theta_{CF}-\theta)&=\sqrt{n}\frac{1}{K}\sum_{k=1}^K\frac{1}{|\mathcal{V}_k|}\sum_{i\in\mathcal{V}_k}\left\{\left(\frac{Z_i}{\pi}\left\{Y_i-h_1^*(X_i)\right\}+h_1^*(X_i)-E\left[\frac{Z}{\pi^2}\left\{Y-h_1^*(X)\right\}\right](Z_i - \pi)-\mu_1\right)\right.\\
    &-\left.\left(\frac{1-Z_i}{1-\pi}\left\{Y_i-h_0^*(X_i)\right\}+h_0^*(X_i)+E\left[\frac{1-Z}{(1-\pi)^2}\left\{Y-h_0^*(X)\right\}\right](Z_i - \pi)-\mu_0\right)\right\}+o_{P_n}(1).
\end{align*}

The asymptotic variance of $\hat\theta_{CF} = \hat\mu_{1, CF}-\hat\mu_{0, CF}$ can be calculated as $1/n$ times the sample variance of the values
\begin{align*}
    &\left\{\left(\frac{Z_i}{\hat\pi_{k(i)}}\left\{Y_i-\hat h_{1,k(i)}(X_i)\right\}+\hat h_{1,k(i)}(X_i)-\frac{1}{|\mathcal{V}_{k(i)}|}\sum_{j\in \mathcal{V}_{k(i)}}\left[\frac{Z_j}{\hat\pi_{k(i)}^2}\left\{Y_j-\hat h_{1,k(i)}(X_j)\right\}\right](Z_i - \hat\pi_{k(i)})\right)\right.\\
    &-\left.\left(\frac{1-Z_i}{1-\hat\pi_{k(i)}}\left\{Y_i-\hat h_{0,k(i)}(X_i)\right\}+\hat h_{0,k(i)}(X_i)+\frac{1}{|\mathcal{V}_{k(i)}|}\sum_{j\in \mathcal{V}_{k(i)}}\left[\frac{1-Z_j}{(1-\hat\pi_{k(i)})^2}\left\{Y_j-\hat h_{0,k(i)}(X_j)\right\}\right](Z_i - \hat\pi_{k(i)})\right)\right\},
\end{align*}
where $k(i)$ indicates to which fold participant $i$ belongs.
It may seem to the reader that one need to weight for the unequal fold sizes, but this has no asymptotic impact.

The uncertainty in the randomization probabilities is taken into account via the terms $$\frac{1}{|\mathcal{V}_{k(i)}|}\sum_{j\in \mathcal{V}_{k(i)}}\left[\frac{Z_j}{\hat\pi_{k(i)}^2}\left\{Y_j-\hat h_{1,k(i)}(X_j)\right\}\right](Z_i - \hat\pi_{k(i)}),$$ and $$\frac{1}{|\mathcal{V}_{k(i)}|}\sum_{j\in \mathcal{V}_{k(i)}}\left[\frac{1-Z_j}{(1-\hat\pi_{k(i)})^2}\left\{Y_j-\hat h_{0,k(i)}(X_j)\right\}\right](Z_i - \hat\pi_{k(i)}).$$ Note that we do not have such a term for the outcome working models as term \eqref{eq:term1.2} converges to zero (due to the randomization probability being a constant). In Appendix \ref{app:cv_tmle}, we prove that TMLE would not need to account for these terms. For time-to-event and ordinal outcomes, \citet{williams2021} have already shown this result, which can be extended straightforwardly to binary and continuous outcomes. In addition, when using the true randomization probability $\pi$ (see Appendix \ref{app:unbiasedness}), even the above error terms for the estimation of the randomization probability can be removed.

\subsubsection{Avoiding cross-fitting when testing the strong null (without additional assumptions)}\label{subsubapp:strongNull}

In this section, we show how cross-fitting can be avoided if interest lies solely in testing the (strong) null hypothesis, that is, $Z$ has no effect on $Y$ (i.e., $Z$ is independent of $Y$). 
As discussed in Section 4.2, we should then estimate $E(Y|Z=1, X)$ and $E(Y|Z=0, X)$ under the null hypothesis that $Z$ is jointly independent of $Y$ and $X$. Specifically, we assume $E(Y|Z=1, X)=E(Y|Z=0, X)=E(Y|X)$ by fitting one model $h(X)$ in all participants only including the covariates $X$ (and not the treatment indicator $Z$).

Under the assumption that $Z$ is independent of $Y$ and $X$, $\mu_1=E(Y|Z=1)$ can be estimated as $\hat\mu_{1, SN}=\frac{1}{n}\sum_{i=1}^{n}\frac{Z_i}{\hat\pi}\left\{Y_i-\hat h(X_i)\right\}+\hat h(X_i)$. Here, `SN' stands for `strong null'. Note that there are no restrictions -- besides blinding -- on how the predictions are obtained, meaning that they are potentially obtained by machine learning algorithms. Following the same reasoning as in Appendix \ref{app:sampleSplitting}, 
\begin{align}
    \frac{1}{\sqrt{n}}\sum_{i=1}^n\left[\frac{Z_i}{\hat{\pi}}\left\{Y_i-\hat h(X_i)\right\}+\hat h(X_i)\right] 
    &= 
    \frac{1}{\sqrt{n}}\sum_{i=1}^n\left[\frac{Z_i}{\pi}\left\{Y_i-\hat h(X_i)\right\}+\hat h(X_i)\right]\label{eq:sampleSplitting_term1}\\
&+\sqrt{n}\left\{-\frac{1}{\pi^2}(\hat \pi - \pi)\right\}E[Z\left\{Y-h^*(X)\right\}]\nonumber\\
&+o_{P_n}(1),\nonumber
\end{align}
where $h^*(X)$ is the probability limit of $\hat h(X)$. Then, in order to handle term \eqref{eq:sampleSplitting_term1}, we define $\hat\mu_{1, SN}^{(\pi)}=\frac{1}{n}\sum_{i=1}^n\left[\frac{Z_i}{\pi}\left\{Y_i-\hat h(X_i)\right\}+\hat h(X_i)\right]$.
Then, it holds that
\begin{align}
\sqrt{n}(\hat\mu_{1, SN}^{(\pi)}-\mu_1)&=\frac{1}{\sqrt{n}}\sum_{i=1}^n\left[\frac{Z_i}{\pi}\left\{Y_i-h^*(X_i)\right\}+h^*(X_i)-\mu_1\right]\nonumber\\
&+\frac{1}{\sqrt{n}}\sum_{i=1}^n\left[\left(\frac{Z_i}{\pi}-1\right)\left\{h^*(X_i)-\hat h(X_i)\right\}\right].\label{eq:sampleSplitting_term1.2}
\end{align}

Then, it holds for term \eqref{eq:sampleSplitting_term1.2} that 
\begin{align*}
    E\left(\frac{1}{n}\sum_{i=1}^{n}\left[\left(\frac{Z_i}{\pi}-1\right)\left\{h^*(X_i)-\hat{h}(X_i)\right\}\right]\right)
    &=E\left\{E\left(\frac{1}{n}\sum_{i=1}^{n}\left[\left(\frac{Z_i}{\pi}-1\right)\left\{h^*(X_i)-\hat{h}(X_i)\right\}\right]\bigg\vert X_1, \dots, X_n, Y_1, \dots, Y_n\right)\right\}\\
    &=E\left(\frac{1}{n}\sum_{i=1}^{n}\left[\left\{h^*(X_i)-\hat{h}(X_i)\right\}E\left(\frac{Z_i}{\pi}-1\bigg\vert X_1, \dots, X_n, Y_1, \dots, Y_n\right)\right]\right)\\
    &=E\left(\frac{1}{n}\sum_{i=1}^{n}\left[\left\{h^*(X_i)-\hat{h}(X_i)\right\}E\left(\frac{Z_i}{\pi}-1\right)\right]\right)\\
    &=0,
\end{align*}
where the third equality is a consequence of $Z$ being jointly independence of ($X$, $Y$) and the last equality follows from $E\left(\frac{Z}{\pi}-1\right)=0$.

Its variance equals 
{\footnotesize
\begin{align*}
    Var\left(\frac{1}{n}\sum_{i=1}^{n}\left[\left(\frac{Z_i}{\pi}-1\right)\left\{h^*(X_i)-\hat{h}(X_i)\right\}\right]\right)
    &=
    E\left\{Var\left(\frac{1}{n}\sum_{i=1}^{n}\left[\left(\frac{Z_i}{\pi}-1\right)\left\{h^*(X_i)-\hat{h}(X_i)\right\}\right]\bigg\vert X_1, \dots, X_n, Y_1, \dots, Y_n\right)\right\}\\
    &+Var\left\{E\left(\frac{1}{n}\sum_{i=1}^{n}\left[\left(\frac{Z_i}{\pi}-1\right)\left\{h^*(X_i)-\hat{h}(X_i)\right\}\right]\bigg\vert X_1, \dots, X_n, Y_1, \dots, Y_n\right)\right\}\\
    &=
    E\left\{Var\left(\frac{1}{n}\sum_{i=1}^{n}\left[\left(\frac{Z_i}{\pi}-1\right)\left\{h^*(X_i)-\hat{h}(X_i)\right\}\right]\bigg\vert X_1, \dots, X_n, Y_1, \dots, Y_n\right)\right\}\\
    &=
    \frac{1}{n^2}E\left(\sum_{i=1}^{n}Var\left[\left(\frac{Z_i}{\pi}-1\right)\left\{h^*(X_i)-\hat{h}(X_i)\right\}\bigg\vert X_1, \dots, X_n, Y_1, \dots, Y_n\right]\right)\\
    &=
    \frac{1}{n^2}E\left(\sum_{i=1}^{n}E\left[\left(\frac{Z_i}{\pi}-1\right)^2\left\{h^*(X_i)-\hat{h}(X_i)\right\}^2\bigg\vert X_1, \dots, X_n, Y_1, \dots, Y_n\right]\right)\\
    &=
    \frac{1}{n^2}E\left[\sum_{i=1}^{n}\left\{h^*(X_i)-\hat{h}(X_i)\right\}^2E\left\{\left(\frac{Z_i}{\pi}-1\right)^2\bigg\vert X_1, \dots, X_n, Y_1, \dots, Y_n\right\}\right]\\
    &=\frac{1}{n}E\left\{\left(\frac{Z}{\pi}-1\right)^2\right\}E\left[\frac{1}{n}\sum_{i=1}^n\left\{h^*(X_i)-\hat{h}(X_i)\right\}^2\right],
\end{align*}
}
where $E\left\{\left(\frac{Z}{\pi}-1\right)^2\right\}$ is bounded and $E\left[\frac{1}{n}\sum_{i=1}^n\left\{h^*(X_i)-\hat{h}(X_i)\right\}^2\right]$  converges to zero, assuming that $\frac{1}{n}\sum_{i=1}^{n}\left\{h^*(X_i)-\hat{h}(X_i)\right\}^2$ is uniformly integrable (see Assumption \ref{as:UniformlyIntegrable}), as $\sqrt{\frac{1}{n}\sum_{i=1}^{n}\left\{h^*(X_i)-\hat{h}(X_i)\right\}^2}=o_{P_n}(1)$. Therefore, application of Chebyshev's inequality shows that term \eqref{eq:sampleSplitting_term1.2} converges in probability to zero.
Consequently,
\begin{align*}
\sqrt{n}(\hat\mu_{1, SN}^{(\pi)}-\mu_1)&=\frac{1}{\sqrt{n}}\sum_{i=1}^n\left[\frac{Z_i}{\pi}\left\{Y_i-h^*(X_i)\right\}+h^*(X_i)-\mu_1\right]+o_{P_n}(1).
\end{align*}
Following a similar reasoning as in Appendix \ref{app:sampleSplitting}, we have
\begin{align*}
    \sqrt{n}(\hat\theta_{SN}-\theta)&=\frac{1}{\sqrt{n}}\sum_{i=1}^n\left\{\left(\frac{Z_i}{\pi}\left\{Y_i-h^*(X_i)\right\}+h^*(X_i)-E\left[\frac{Z}{\pi^2}\left\{Y-h^*(X)\right\}\right](Z_i - \pi)-\mu_1\right)\right.\\
    &-\left.\left(\frac{1-Z_i}{1-\pi}\left\{Y_i-h^*(X_i)\right\}+h^*(X_i)+E\left[\frac{1-Z}{(1-\pi)^2}\left\{Y-h^*(X)\right\}\right](Z_i - \pi)-\mu_0\right)\right\}+o_{P_n}(1),
\end{align*}
which can be simplified to
\begin{align*}
    \sqrt{n}(\hat\theta_{SN}-\theta)&=\frac{1}{\sqrt{n}}\sum_{i=1}^n\left(\frac{Z_i-\pi}{\pi(1-\pi)}\left\{Y_i-h^*(X_i)\right\}-E\left[\left(\frac{Z}{\pi^2}+\frac{1-Z}{(1-\pi)^2}\right)\left\{Y-h^*(X)\right\}\right](Z_i - \pi)-\left(\mu_1-\mu_0\right)\right)\\
    &+o_{P_n}(1).
\end{align*}
Inference can be conducted by calculating the standard error of $\hat\theta_{SN}$ as $1/n$ times the sample variance of
$$\frac{Z_i-\hat\pi}{\hat\pi(1-\hat\pi)}\left\{Y_i-\hat{h}(X_i)\right\}-\frac{1}{n}\sum_{j=1}^n\left[\left(\frac{Z_j}{\hat\pi^2}+\frac{1-Z_j}{(1-\hat\pi)^2}\right)\left\{Y_j-\hat{h}(X_j)\right\}\right](Z_i - \hat\pi).$$

\subsubsection{Theorem 4.2: Finite sample unbiasedness}\label{app:unbiasedness}
We prove that $\hat\theta_{CF}$ is exactly unbiased, even in finite samples, when the the known randomization probability $\pi$ (e.g., 0.5 for a 1:1 randomized trial) is plugged-in for the propensity score. Thus we show that 
$$
\hat\theta_k = \frac{1}{|\mathcal{V}_k|}\sum_{i\in \mathcal{V}_k}\frac{Z_i}{\pi}\left\{Y_i-\hat h_{1,k}(X_i)\right\}+\hat h_{1,k}(X_i)-\left[\frac{1-Z_i}{1-\pi}\left\{Y_i-\hat h_{0,k}(X_i)\right\}+\hat h_{0,k}(X_i)\right],
$$
has mean $E\left\{Y(1)-Y(0)\right\}$ for each $k=1,\dots,K$ with $Y(j)$ denoting the counterfactual outcome under $Z=j$ for $j=0,1$. 
Following a similar proof as for term \eqref{eq:term1.2}, we have
\begin{align*}
    &E\left(\frac{1}{|\mathcal{V}_k|}\sum_{i\in \mathcal{V}_k}\frac{Z_i}{\pi}\left\{Y_i-\hat h_{1,k}(X_i)\right\}+\hat h_{1,k}(X_i)-\left[\frac{1-Z_i}{1-\pi}\left\{Y_i-\hat h_{0,k}(X_i)\right\}+\hat h_{0,k}(X_i)\right]\right)\\
    &=E\left[\frac{1}{|\mathcal{V}_k|}\sum_{i\in \mathcal{V}_k}\frac{Z_i}{\pi}Y_i-\frac{1-Z_i}{1-\pi}Y_i+\left\{1-\frac{Z_i}{\pi}\right\}\hat h_{1,k}(X_i)-\left\{1-\frac{1-Z_i}{1-\pi}\right\}\hat h_{0,k}(X_i)\right]\\
    &=E\left[\frac{1}{|\mathcal{V}_k|}\sum_{i\in \mathcal{V}_k}\frac{Z_i}{\pi}Y_i-\frac{1-Z_i}{1-\pi}Y_i\right]+E\left[\frac{1}{|\mathcal{V}_k|}\sum_{i\in \mathcal{V}_k}\left\{1-\frac{Z_i}{\pi}\right\}\hat h_{1,k}(X_i)-\left\{1-\frac{1-Z_i}{1-\pi}\right\}\hat h_{0,k}(X_i)\right]\\
    &= E\left[Y(1)-Y(0)\right]+E\left[\frac{1}{|\mathcal{V}_k|}\sum_{i\in \mathcal{V}_k}\left\{1-\frac{Z_i}{\pi}\right\}\hat h_{1,k}(X_i)-\left\{1-\frac{1-Z_i}{1-\pi}\right\}\hat h_{0,k}(X_i)\right],
\end{align*}
because of consistency and randomization. Relying on sample splitting, the second expectation can be rewritten as
\begin{align*}
    &E\left[\frac{1}{|\mathcal{V}_k|}\sum_{i\in \mathcal{V}_k}\left\{1-\frac{Z_i}{\pi}\right\}\hat h_{1,k}(X_i)-\left\{1-\frac{1-Z_i}{1-\pi}\right\}\hat h_{0,k}(X_i)\right]\\
    &=E\left(E\left[\frac{1}{|\mathcal{V}_k|}\sum_{i\in \mathcal{V}_k}\left\{1-\frac{Z_i}{\pi}\right\}\hat h_{1,k}(X_i)-\left\{1-\frac{1-Z_i}{1-\pi}\right\}\hat h_{0,k}(X_i)\bigg\vert\mathbf{O}^{\bar k}\right]\right)\\
    &=E\left(E\left[\left\{1-\frac{Z}{\pi}\right\}\hat h_{1,k}(X)-\left\{1-\frac{1-Z}{1-\pi}\right\}\hat h_{0,k}(X)\bigg\vert\mathbf{O}^{\bar k}\right]\right)\\
    &=E\left(1-\frac{Z}{\pi}\right)E\left[E\left\{\hat h_{1,k}(X)\bigg\vert\mathbf{O}^{\bar k}\right\}\right]-E\left(1-\frac{1-Z}{1-\pi}\right)E\left[E\left\{\hat h_{0,k}(X)\bigg\vert\mathbf{O}^{\bar k}\right\}\right]\\
    &=0,
\end{align*}
as $E\left(\frac{Z}{\pi}-1\right)=0$ and $E\left(1-\frac{1-Z}{1-\pi}\right)=0$ because of randomization.

Under the strong null, cross-fitting can even be avoided. Specifically, under the strong null, $\hat\theta_{SN}=\hat\mu_{1, SN}-\hat\mu_{0, SN}$ also has the advantage of being exactly unbiased, even in finite samples as long as the known randomization probability $\pi$ (e.g., 0.5 for a 1:1 randomized trial) is used. A proof for this result follows a similar reasoning as in Appendix \ref{subsubapp:strongNull} by using the independence of $Z$ and $\hat{h}(X)$.

\subsection{Theorem 3.1: Data-adaptive covariate adjustment without cross-fitting via parametric outcome models}\label{app:withoutsamplesplitting}

In this section we prove that we can avoid cross-fitting when a generalized linear model with canonical link for the outcome is fitted via maximum likelihood, possibly in a data-adaptive way (using consistent variable selection, regardless of loss function). We denote the regression coefficients obtained by the data-adaptive procedure in Section 3 as $\tilde{\boldsymbol\gamma}$ for estimation of $\mu_1$, and $\tilde{\boldsymbol\eta}$ for estimation of $\mu_0$.

The ``Model fitting'' step (Step 1b) in Section 3 ensures that
\begin{align}
        \frac{1}{\sqrt{n}}\sum_{i=1}^n\left[\frac{Z_i}{\hat{\pi}}\left\{Y_i- h_{1}(X_i;\tilde{\boldsymbol\gamma})\right\}+h_{1}(X_i;\tilde{\boldsymbol\gamma})\right] &= \frac{1}{\sqrt{n}}\sum_{i=1}^n\left[\frac{Z_i}{\pi}\left\{Y_i-h_{1}(X_i;\tilde{\boldsymbol\gamma})\right\}+h_{1}(X_i;\tilde{\boldsymbol\gamma})\right]\nonumber\\
&+\frac{1}{n}\sum_{i=1}^nZ_i\left\{Y_i-h_{1}(X_i;\tilde{\boldsymbol\gamma})\right\}\left(\frac{1}{\hat{\pi}}-\frac{1}{\pi}\right)\sqrt{n}\nonumber\\
&=\frac{1}{\sqrt{n}}\sum_{i=1}^n\left[\frac{Z_i}{\pi}\left\{Y_i-h_{1}(X_i;\tilde{\boldsymbol\gamma})\right\}+h_{1}(X_i;\tilde{\boldsymbol\gamma})\right],\label{eq:outcRegr_simple}
\end{align}
as $\frac{1}{n}\sum_{i=1}^nZ_i\left\{Y_i-h_{1}(X_i;\tilde{\boldsymbol\gamma})\right\}=0$ for generalized linear models with a canonical link function and intercept, fitted via maximum likelihood.

In order to handle term \eqref{eq:outcRegr_simple}, we define $\hat\mu_{1, DA}^{(\pi)}=\frac{1}{n}\sum_{i=1}^n\left[\frac{Z_i}{\pi}\left\{Y_i-h_{1}(X_i;\tilde{\boldsymbol\gamma})\right\}+h_{1}(X_i;\tilde{\boldsymbol\gamma})\right]$. Then, it holds that
\begin{align}
\sqrt{n}(\hat\mu_{1, DA}^{(\pi)}-\mu_1)&=\frac{1}{\sqrt{n}}\sum_{i=1}^n\left[\frac{Z_i}{\pi}\left\{Y_i-h_{1}(X_i;\boldsymbol\gamma^*)\right\}+h_{1}(X_i;\boldsymbol\gamma^*)-\mu_1\right]\nonumber\\
&+\frac{1}{\sqrt{n}}\sum_{i=1}^n\left[\left(\frac{Z_i}{\pi}-1\right)\left\{h_{1}(X_i;\boldsymbol\gamma^*)-h_{1}(X_i;\tilde{\boldsymbol\gamma})\right\}\right].\label{eq:outcRegr_pi}
\end{align}
Next, following a Taylor expansion, term \eqref{eq:outcRegr_pi} can be rewritten as
\begin{align*}
    \frac{1}{\sqrt{n}}\sum_{i=1}^n\left[\left(\frac{Z_i}{\pi}-1\right)\frac{\partial}{\partial \boldsymbol\gamma}h_1(X_i;\boldsymbol\gamma)\biggr\rvert_{\boldsymbol\gamma=\boldsymbol\gamma^*}\right](\tilde{\boldsymbol\gamma}-\boldsymbol\gamma^*)+O_{P_n}(\sqrt{n}||\boldsymbol\gamma^*-\tilde{\boldsymbol\gamma}||^2_2),
\end{align*}
assuming that $\max_{j\leq p,j'\leq p, i\leq n}\left|\frac{\partial^2}{\partial \gamma_j\partial \gamma_{j'}}h_1(X_i;\boldsymbol\gamma)\biggr\rvert_{(\gamma_j=\gamma^*_j, \gamma_{j'}=\gamma^*_{j'})}\right|$ is bounded, with $\gamma_j$ and $\gamma^*_{j}$ denoting component $j$ ($j\in 1, \dots, p$) of $\boldsymbol\gamma$ and $\boldsymbol\gamma^*$, respectively.

Define $\tilde\Sigma_j=\frac{1}{n}\sum_{i=1}^n\left[\left(\frac{Z_i}{\pi}-1\right)\left(\frac{\partial}{\partial \gamma_j}h_1(X_i;\boldsymbol\gamma)\biggr\rvert_{\gamma_j=\gamma^*_j}\right)\right]^2$. Then, by H\"older's inequality
\begin{align*}
    &\left|\frac{1}{\sqrt{n}}\sum_{i=1}^n\left[\left(\frac{Z_i}{\pi}-1\right)\frac{\partial}{\partial \boldsymbol\gamma}h_1(X_i;\boldsymbol\gamma)\biggr\rvert_{\boldsymbol\gamma=\boldsymbol\gamma^*}\right](\tilde{\boldsymbol\gamma}-\boldsymbol\gamma^*)\right|\\
    &\leq \left(\max_{j\leq p}\tilde\Sigma_j^{1/2}\right) \max_{j\leq p}\left|\frac{1}{\sqrt{n}}\sum_{i=1}^n\frac{\left(\frac{Z_i}{\pi}-1\right)\left(\frac{\partial}{\partial \gamma_j}h_1(X_i;\boldsymbol\gamma)\biggr\rvert_{\gamma_j=\gamma^*_j}\right)}{\tilde\Sigma_j^{1/2}}\right|\left|\left|\tilde{\boldsymbol\gamma}-\boldsymbol\gamma^*\right|\right|_1\\
    &\leq O(1) O_{P_n}(\sqrt{\log(p)})O_{P_n}\left(s_\gamma\sqrt{\log(p\lor n)/n}\right)\\
    &=o_{P_n}(1),
\end{align*}
where the second inequality follows from the bound on the center factor by applying the moderate deviation theory for self-normalized sums of \cite{delapena2009}, and in particular Lemma 5 in \cite{belloni2012sparse} and the verification of Assumption 3 in \cite{farrell2015robust}. 

Consequently, 
\begin{align*}
\sqrt{n}(\hat\mu_{1, DA}^{(\pi)}-\mu_1)&=\frac{1}{\sqrt{n}}\sum_{i=1}^n\left[\frac{Z_i}{\pi}\left\{Y_i-h_{1}(X_i;\boldsymbol\gamma^*)\right\}+h_{1}(X_i;\boldsymbol\gamma^*)-\mu_1\right] + o_{P_n}(1),
\end{align*}
and thus
\begin{align*}
    \frac{1}{\sqrt{n}}\sum_{i=1}^n\left[\frac{Z_i}{\pi}\left\{Y_i- h_{1}(X_i;\tilde{\boldsymbol\gamma})\right\}+h_{1}(X_i;\tilde{\boldsymbol\gamma})\right] &= \frac{1}{\sqrt{n}}\sum_{i=1}^n\left[\frac{Z_i}{\pi}\left\{Y_i-h_{1}(X_i;\boldsymbol\gamma^*)\right\}+h_{1}(X_i;\boldsymbol\gamma^*)\right]+ o_{P_n}(1).
\end{align*}
Finally, we have
\begin{align}
        \frac{1}{\sqrt{n}}\sum_{i=1}^n\left[\frac{Z_i}{\hat{\pi}}\left\{Y_i- h_{1}(X_i;\tilde{\boldsymbol\gamma})\right\}+h_{1}(X_i;\tilde{\boldsymbol\gamma})\right] &= \frac{1}{\sqrt{n}}\sum_{i=1}^n\left[\frac{Z_i}{\pi}\left\{Y_i-h_{1}(X_i;\boldsymbol\gamma^*)\right\}+h_{1}(X_i;\boldsymbol\gamma^*)\right]+ o_{P_n}(1),
\end{align}
which leads to
\begin{align}
    \sqrt{n}(\hat\mu_{1, DA}-\mu_1)&=\frac{1}{\sqrt{n}}\sum_{i=1}^n\left[\frac{Z_i}{\pi}\left\{Y_i-h_{1}(X_i;\boldsymbol\gamma^*)\right\}+h_{1}(X_i;\boldsymbol\gamma^*)-\mu_1\right] + o_{P_n}(1).\label{eq:est_trtEffect_parametric_IF}
\end{align}

Following a similar reasoning for $\hat\mu_{0, DA}$, we have 
\begin{align*}
    \sqrt{n}(\hat\theta_{DA}-\theta)&=\frac{1}{\sqrt{n}}\sum_{i=1}^n\left(\left[\frac{Z_i}{\pi}\left\{Y_i-h_{1}(X_i;\boldsymbol\gamma^*)\right\}+h_{1}(X_i;\boldsymbol\gamma^*)-\mu_1\right]-\left[\frac{1-Z_i}{1-\pi}\left\{Y_i-h_{0}(X_i;\boldsymbol\eta^*)\right\}+h_{0}(X_i;\boldsymbol\eta^*)-\mu_0\right]\right)\\
    &+ o_{P_n}(1).
\end{align*}

The asymptotic variance of $\hat\theta_{DA} = \hat\mu_{1, DA}-\hat\mu_{0, DA}$ can be calculated as $1/n$ times the sample variance of the values
\begin{align*}
    \frac{Z_i}{\hat\pi}(Y_i-h_1(X_i; \tilde{\boldsymbol\gamma}))+h_1(X_i; \tilde{\boldsymbol\gamma}) - \left[\frac{1-Z_i}{1-\hat\pi}(Y_i-h_0(X_i; \tilde{\boldsymbol\eta}))+h_0(X_i; \tilde{\boldsymbol\eta})\right].
\end{align*}
As opposed to the asymptotic variance for $\hat\theta_{CF} = \hat\mu_{1, CF}-\hat\mu_{0, CF}$, there is no need to take into account the uncertainty in the randomization probabilities as there are no terms like
$\frac{1}{n}\sum_{j=1}^n\left[\frac{Z_j}{\hat\pi^2}\left\{Y_j-\hat h_{1}(X_j; \tilde{\boldsymbol\gamma})\right\}\right](Z_i - \hat\pi)$
and
$\frac{1}{n}\sum_{j=1}^n\left[\frac{1-Z_j}{(1-\hat\pi)^2}\left\{Y_j-\hat h_{0}(X_j; \tilde{\boldsymbol\eta})\right\}\right](Z_i - \hat\pi)$ for participant $i$.

\subsection{Targeted maximum likelihood estimation}\label{app:tmle}
The data-adaptive approach discussed in Section 3 is closely related to targeted maximum likelihood estimation (TMLE) \citep[see e.g.,][]{rosenblum2010simple}. 
In this appendix, we show that the validity of the inference based on this TMLE approach is asymptotically insensitive to selection mistakes. In particular, standard errors obtained using the aforementioned calculation 
are unbiased in large samples, even when the chosen model is misspecified. This has been conjectured before by \cite{moore2009covariate}, and has more recently been shown by \citet{williams2021} for time-to-event and ordinal outcomes. Their result can be extended directly to continuous and binary outcomes. Such a general proof is given in Appendix \ref{app:cv_tmle}. This is also in line with the findings in \cite{benkeser2017doubly}; their approach to obtain doubly robust non-parametric inference for the average treatment effect reduces to standard TMLE -- as described here -- in randomized clinical trials. 

\subsubsection{Cross-validated targeted minimum-loss-based estimation}\label{app:cv_tmle}
In this appendix, we discuss cross-validated targeted minimum-loss-based estimation (CV-TMLE), another imputation estimator that ensures valid inference when data-adaptive prediction models are used, irrespective of model misspecification. Its simplicity makes it an appealing estimator for valid inference and estimation in randomized trials.
This approach first constructs an initial prediction $\hat{h}_{1,k}(X_i)$ for participant $i$ belonging to a certain fold $k=1, \dots, K$ based on the $K-1$ remaining folds, which is then updated 
in a clever way to $\hat{h}_{1,k}^{(u)}(X_i; \hat{\epsilon})$ so that \citep{zheng2011cross}
$$\frac{1}{n}\sum_{i=1}^n\left[Z_i\left\{Y_i-\hat{h}_{1,k}^{(u)}(X_i; \hat{\epsilon})\right\}\right]=0.$$
This ensures that
$$ \frac{1}{K}\sum_{k=1}^K\frac{1}{|\mathcal{V}_k |}\sum_{i\in\mathcal{V}_k}\left[Z_i\left\{Y_i-\hat{h}_{1,k}^{(u)}(X_i; \hat{\epsilon})\right\}\right]=0, $$
if $n/K=|\mathcal{V}_k|$ for all $k\leq K$, and
$$ \frac{1}{K}\sum_{k=1}^K\frac{1}{|\mathcal{V}_k |}\sum_{i\in\mathcal{V}_k}\left[Z_i\left\{Y_i-\hat{h}_{1,k}^{(u)}(X_i; \hat{\epsilon})\right\}\right]=o_{P_n}(1), $$
otherwise (assuming $K$ does not grow with $n$). Specifically, the outcome model fit
$\hat{h}_{1,k}(X)$ is updated by fitting the following regression model with single
parameter $\epsilon$ and offset based on the initial predictions $\hat{h}_{1,k}(X)$:
$$E(Y|Z=1, X)= g^{-1}\left\{g(\hat h_{1,k}(X))+\epsilon\right\}, $$
with $g$ the considered canonical link function (e.g., identity for continuous outcome and logit for binary outcome). This can be done using standard statistical software for fitting a generalized linear regression (with canonical link) of $Y$ with an intercept (denoted as $\epsilon$) using offset $g(\hat h_{1,k}(X))$ among participants with $Z=1$ pooled over all folds.
In fact, the updating step can also be performed on the validation sample, which leads to $K$ different estimates $\hat\epsilon_1, \dots, \hat\epsilon_K$ for $\epsilon$ (i.e., one for each fold). 

This results in the imputation estimator
\begin{align}
\frac{1}{n}\sum_{i=1}^n\hat{h}_{1,k}^{(u)}(X_i;\hat\epsilon).\label{eq:cvtmle}
\end{align}
Alternatively, one can consider $\hat\mu_{1, TMLE}=\frac{1}{K}\sum_{k=1}^K\hat\mu_{1,k, TMLE}$ with $\hat{\mu}_{1,k, TMLE}=\frac{1}{|\mathcal{V}_k |}\sum_{i\in\mathcal{V}_k}\hat{h}_{1,k}^{(u)}(X_i;\hat\epsilon)$ for fold $k$, which equals \eqref{eq:cvtmle} if $n/K=|\mathcal{V}_k|$ for all $k\leq K$ and is equal up to a term that's asymptotically negligible otherwise. In what follows, we therefore focus on $\hat\mu_{1, TMLE}=\frac{1}{K}\sum_{k=1}^K\hat\mu_{1,k,TMLE}$ to be consistent with previous sections. 

The update of the predictions is the reason why the proof in Appendix \ref{app:sampleSplitting} no longer holds for CV-TMLE as the predictions $\hat{h}_{1,k}^{(u)}(X; \hat{\epsilon})$ are no longer obtained from a `separate sample'. In particular, an update of the proof for term \eqref{eq:term1.2} is required. 

To this end, we assume that $\hat\epsilon$ converges in probability to a limit $\epsilon^*$ in the sense that $|\hat\epsilon-\epsilon^*|=o_{P_n}(1)$ since $\hat\epsilon$ is obtained with maximum likelihood estimation, and that $\hat h_{1,k}^{(u)}(X_i;\epsilon^*)$ converges in probability to $h_{1}^*(X_i;\epsilon^*)$ in the sense that
$\sqrt{\frac{1}{|\mathcal{V}_k |}\sum_{i\in\mathcal{V}_k}\left\{h_{1}^*(X_i;\epsilon^*)-\hat h_{1,k}^{(u)}(X_i;\epsilon^*)\right\}^2}=o_{P_n}(1)$, $\forall k\in\{1, \dots, K\}$ (see also Assumption \ref{as:convergenceSampleSplitting}). 

First, the `clever' update ensures that
\begin{align*}
\frac{1}{K}\sum_{k=1}^K\frac{1}{|\mathcal{V}_k |}\sum_{i\in\mathcal{V}_k}\left[\frac{Z_i}{\hat \pi}\left\{Y_i-\hat h_{1,k}^{(u)}(X_i;\hat\epsilon)\right\}+\hat h_{1,k}^{(u)}(X_i;\hat\epsilon)\right]
&=\frac{1}{K}\sum_{k=1}^K\frac{1}{|\mathcal{V}_k |}\sum_{i\in\mathcal{V}_k}\left[\frac{Z_i}{\pi}\left\{Y_i-\hat h_{1,k}^{(u)}(X_i;\hat\epsilon)\right\}+\hat h_{1,k}^{(u)}(X_i;\hat\epsilon)\right]\\
&+\frac{1}{K}\sum_{k=1}^K\frac{1}{|\mathcal{V}_k |}\sum_{i\in\mathcal{V}_k}Z_i\left\{Y_i-\hat h_{1,k}^{(u)}(X_i;\hat\epsilon)\right\}\left(\frac{1}{\hat \pi}-\frac{1}{\pi}\right)\\
&=\frac{1}{K}\sum_{k=1}^K\frac{1}{|\mathcal{V}_k |}\sum_{i\in\mathcal{V}_k}\left[\frac{Z_i}{\pi}\left\{Y_i-\hat h_{1,k}^{(u)}(X_i;\hat\epsilon)\right\}+\hat h_{1,k}^{(u)}(X_i;\hat\epsilon)\right],
\end{align*}
if $n/K=|\mathcal{V}_k|$ for all $k\leq K$; otherwise, the left hand side equals $$\frac{1}{K}\sum_{k=1}^K\frac{1}{|\mathcal{V}_k |}\sum_{i\in\mathcal{V}_k}\left[\frac{Z_i}{\pi}\left\{Y_i-\hat h_{1,k}^{(u)}(X_i;\hat\epsilon)\right\}+\hat h_{1,k}^{(u)}(X_i;\hat\epsilon)\right]+o_{P_n}(1).$$

In line with the previous proofs, we focus on 
\begin{align}
\frac{1}{\sqrt{|\mathcal{V}_k|}}\sum_{i\in\mathcal{V}_k}\left[\frac{Z_i}{\pi}\left\{Y_i-\hat h_{1,k}^{(u)}(X_i;\hat\epsilon)\right\}+\hat h_{1,k}^{(u)}(X_i;\hat\epsilon)\right]\label{eq:cvtmle_term1}.
\end{align}
In order to handle term \eqref{eq:cvtmle_term1}, we define $\hat\mu_{1,k, TMLE}^{(\pi)}=\frac{1}{|\mathcal{V}_k|}\sum_{i\in\mathcal{V}_k}\left[\frac{Z_i}{\pi}\left\{Y_i-\hat h_{1,k}^{(u)}(X_i;\hat\epsilon)\right\}+\hat h_{1,k}^{(u)}(X_i;\hat\epsilon)\right]$.
Then, it holds that
\begin{align}
\sqrt{|\mathcal{V}_k|}(\hat\mu_{1,k, TMLE}^{(\pi)}-\mu_1)&=\frac{1}{\sqrt{|\mathcal{V}_k|}}\sum_{i\in\mathcal{V}_k}\left[\frac{Z_i}{\pi}\left\{Y_i-h_1^*(X_i;\epsilon^*)\right\}+h^*_1(X_i;\epsilon^*)-\mu_1\right]\nonumber\\
&+\frac{1}{\sqrt{|\mathcal{V}_k|}}\sum_{i\in\mathcal{V}_k}\left[\left(\frac{Z_i}{\pi}-1\right)\left\{h^*_1(X_i;\epsilon^*)-\hat h_{1,k}^{(u)}(X_i;\hat\epsilon)\right\}\right].\label{eq:tmle_term1.2}
\end{align}
Next, for term \eqref{eq:tmle_term1.2} it holds that
\begin{align}
    \frac{1}{\sqrt{|\mathcal{V}_k |}}\sum_{i\in\mathcal{V}_k}\left[\left(\frac{Z_i}{\pi}-1\right)\left\{h^*_1(X_i;\epsilon^*)-\hat h_{1,k}^{(u)}(X_i; \hat\epsilon)\right\}\right]&=\frac{1}{\sqrt{|\mathcal{V}_k|}}\sum_{i\in\mathcal{V}_k}\left[\left(\frac{Z_i}{\pi}-1\right)\left\{h^*_1(X_i;\epsilon^*)-\hat h_{1,k}^{(u)}(X_i; \epsilon^*)\right\}\right]\label{eq:term_tmle.1}\\
    &+\frac{1}{\sqrt{|\mathcal{V}_k|}}\sum_{i\in\mathcal{V}_k}\left[\left(\frac{Z_i}{\pi}-1\right)\left\{\hat h_{1,k}^{(u)}(X_i; \epsilon^*)-\hat h_{1,k}^{(u)}(X_i; \hat\epsilon)\right\}\right]\label{eq:term_tmle.2}.
\end{align}

Under the assumption that $\sqrt{\frac{1}{|\mathcal{V}_k |}\sum_{i\in\mathcal{V}_k}\left\{h_{1}^*(X_i;\epsilon^*)-\hat h_{1,k}^{(u)}(X_i;\epsilon^*)\right\}^2}=o_{P_n}(1)$, one can prove that term \eqref{eq:term_tmle.1} converges in probability to zero by following a similar approach as for term \eqref{eq:term1.2}. 
Moreover, following a Taylor expansion for term \eqref{eq:term_tmle.2} gives
\begin{align}
    \sqrt{|\mathcal{V}_k|}(\epsilon^*-\hat{\epsilon})\frac{1}{|\mathcal{V}_k|}\sum_{i\in\mathcal{V}_k}\left(\frac{Z_i}{\pi}-1\right)\frac{\partial}{\partial \epsilon}\hat{h}_{1,k}^{(u)}(X_i;\epsilon)\biggr\rvert_{\epsilon=\epsilon^*}+O_{P_n}\left(\sqrt{|\mathcal{V}_k|}|\epsilon^*-\hat{\epsilon}|^2\right).\label{eq:TMLE_taylor}
\end{align}

For the main term in Equation \eqref{eq:TMLE_taylor}, we have
\begin{align*}
    E\left[E\left\{\frac{1}{|\mathcal{V}_k|}\sum_{i\in\mathcal{V}_k}\left(\frac{Z_i}{\pi}-1\right)\frac{\partial}{\partial \epsilon}\hat{h}_{1,k}^{(u)}(X_i;\epsilon)\biggr\rvert_{\epsilon=\epsilon^*}\bigg\vert\mathbf{O}^{\bar k}\right\}\right]&=E\left[E\left\{\left(\frac{Z}{\pi}-1\right)\frac{\partial}{\partial \epsilon}\hat{h}_{1,k}^{(u)}(X;\epsilon)\biggr\rvert_{\epsilon=\epsilon^*}\bigg\vert\mathbf{O}^{\bar k}\right\}\right]\\
    &=E\left(\frac{Z}{\pi}-1\right)E\left[E\left\{\frac{\partial}{\partial \epsilon}\hat{h}_{1,k}^{(u)}(X;\epsilon)\biggr\rvert_{\epsilon=\epsilon^*}\bigg\vert\mathbf{O}^{\bar k}\right\}\right]\\
    &=0.
\end{align*}
By the law of total variance
\begin{align*}
    Var\left[\frac{1}{|\mathcal{V}_k|}\sum_{i\in\mathcal{V}_k}\left(\frac{Z_i}{\pi}-1\right)\frac{\partial}{\partial \epsilon}\hat{h}_{1,k}^{(u)}(X_i;\epsilon)\biggr\rvert_{\epsilon=\epsilon^*}\right]
    &=E\left[Var\left\{\frac{1}{|\mathcal{V}_k|}\sum_{i\in\mathcal{V}_k}\left(\frac{Z_i}{\pi}-1\right)\frac{\partial}{\partial \epsilon}\hat{h}_{1,k}^{(u)}(X_i;\epsilon)\biggr\rvert_{\epsilon=\epsilon^*}\bigg\vert\mathbf{O}^{\bar k}\right\}\right]\\
    &=E\left[\frac{1}{|\mathcal{V}_k|}Var\left\{\left(\frac{Z}{\pi}-1\right)\frac{\partial}{\partial \epsilon}\hat{h}_{1,k}^{(u)}(X;\epsilon)\biggr\rvert_{\epsilon=\epsilon^*}\bigg\vert\mathbf{O}^{\bar k}\right\}\right]\\
    &=\frac{1}{|\mathcal{V}_k|}E\left(E\left[\left\{\left(\frac{Z}{\pi}-1\right)\frac{\partial}{\partial \epsilon}\hat{h}_{1,k}^{(u)}(X;\epsilon)\biggr\rvert_{\epsilon=\epsilon^*}\right\}^2\bigg\vert\mathbf{O}^{\bar k}\right]\right)\\
    &=\frac{1}{|\mathcal{V}_k|}E\left\{\left(\frac{Z}{\pi}-1\right)^2\right\}E\left(E\left[\left\{\frac{\partial}{\partial \epsilon}\hat{h}_{1,k}^{(u)}(X;\epsilon)\biggr\rvert_{\epsilon=\epsilon^*}\right\}^2\bigg\vert\mathbf{O}^{\bar k}\right]\right)\\
    &=\frac{1}{|\mathcal{V}_k|}E\left\{\left(\frac{Z}{\pi}-1\right)^2\right\}E\left[\left\{\frac{\partial}{\partial \epsilon}\hat{h}_{1,k}^{(u)}(X;\epsilon)\biggr\rvert_{\epsilon=\epsilon^*}\right\}^2\right],
\end{align*}

Under the assumption that $E\left[\left\{\frac{\partial}{\partial \epsilon}\hat{h}_{1,k}^{(u)}(X;\epsilon)\biggr\rvert_{\epsilon=\epsilon^*}\right\}^2\right]$ is bounded (which is satisfied for generalized linear models with common canonical link when $X$ is bounded), application of Chebyshev's inequality shows that 
\begin{align*}
    \left|\frac{1}{|\mathcal{V}_k|}\sum_{i\in\mathcal{V}_k}\left(\frac{Z_i}{\pi}-1\right)\frac{\partial}{\partial \epsilon}\hat{h}_{1,k}^{(u)}(X_i;\epsilon)\biggr\rvert_{\epsilon=\epsilon^*}\right|=O_{P_n}\left(\frac{1}{\sqrt{|\mathcal{V}_k|}}\right).
\end{align*}
Then,
\begin{align*}
    \left|\sqrt{|\mathcal{V}_k|}(\epsilon^*-\hat{\epsilon})\frac{1}{|\mathcal{V}_k|}\sum_{i\in\mathcal{V}_k}\left(\frac{Z_i}{\pi}-1\right)\frac{\partial}{\partial \epsilon}\hat{h}_{1,k}^{(u)}(X_i;\epsilon)\biggr\rvert_{\epsilon=\epsilon^*}\right|
    &\leq \sqrt{|\mathcal{V}_k|}\left|\epsilon^*-\hat{\epsilon}\right|\left|\frac{1}{|\mathcal{V}_k|}\sum_{i\in\mathcal{V}_k}\left(\frac{Z_i}{\pi}-1\right)\frac{\partial}{\partial \epsilon}\hat{h}_{1,k}^{(u)}(X_i;\epsilon)\biggr\rvert_{\epsilon=\epsilon^*}\right|\\
    &=\sqrt{|\mathcal{V}_k|}o_{P_n}(1)O_{P_n}\left(\frac{1}{\sqrt{|\mathcal{V}_k|}}\right)\\
    &=o_{P_n}(1),
\end{align*}
as $\left|\epsilon^*-\hat{\epsilon}\right|=o_{P_n}(1)$ (see e.g., \cite{zheng2011cross}).
Moreover, following a similar reasoning, it holds that $O_{P_n}\left(\sqrt{|\mathcal{V}_k|}|\epsilon^*-\hat{\epsilon}|^2\right)=o_{P_n}(1)$ in Equation \eqref{eq:TMLE_taylor}, as $|\hat\epsilon-\epsilon^*|=o_{P_n}(1)$ and assuming $\hat{h}_{1,k}^{(u)}(X_i;\epsilon)$ is an analytic function which ensures that $\frac{1}{k!}\left|\frac{\partial^k}{\partial \epsilon^k}\hat{h}_{1,k}^{(u)}(X_i;\epsilon)\biggr\rvert_{\epsilon=\epsilon^*}\right|$ is bounded for every compact set \citep{krantz2002primer}. Note that this assumption is satisfied for generalized linear models with common canonical link such as linear regression, logistic regression, and Poisson regression over a compact parameter space.

Then,
\begin{align*}
    \sqrt{n}(\hat\mu_{1, TMLE}-\mu_1)&=\sqrt{n}\frac{1}{K}\sum_{k=1}^K\frac{1}{|\mathcal{V}_k|}\sum_{i\in\mathcal{V}_k}\left\{\left(\frac{Z_i}{\pi}\left\{Y_i-h_1^*(X_i;\epsilon^*)\right\}+h_1^*(X_i;\epsilon^*)-\mu_1\right)\right\}+o_{P_n}(1).
\end{align*}
We can follow a similar reasoning for $\hat\mu_{0, TMLE}$.

\subsubsection{Targeted maximum likelihood estimation without cross-fitting: Outcome regression fitted in a parametric model}
A proof for the validity of TMLE without cross-fitting can be obtained by following a similar reasoning as in Appendix \ref{app:withoutsamplesplitting} and \ref{app:cv_tmle}, under similar assumptions as in  \ref{app:withoutsamplesplitting} (e.g., Assumption \ref{as:ultraSparsity} and Assumption \ref{as:ratescoeff}).

\subsection{Theorem 6.1: Parametric estimation of propensity score}\label{app:par_ps}
Till now, we have estimated the propensity score $\pi=P(Z=1)$ independently of baseline variables $X$. In this section, we prove that the results above hold in more generality when the treatment probability is estimated within a parametric model (e.g., using a logistic regression model). This might be of interest in practice as estimating the propensity score in a parametric model may yield larger efficiency gains. Specifically, we define $P(Z=1|X_{PS})=p(X_{PS}; \boldsymbol\beta_0)$ where $p(X_{PS}; \boldsymbol\beta)$ is a known function evaluated at a parameter $\boldsymbol\beta$ with population value $\boldsymbol\beta_0$ and $X_{PS}\subseteq X$ (which is pre-determinged at the design stage and does not grow with $n$). For example, as $Z$ is binary, its association with $X_{PS}$ can be modeled via a logistic regression.

\subsubsection{Data-adaptive covariate adjustment with cross-fitting}\label{app:par_samplesplitting}
Since $\pi=p(X_{PS}; \boldsymbol\beta_0)$ we can then follow a similar reasoning as before, except for Term \eqref{eq:term2.1} and Term \eqref{eq:term2.2} as these involve $p(X_{PS}; \hat{\boldsymbol\beta})$, with $\hat{\boldsymbol\beta}$ an estimate of $\boldsymbol\beta_0$ obtained via estimation within a parametric model (e.g., using logistic regression). 

First, for Term \eqref{eq:term2.1}, we have 
\begin{align*}
\frac{1}{|\mathcal{V}_k|}\sum_{i\in \mathcal{V}_k}&Z_i\left\{Y_i-h_{1}^*(X_i)\right\}\left(\frac{1}{p(X_{PS,i}; \hat{\boldsymbol\beta})}-\frac{1}{p(X_{PS,i};\boldsymbol\beta_0)}\right)\sqrt{|\mathcal{V}_k|}\\
&=-\frac{1}{|\mathcal{V}_k|}\sum_{i\in \mathcal{V}_k}Z_i\left\{Y_i-h_{1}^*(X_i)\right\}\frac{1}{p(X_{PS,i};\boldsymbol\beta_0)^2}\frac{\partial}{\partial \boldsymbol\beta}p(X_{PS,i};\boldsymbol\beta)\biggr\rvert_{\boldsymbol\beta=\boldsymbol\beta_0}\sqrt{|\mathcal{V}_k|}(\hat{\boldsymbol\beta}-\boldsymbol\beta_0)+o_{P_n}(1),
\end{align*}
where the $o_{P_n}(1)$ term follows from the fact that $\hat{\boldsymbol\beta}$ is a root-$n$ consistent maximum likelihood estimator of a parameter in a parametric model and some regularity conditions (e.g., uniform convergence of the sample mean of the second-order derivatives). 
Under suitable
regularity conditions \citep[see e.g.,][]{newey1994large}, it follows from the uniform weak law of large numbers that this equals 
\begin{align*}
    -E\left[Z\left\{Y-h_{1}^*(X)\right\}\frac{1}{p(X_{PS};\boldsymbol\beta_0)^2}\frac{\partial}{\partial \beta}p(X_{PS};\boldsymbol\beta)\biggr\rvert_{\boldsymbol\beta=\boldsymbol\beta_0}\right]\sqrt{|\mathcal{V}_k|}(\hat{\boldsymbol\beta}-\boldsymbol\beta_0)+o_{P_n}(1).
\end{align*}
Allowing for misspecification of the outcome model, the product of the first three terms is $O_{P_n}(1)$ as $\hat{\boldsymbol\beta}$ is a root-$n$ consistent maximum likelihood estimator of a parameter in a parametric model (i.e., $|\hat \beta-\beta_0|=O_{P_n}(n^{-1/2})$).

Next, we define $s(X_{PS};\boldsymbol\beta)$ as the score function of $\boldsymbol\beta$. It then follows from \cite{tsiatis2006semiparametric} that
\begin{align*}
    \sqrt{|\mathcal{V}_k|}(\hat{\boldsymbol\beta}-\boldsymbol\beta_0)=-\left[E\left\{\frac{\partial s(X_{PS};\boldsymbol\beta_0)}{\partial \boldsymbol\beta}\right\}\right]^{-1}\left\{\frac{1}{\sqrt{|\mathcal{V}_k|}}\sum_{i\in \mathcal{V}_k}s(X_{PS,i};\boldsymbol\beta_0)\right\}+o_{P_n}(1).
\end{align*}
Thus, Term \eqref{eq:term2.1} can be rewritten as
\begin{align*}
    E\left[Z\left\{Y-h_{1}^*(X)\right\}\frac{1}{p(X_{PS};\boldsymbol\beta_0)^2}\frac{\partial}{\partial \boldsymbol\beta}p(X_{PS};\boldsymbol\beta)\biggr\rvert_{\boldsymbol\beta=\boldsymbol\beta_0}\right]\left[E\left\{\frac{\partial s(X_{PS};\boldsymbol\beta_0)}{\partial \boldsymbol\beta}\right\}\right]^{-1}\left\{\frac{1}{\sqrt{|\mathcal{V}_k|}}\sum_{i\in \mathcal{V}_k}s(X_{PS,i};\boldsymbol\beta_0)\right\}+o_{P_n}(1).
\end{align*}

Next, from the the Cauchy Schwarz inequality, it follows that Term \eqref{eq:term2.2} yields
{\small
\begin{align*}
    &\sqrt{|\mathcal{V}_k|}\left|\frac{1}{|\mathcal{V}_k|}\sum_{i\in \mathcal{V}_k}Z_i\left\{h_{1}^*(X_i)-\hat h_{1,k}(X_i)\right\}\left\{\frac{1}{p(X_{PS,i}; \hat{\boldsymbol\beta})}-\frac{1}{p(X_{PS,i};\boldsymbol\beta_0)}\right\}\right|\\
    &\leq \sqrt{|\mathcal{V}_k|}\left(\max_{i\in \mathcal{V}_k}\frac{1}{p(X_{PS,i};\boldsymbol\beta_0)p(X_{PS,i}; \hat{\boldsymbol\beta})}\right)\sqrt{\frac{1}{|\mathcal{V}_k|}\sum_{i\in \mathcal{V}_k}Z_i\left\{h_1^*(X_i)-\hat h_{1,k}(X_i)\right\}^2}\sqrt{\frac{1}{|\mathcal{V}_k|}\sum_{i\in \mathcal{V}_k}\left\{p(X_{PS,i}; \hat{\boldsymbol\beta})-p(X_{PS,i};\boldsymbol\beta_0)\right\}^2}\\
    &=o_{P_n}(1),
\end{align*}
}
as $1>1-\rho \geq p(X_{PS,i};\boldsymbol\beta_0)\geq\rho>0$ for $i=1, \dots n$, Assumption \ref{as:convergenceSampleSplitting}, and the fact that $\hat{\boldsymbol\beta}$ is a root-$n$ consistent maximum likelihood estimator of a parameter in a parametric model $p(X_{PS};\boldsymbol\beta)$ which is assumed to be continuous in $\beta$.

Consequently,
\begin{align*}
    \sqrt{|\mathcal{V}_k|}(\hat{\mu}_{1,k}-\mu_1)&=
    \frac{1}{\sqrt{|\mathcal{V}_k|}}\sum_{i\in \mathcal{V}_k}\left(\vphantom{E\left[A\left\{Y-m^*(L)\right\}\frac{1}{p(L;\boldsymbol\beta_0)^2}\frac{\partial}{\partial \boldsymbol\beta}p(L;\boldsymbol\beta)\biggr\rvert_{\boldsymbol\beta=\boldsymbol\beta_0}\right]}\frac{Z_i}{p(X_{PS,i};\boldsymbol\beta_0)}\left\{Y_i-h_1^*(X_i)\right\}+h_1^*(X_i)\right.\\
    &\left.+E\left[Z\left\{Y-h_{1}^*(X)\right\}\frac{1}{p(X_{PS};\boldsymbol\beta_0)^2}\frac{\partial}{\partial \boldsymbol\beta}p(X_{PS};\boldsymbol\beta)\biggr\rvert_{\boldsymbol\beta=\boldsymbol\beta_0}\right]\left[E\left\{\frac{\partial s(X_{PS};\boldsymbol\beta_0)}{\partial \boldsymbol\beta}\right\}\right]^{-1}s(X_{PS,i};\boldsymbol\beta_0)-\mu_1\right)\\&
    +o_{P_n}(1).
\end{align*}

We can follow a similar reasoning for $\hat\mu_{0, k}$.

\subsubsection{Outcome regression fitted in a parametric model with canonical link}\label{app:par_canonical}
Recall that, when parametrically modeling the association of $Z$ with $X_{PS}$, ``Step 1b: Model fitting'' is performed as follows:
\begin{itemize}[left=2em]
	\item[\textbf{Step 1b:}]\textbf{Model fitting} Fit a generalized linear regression model with a canonical link (e.g., logistic, linear, \dots) and intercept via maximum likelihood that regresses the outcome $Y$ on all covariates selected in Step 1a among the treated participants with $Z=1$, using weights  $1/p(X_{PS}; \hat{\boldsymbol\beta})$. 
\end{itemize}

This step then ensures that
\begin{align}
       \frac{1}{\sqrt{n}}\sum_{i=1}^n&\left[\frac{Z_i}{p(X_{PS,i}; \hat{\boldsymbol\beta})}\left\{Y_i- h_{1}(X_i;\tilde{\boldsymbol\gamma})\right\}+h_{1}(X_i;\tilde{\boldsymbol\gamma})\right]\nonumber\\ &= \frac{1}{\sqrt{n}}\sum_{i=1}^n\left[\frac{Z_i}{p(X_{PS,i}; \boldsymbol\beta_0)}\left\{Y_i-h_{1}(X_i;\tilde{\boldsymbol\gamma})\right\}+h_{1}(X_i;\tilde{\boldsymbol\gamma})\right]\nonumber\\
&+\frac{1}{n}\sum_{i=1}^nZ_i\left\{Y_i-h_{1}(X_i;\tilde{\boldsymbol\gamma})\right\}\left(\frac{1}{p(X_{PS,i}; \hat{\boldsymbol\beta})}-\frac{1}{p(X_i; \boldsymbol\beta_0)}\right)\sqrt{n}\nonumber\\
&=\frac{1}{\sqrt{n}}\sum_{i=1}^n\left[\frac{Z_i}{p(X_{PS,i}; \boldsymbol\beta_0)}\left\{Y_i-h_{1}(X_i;\tilde{\boldsymbol\gamma})\right\}+h_{1}(X_i;\tilde{\boldsymbol\gamma})\right]\label{eq:DA_par_1}\\
&+\frac{1}{n}\sum_{i=1}^nZ_i\left\{Y_i-h_{1}(X_i;\boldsymbol\gamma^*)\right\}\left(\frac{1}{p(X_{PS,i}; \hat{\boldsymbol\beta})}-\frac{1}{p(X_{PS,i}; \boldsymbol\beta_0)}\right)\sqrt{n}\label{eq:DA_par_2}\\
&+\frac{1}{n}\sum_{i=1}^nZ_i\left\{h_{1}(X_i;\boldsymbol\gamma^*)-h_{1}(X_i;\tilde{\boldsymbol\gamma})\right\}\left(\frac{1}{p(X_{PS,i}; \hat{\boldsymbol\beta})}-\frac{1}{p(X_{PS,i}; \boldsymbol\beta_0)}\right)\sqrt{n}.\label{eq:DA_par_3}
\end{align}
Term \eqref{eq:DA_par_1} is equal to Term \eqref{eq:outcRegr_simple} as $\pi = p(X_{PS}; \boldsymbol\beta_0)$, and can therefore be handled in the same way. It then follows that Term \eqref{eq:DA_par_1} equals
$$\frac{1}{\sqrt{n}}\sum_{i=1}^n\left[\frac{Z_i}{p(X_{PS,i}; \boldsymbol\beta_0)}\left\{Y_i-h_{1}(X_i;\boldsymbol\gamma^*)\right\}+h_{1}(X_i;\boldsymbol\gamma^*)\right]+o_{P_n}(1).$$

Following a similar reasoning as in Appendix \ref{app:par_samplesplitting}, Term \eqref{eq:DA_par_2} can be rewritten as
\begin{align*}
    -\frac{1}{n}\sum_{i=1}^nZ_i\left\{Y_i-h_{1}(X_i;\boldsymbol\gamma^* )\right\}\frac{1}{p(X_{PS,i};\boldsymbol\beta_0)^2}\frac{\partial}{\partial \beta}p(X_{PS,i};\boldsymbol\beta)\biggr\rvert_{\boldsymbol\beta=\boldsymbol\beta_0}\sqrt{n}(\hat{\boldsymbol\beta}-\boldsymbol\beta_0)+o_{P_n}(1).
\end{align*}
Here, $\frac{1}{p(X_{PS,i};\boldsymbol\beta_0)^2}\frac{\partial}{\partial \boldsymbol\beta}p(X_{PS,i};\boldsymbol\beta)\biggr\rvert_{\boldsymbol\beta=\boldsymbol\beta_0}$ is a constant times vector 
$\left(\begin{array}{c c}
    1&
    X_{PS,i}'\end{array}\right)$. Therefore, we have
\begin{align*}
    &\left|\frac{1}{n}\sum_{i=1}^nZ_i\left\{Y_i-h_{1}(X_i;\boldsymbol\gamma^* )\right\}\frac{1}{p(X_{PS,i};\boldsymbol\beta_0)^2}\frac{\partial}{\partial \boldsymbol\beta}p(X_{PS,i};\boldsymbol\beta)\biggr\rvert_{\boldsymbol\beta=\boldsymbol\beta_0}\sqrt{n}(\hat{\boldsymbol\beta}-\boldsymbol\beta_0)\right|\\
    &\leq O(1)\sqrt{n}\left|\left|\hat{\boldsymbol\beta}-\boldsymbol\beta_0\right|\right|_1\left|\left|\frac{1}{n}\sum_{i=1}^nZ_i\left\{Y_i-h_{1}(X_i;\boldsymbol\gamma^* )\right\}\left(\begin{array}{c c}
    1&
    X_{PS,i}'\end{array}\right)\right|\right|_\infty\\
    &\leq O(1)\sqrt{n}O_{P_n}(n^{-1/2})O_{P_n}(\sqrt{\log(p\lor n)/n})\\
    &= o_{P_n}(1),
\end{align*}
where the last inequality follows from Assumption \ref{as:ultraSparsity}, Assumption \ref{ass:concGradients} and $||\hat{ \boldsymbol\beta}-\boldsymbol\beta_0||_1=O_{P}(n^{-1/2})$ when estimating the propensity score via maximum likelihood estimation.

Next, from the the Cauchy Schwarz inequality, it follows that Term \eqref{eq:DA_par_3} yields
{\small
\begin{align*}
    \sqrt{n}&\left|\frac{1}{n}\sum_{i=1}^nZ_i\left\{h_{1}(X_i;\boldsymbol\gamma^*)-h_{1}(X_i;\tilde{\boldsymbol\gamma})\right\}\left(\frac{1}{p(X_{PS,i}; \hat{\boldsymbol\beta})}-\frac{1}{p(X_{PS,i}; \boldsymbol\beta_0)}\right)\right|\\
    &\leq \sqrt{n}\left(\max_{i\leq n}\frac{1}{p(X_{PS,i};\beta_0)p(X_{PS,i}; \hat{\boldsymbol\beta})}\right)\sqrt{\frac{1}{n}\sum_{i=1}^nZ_i\left\{h_{1}(X_i;\boldsymbol\gamma^*)-h_{1}(X_i;\tilde{\boldsymbol\gamma})\right\}^2}\sqrt{\frac{1}{|\mathcal{V}_k|}\sum_{i\in \mathcal{V}_k}\left\{p(X_{PS,i}; \hat{\boldsymbol\beta})-p(X_{PS,i};\boldsymbol\beta_0)\right\}^2}\\
    &=o_{P_n}(1),
\end{align*}
}
as $1>p(X_{PS,i};\boldsymbol\beta_0)\geq\rho>0$ for $i=1, \dots n$, Assumption \ref{as:convergence} and the fact that $\hat{\boldsymbol\beta}$ is a root-$n$ consistent maximum likelihood estimator of a parameter in a parametric model.

\subsubsection{Targeted maximum likelihood estimation}\label{app:par_tmle}
To allow the `clever covariate' to depend on baseline covariates $X_{PS}$, we extend the notation of our outcome models. In particular, we add an argument for the parameter $\boldsymbol\beta$ which indexes the propensity score model, and consider a vector $\boldsymbol\epsilon = (\epsilon_1, \boldsymbol\epsilon_2')'$. The probability limit of $\hat{\boldsymbol\epsilon}$ is denoted by $\boldsymbol\epsilon^*$.

For each $k\in{1,\dots,K}$, we then consider
\begin{align}
\frac{1}{\sqrt{|\mathcal{V}_k|}}&\sum_{i\in\mathcal{V}_k}\left[\frac{Z_i}{p(X_{PS,i}; \hat{\boldsymbol\beta})}\left\{Y_i-\hat h_{1,k}^{(u)}(X_i;\hat{\boldsymbol\epsilon}, \hat{\boldsymbol\beta})\right\}+\hat h_{1,k}^{(u)}(X_i;\hat{\boldsymbol\epsilon}, \hat{\boldsymbol\beta})\right]\nonumber\\
&=\frac{1}{\sqrt{|\mathcal{V}_k|}}\sum_{i\in\mathcal{V}_k}\left[\frac{Z_i}{p(X_{PS,i};\boldsymbol\beta_0)}\left\{Y_i-\hat h_{1,k}^{(u)}(X_i;\hat{\boldsymbol\epsilon}, \hat{\boldsymbol\beta})\right\}+\hat h_{1,k}^{(u)}(X_i;\hat{\boldsymbol\epsilon}, \hat{\boldsymbol\beta})\right]\label{eq:tmle_par_term1}\\
&+\frac{1}{|\mathcal{V}_k|}\sum_{i\in\mathcal{V}_k}Z_i\left\{Y_i-\hat h_{1,k}^{(u)}(X_i;\hat{\boldsymbol\epsilon}, \hat{\boldsymbol\beta})\right\}\left(\frac{1}{p(X_{PS,i}; \hat{\boldsymbol\beta})}-\frac{1}{p(X_{PS,i};\boldsymbol\beta_0)}\right)\sqrt{|\mathcal{V}_k|},\label{eq:tmle_par_term2}
\end{align}
where $\boldsymbol{\epsilon}=(\epsilon_1, \boldsymbol{\epsilon}_2')'$.

For Term \eqref{eq:tmle_par_term1}, we follow a similar reasoning as for Term \eqref{eq:cvtmle_term1}. In particular, defining $\hat\mu_{1,k, TMLE}^{(\pi)}=\frac{1}{|\mathcal{V}_k|}\sum_{i\in\mathcal{V}_k}\left[\frac{Z_i}{p(X_{PS,i};\boldsymbol\beta_0)}\left\{Y_i-\hat h_{1,k}^{(u)}(X_i;\hat{\boldsymbol\epsilon}, \hat{\boldsymbol\beta})\right\}+\hat h_{1,k}^{(u)}(X_i;\hat{\boldsymbol\epsilon}, \hat{\boldsymbol\beta})\right]$, it holds that
\begin{align*}
\sqrt{|\mathcal{V}_k|}(\hat\mu_{1,k, TMLE}^{(\pi)}-\mu_1)&=\frac{1}{\sqrt{|\mathcal{V}_k|}}\sum_{i\in\mathcal{V}_k}\left[\frac{Z_i}{p(X_{PS,i};\boldsymbol\beta_0)}\left\{Y_i-h_1^*(X_i;\boldsymbol\epsilon^*, \boldsymbol\beta_0)\right\}+h^*_1(X_i;\boldsymbol\epsilon^*, \boldsymbol\beta_0)-\mu_1\right]\\
&+\frac{1}{\sqrt{|\mathcal{V}_k|}}\sum_{i\in\mathcal{V}_k}\left[\left(\frac{Z_i}{p(X_{PS,i};\boldsymbol\beta_0)}-1\right)\left\{h^*_1(X_i;\boldsymbol\epsilon^*, \boldsymbol\beta_0)-\hat h_{1,k}^{(u)}(X_i;\hat{\boldsymbol\epsilon}, \hat{\boldsymbol\beta})\right\}\right],
\end{align*}
where we have for the last term that
\begin{align*}
    \frac{1}{\sqrt{|\mathcal{V}_k|}}&\sum_{i\in\mathcal{V}_k}\left[\left(\frac{Z_i}{p(X_{PS,i};\boldsymbol\beta_0)}-1\right)\left\{h^*_1(X_i;\boldsymbol\epsilon^*, \boldsymbol\beta_0)-\hat h_{1,k}^{(u)}(X_i;\hat{\boldsymbol\epsilon}, \hat{\boldsymbol\beta})\right\}\right]\\
    &=\frac{1}{\sqrt{|\mathcal{V}_k|}}\sum_{i\in\mathcal{V}_k}\left[\left(\frac{Z_i}{p(X_{PS,i};\boldsymbol\beta_0)}-1\right)\left\{h^*_1(X_i;\boldsymbol\epsilon^*, \boldsymbol\beta_0)-\hat h_{1,k}^{(u)}(X_i;\boldsymbol\epsilon^*, \boldsymbol\beta_0)\right\}\right]\\
    &+\frac{1}{\sqrt{|\mathcal{V}_k|}}\sum_{i\in\mathcal{V}_k}\left[\left(\frac{Z_i}{p(X_{PS,i};\boldsymbol\beta_0)}-1\right)\left\{\hat h_{1,k}^{(u)}(X_i;\boldsymbol\epsilon^*, \boldsymbol\beta_0)-\hat h_{1,k}^{(u)}(X_i;\hat{\boldsymbol\epsilon}, \boldsymbol\beta_0)\right\}\right]\\
    &+\frac{1}{\sqrt{|\mathcal{V}_k|}}\sum_{i\in\mathcal{V}_k}\left[\left(\frac{Z_i}{p(X_{PS,i};\boldsymbol\beta_0)}-1\right)\left\{\hat h_{1,k}^{(u)}(X_i;\hat{\boldsymbol\epsilon}, \boldsymbol\beta_0)-\hat h_{1,k}^{(u)}(X_i;\hat{\boldsymbol\epsilon}, \hat{\boldsymbol\beta})\right\}\right].
\end{align*}
We can show that the first two terms converge to zero in a similar way as for Term \eqref{eq:term_tmle.1} and Term \eqref{eq:term_tmle.2}, respectively. For the last term, we have
\begin{align*}
    \frac{1}{\sqrt{|\mathcal{V}_k|}}&\sum_{i\in\mathcal{V}_k}\left[\left(\frac{Z_i}{p(X_{PS,i};\boldsymbol\beta_0)}-1\right)\left\{\hat h_{1,k}^{(u)}(X_i;\hat{\boldsymbol\epsilon}, \boldsymbol\beta_0)-\hat h_{1,k}^{(u)}(X_i;\hat{\boldsymbol\epsilon}, \hat{\boldsymbol\beta})\right\}\right]\\
    &= \frac{1}{|\mathcal{V}_k|}\sum_{i\in\mathcal{V}_k}\left[\left(\frac{Z_i}{p(X_{PS,i};\boldsymbol\beta_0)}-1\right)\frac{\partial}{\partial \boldsymbol\beta}\hat h_{1,k}^{(u)}(X_i;\hat{\boldsymbol\epsilon}, \boldsymbol\beta)\biggr\rvert_{\boldsymbol{\boldsymbol\beta}=\boldsymbol\beta_0}\right]\sqrt{|\mathcal{V}_k|}(\boldsymbol\beta_0-\hat{\boldsymbol\beta}) +\sqrt{|\mathcal{V}_k|}||\boldsymbol\beta_0-\hat{\boldsymbol\beta}||^2_2O_{P_n}(1),
\end{align*}
where $||\cdot||^2_2$ denotes the Euclidian norm. The remainder term $\sqrt{|\mathcal{V}_k|}||\boldsymbol\beta_0-\hat{\boldsymbol\beta}||^2_2O_{P_n}(1)$ converges to zero in probability. The first term converges to zero as $\sqrt{|\mathcal{V}_k|}(\boldsymbol\beta_0-\hat{\boldsymbol\beta})=O_p(1)$ and \\$\left| \frac{1}{|\mathcal{V}_k|}\sum_{i\in\mathcal{V}_k}\left[\left(\frac{Z_i}{p(X_{PS,i};\boldsymbol\beta_0)}-1\right)\frac{\partial}{\partial \beta}h_{1,k}^{(u)}(X_i;\hat{\boldsymbol\epsilon}, \boldsymbol\beta)\biggr\rvert_{\boldsymbol\beta=\boldsymbol\beta_0}\right] \right|=o_p(1)$ by a similar reasoning as for Term \eqref{eq:term_tmle.2} (applying Chebyshev's inequality for a finite-dimensional vector) assuming \\$E\left(\left|\left|\frac{\partial}{\partial \beta}h_{1,k}^{(u)}(X_i;\hat{\boldsymbol\epsilon}, \boldsymbol\beta)\biggr\rvert_{\boldsymbol\beta=\boldsymbol\beta_0}\right|\right|_2\right)$ is bounded.

Consequently, 
\begin{align*}
    \sqrt{|\mathcal{V}_k|}(\hat\mu_{1,k, TMLE}^{(\pi)}-\mu_1)&=\frac{1}{\sqrt{|\mathcal{V}_k|}}\sum_{i\in\mathcal{V}_k}\left[\frac{Z_i}{p(X_{PS,i};\boldsymbol\beta_0)}\left\{Y_i-h_1^*(X_i;\boldsymbol\epsilon^*, \boldsymbol\beta_0)\right\}+h^*_1(X_i;\boldsymbol\epsilon^*, \boldsymbol\beta_0)-\mu_1\right]+o_{P_n}(1).
\end{align*}

Term \eqref{eq:tmle_par_term2} can be rewritten as follows
\begin{align}
    \frac{1}{|\mathcal{V}_k|}&\sum_{i\in\mathcal{V}_k}Z_i\left\{Y_i-\hat h_{1,k}^{(u)}(X_i;\hat{\boldsymbol\epsilon}, \hat{\boldsymbol\beta})\right\}\left(\frac{1}{p(X_{PS,i}; \hat{\boldsymbol\beta})}-\frac{1}{p(X_{PS,i};\boldsymbol\beta_0)}\right)\sqrt{|\mathcal{V}_k|}\nonumber\\
    &= \frac{1}{|\mathcal{V}_k|}\sum_{i\in\mathcal{V}_k}Z_i\left\{Y_i-h^*_1(X_i;\boldsymbol\epsilon^*, \boldsymbol\beta_0)\right\}\left(\frac{1}{p(X_{PS,i}; \hat{\boldsymbol\beta})}-\frac{1}{p(X_{PS,i};\boldsymbol\beta_0)}\right)\sqrt{|\mathcal{V}_k|}\label{eq:tmle_par_term2a}\\
    &+ \frac{1}{|\mathcal{V}_k|}\sum_{i\in\mathcal{V}_k}Z_i\left\{h^*_1(X_i;\boldsymbol\epsilon^*, \boldsymbol\beta_0)-\hat h_{1,k}^{(u)}(X_i;\boldsymbol\epsilon^*, \boldsymbol\beta_0)\right\}\left(\frac{1}{p(X_{PS,i}; \hat{\boldsymbol\beta})}-\frac{1}{p(X_{PS,i};\boldsymbol\beta_0)}\right)\sqrt{|\mathcal{V}_k|}\label{eq:tmle_par_term2b}\\
    &+ \frac{1}{|\mathcal{V}_k|}\sum_{i\in\mathcal{V}_k}Z_i\left\{\hat h_{1,k}^{(u)}(X_i;\boldsymbol\epsilon^*, \boldsymbol\beta_0)-\hat h_{1,k}^{(u)}(X_i;\hat{\boldsymbol\epsilon}, \boldsymbol\beta_0)\right\}\left(\frac{1}{p(X_{PS,i}; \hat{\boldsymbol\beta})}-\frac{1}{p(X_{PS,i};\boldsymbol\beta_0)}\right)\sqrt{|\mathcal{V}_k|}\label{eq:tmle_par_term2c}\\
    &+ \frac{1}{|\mathcal{V}_k|}\sum_{i\in\mathcal{V}_k}Z_i\left\{\hat h_{1,k}^{(u)}(X_i;\hat{\boldsymbol\epsilon}, \boldsymbol\beta_0)-\hat h_{1,k}^{(u)}(X_i;\hat{\boldsymbol\epsilon}, \hat{\boldsymbol\beta})\right\}\left(\frac{1}{p(X_{PS,i}; \hat{\boldsymbol\beta})}-\frac{1}{p(X_{PS,i};\boldsymbol\beta_0)}\right)\sqrt{|\mathcal{V}_k|}\label{eq:tmle_par_term2d}.
\end{align}

Following a similar reasoning as in Appendix \ref{app:par_samplesplitting}, Term \eqref{eq:tmle_par_term2a} can be rewritten as
\begin{align*}
    E\left[Z\left\{Y-h^*_1(X_i;\boldsymbol\epsilon^*, \boldsymbol\beta_0)\right\}\frac{1}{p(X_{PS};\boldsymbol\beta_0)^2}\frac{\partial}{\partial \beta}p(X_{PS};\boldsymbol\beta)\biggr\rvert_{\beta=\beta_0}\right]\left[E\left\{\frac{\partial s(X_{PS};\boldsymbol\beta_0)}{\partial \boldsymbol\beta}\right\}\right]^{-1}\left\{\frac{1}{\sqrt{|\mathcal{V}_k|}}\sum_{i\in \mathcal{V}_k}s(X_{PS,i};\boldsymbol\beta_0)\right\}+o_{P_n}(1),
\end{align*}
which converges to zero as $E\left[Z\left\{Y-h^*_1(X_i;\boldsymbol\epsilon^*, \boldsymbol\beta_0)\right\}\frac{1}{p(X_{PS};\boldsymbol\beta_0)^2}\frac{\partial}{\partial \boldsymbol\beta}p(X_{PS};\boldsymbol\beta)\biggr\rvert_{\boldsymbol\beta=\boldsymbol\beta_0}\right]=0$ due to the fact that the targeting step included the baseline covariates $X_{PS}$.
Term \eqref{eq:tmle_par_term2b} also converges to zero by a similar reasoning as for Term \eqref{eq:term2.2} with parametrically estimated propensity scores.

Next, from the the Cauchy Schwarz inequality, it follows that Term \eqref{eq:tmle_par_term2c} yields
{\small
\begin{align*}
    &\sqrt{|\mathcal{V}_k|}\left|\frac{1}{|\mathcal{V}_k|}\sum_{i\in\mathcal{V}_k}Z_i\left\{\hat h_{1,k}^{(u)}(X_i;\boldsymbol\epsilon^*, \boldsymbol\beta_0)-\hat h_{1,k}^{(u)}(X_i;\hat{\boldsymbol\epsilon}, \boldsymbol\beta_0)\right\}\left(\frac{1}{p(X_{PS,i}; \hat{\boldsymbol\beta})}-\frac{1}{p(X_{PS,i};\boldsymbol\beta_0)}\right)\right|\\
    &\leq \sqrt{|\mathcal{V}_k|}\left(\max_{i\in\mathcal{V}_k}\frac{1}{p(X_i;\boldsymbol\beta_0)p(X_{PS,i}; \hat{\boldsymbol\beta})}\right)\sqrt{\frac{1}{|\mathcal{V}_k|}\sum_{i\in\mathcal{V}_k}\left\{\hat h_{1,k}^{(u)}(X_i;\boldsymbol\epsilon^*, \boldsymbol\beta_0)-\hat h_{1,k}^{(u)}(X_i;\hat{\boldsymbol\epsilon}, \boldsymbol\beta_0)\right\}^2}\sqrt{\frac{1}{|\mathcal{V}_k|}\sum_{i\in\mathcal{V}_k}\left\{p(X_{PS,i}; \hat{\boldsymbol\beta})-p(X_{PS,i};\boldsymbol\beta_0)\right\}^2}\\
    &=o_{P_n}(1),
\end{align*}
}
as $1>p(X_{PS,i};\boldsymbol\beta_0)\geq\rho>0$ for $i=1, \dots n$, 
the fact that $\hat{\boldsymbol\beta}$ are root-$n$ consistent maximum likelihood estimators of parameters in a parametric model $p(X_{PS};\boldsymbol\beta)$ which is assumed to be continuous in $\beta$, and because $\hat{\boldsymbol\epsilon}$ converges in probability to $\boldsymbol\epsilon^*$ (with $\hat h_{1,k}^{(u)}(X;\boldsymbol\epsilon, \boldsymbol\beta_0)$ continuous in $\epsilon$).

For Term \eqref{eq:tmle_par_term2d}, we have
\begin{align*}
    \frac{1}{|\mathcal{V}_k|}&\sum_{i\in\mathcal{V}_k}Z_i\left\{\hat h_{1,k}^{(u)}(X_i;\hat{\boldsymbol\epsilon}, \boldsymbol\beta_0)-\hat h_{1,k}^{(u)}(X_i;\hat{\boldsymbol\epsilon}, \hat{\boldsymbol\beta})\right\}\left(\frac{1}{p(X_{PS,i}; \hat{\boldsymbol\beta})}-\frac{1}{p(X_{PS,i};\boldsymbol\beta_0)}\right)\sqrt{|\mathcal{V}_k|}\\
    &=\frac{1}{|\mathcal{V}_k|}\sum_{i\in\mathcal{V}_k}\left[\left(\frac{1}{p(X_{PS,i}; \hat{\boldsymbol\beta})}-\frac{1}{p(X_{PS,i};\boldsymbol\beta_0)}\right)\frac{\partial}{\partial \boldsymbol\beta}\hat h_{1,k}^{(u)}(X_i;\hat{\boldsymbol\epsilon}, \boldsymbol\beta)\biggr\rvert_{\boldsymbol\beta=\boldsymbol\beta_0}\right]\sqrt{|\mathcal{V}_k|}(\boldsymbol\beta_0-\hat{\boldsymbol\beta})\\ 
    &+\sqrt{|\mathcal{V}_k|}||\boldsymbol\beta_0-\hat{\boldsymbol\beta}||^2_2O_{P_n}(1),
\end{align*}
where $||\cdot||^2_2$ denotes the Euclidian norm. The remainder term $\sqrt{|\mathcal{V}_k|}||\boldsymbol\beta_0-\hat{\boldsymbol\beta}||^2_2O_{P_n}(1)$ converges to zero in probability. Denote $\beta_j$ and $\beta_{0,j}$ for component $j$ ($j\in 1, \dots, p$) of $\boldsymbol\beta$ and $\boldsymbol\beta_0$, respectively. Next, by H\"older's inequality
{\footnotesize
\begin{align*}
    &\left|\frac{1}{|\mathcal{V}_k|}\sum_{i\in\mathcal{V}_k}\left[\left(\frac{1}{p(X_{PS,i}; \hat{\boldsymbol\beta})}-\frac{1}{p(X_{PS,i};\boldsymbol\beta_0)}\right)\frac{\partial}{\partial \boldsymbol\beta}\hat h_{1,k}^{(u)}(X_i;\hat{\boldsymbol\epsilon}, \boldsymbol\beta)\biggr\rvert_{\boldsymbol\beta=\boldsymbol\beta_0}\right]\sqrt{|\mathcal{V}_k|}(\boldsymbol\beta_0-\hat{\boldsymbol\beta})\right|\\
    &\leq \sqrt{|\mathcal{V}_k|} \left|\left|\boldsymbol\beta_0-\hat{\boldsymbol\beta}\right|\right|_1
    \left|\left|\frac{1}{|\mathcal{V}_k|}\sum_{i\in\mathcal{V}_k}\left[\left(\frac{1}{p(X_{PS,i}; \hat{\boldsymbol\beta})}-\frac{1}{p(X_{PS,i};\boldsymbol\beta_0)}\right)\frac{\partial}{\partial \boldsymbol\beta}\hat h_{1,k}^{(u)}(X_i;\hat{\boldsymbol\epsilon}, \boldsymbol\beta)\biggr\rvert_{\boldsymbol\beta=\boldsymbol\beta_0}\right]\right|\right|_\infty\\
    &\leq \sqrt{|\mathcal{V}_k|} \left|\left|\boldsymbol\beta_0-\hat{\boldsymbol\beta}\right|\right|_1\max_{j\leq p, i\in\mathcal{V}_k}\left|\frac{\partial}{\partial \beta_j}\hat h_{1,k}^{(u)}(X_i;\hat{\boldsymbol\epsilon}, \boldsymbol\beta)\biggr\rvert_{\beta_j=\beta_{0,j}}\right|
    \frac{1}{|\mathcal{V}_k|}\sum_{i\in\mathcal{V}_k}\left|\left(\frac{1}{p(X_{PS,i}; \hat{\boldsymbol\beta})}-\frac{1}{p(X_{PS,i};\boldsymbol\beta_0)}\right)\right|\\
    &\leq \sqrt{|\mathcal{V}_k|} \left|\left|\boldsymbol\beta_0-\hat{\boldsymbol\beta}\right|\right|_1\max_{j\leq p, i\in\mathcal{V}_k}\left|\frac{\partial}{\partial \beta_j}\hat h_{1,k}^{(u)}(X_i;\hat{\boldsymbol\epsilon}, \boldsymbol\beta)\biggr\rvert_{\beta_j=\beta_{0,j}}\right|\left(\max_{i\in\mathcal{V}_k}\frac{1}{p(X_{PS,i};\boldsymbol\beta_0)p(X_i; \hat{\boldsymbol\beta})}\right)
    \frac{1}{|\mathcal{V}_k|}\sum_{i\in\mathcal{V}_k}\left|p(X_{PS,i}; \hat{\boldsymbol\beta})-p(X_{PS,i};\boldsymbol\beta_0)\right|\\
    &\leq \sqrt{|\mathcal{V}_k|} \left|\left|\boldsymbol\beta_0-\hat{\boldsymbol\beta}\right|\right|_1\max_{j\leq p, i\in\mathcal{V}_k}\left|\frac{\partial}{\partial \beta_j}\hat h_{1,k}^{(u)}(X_i;\hat{\boldsymbol\epsilon}, \boldsymbol\beta)\biggr\rvert_{\beta_j=\beta_{0,j}}\right|\left(\max_{i\in\mathcal{V}_k}\frac{1}{p(X_{PS,i};\boldsymbol\beta_0)p(X_{PS,i}; \hat{\boldsymbol\beta})}\right)
    \frac{\sqrt{|\mathcal{V}_k|}}{|\mathcal{V}_k|}\sqrt{\sum_{i\in\mathcal{V}_k}\left\{p(X_{PS,i}; \hat{\boldsymbol\beta})-p(X_{PS,i};\boldsymbol\beta_0)\right\}^2}\\
    &\leq \sqrt{|\mathcal{V}_k|} \left|\left|\boldsymbol\beta_0-\hat{\boldsymbol\beta}\right|\right|_1\max_{j\leq p, i\in\mathcal{V}_k}\left|\frac{\partial}{\partial \beta_j}\hat h_{1,k}^{(u)}(X_i;\hat{\boldsymbol\epsilon}, \boldsymbol\beta)\biggr\rvert_{\beta_j=\beta_{0,j}}\right|
    \left(\max_{i\in\mathcal{V}_k}\frac{1}{p(X_{PS,i};\boldsymbol\beta_0)p(X_{PS,i}; \hat{\boldsymbol\beta})}\right)\sqrt{\frac{1}{|\mathcal{V}_k|}\sum_{i\in\mathcal{V}_k}\left\{p(X_{PS,i}; \hat{\boldsymbol\beta})-p(X_{PS,i};\boldsymbol\beta_0)\right\}^2}\\
    &=o_{P_n}(1),
\end{align*}
}
as $\left|\left|\boldsymbol\beta_0-\hat{\boldsymbol\beta}\right|\right|_1=O_{P_n}\left(\frac{1}{\sqrt{n}}\right)$, $1>p(L_i;\boldsymbol\beta_0)\geq\rho>0$, the fact that $\hat{\boldsymbol\beta}$ is an estimate of a parameter in a parametric model obtained with maximum likelihood estimation and assuming that $\max_{j\leq p, i\in\mathcal{V}_k}\left|\frac{\partial}{\partial \beta_j}\hat h_{1,k}^{(u)}(X_i;\hat{\boldsymbol\epsilon}, \boldsymbol{\beta})\biggr\rvert_{\beta_j=\beta_{0,j}}\right|$ is bounded.

This then leads to 
\begin{align*}
    \sqrt{n}(\hat\mu_{1, TMLE}-\mu_1)&=\sqrt{n}\frac{1}{K}\sum_{k=1}^K\frac{1}{|\mathcal{V}_k|}\sum_{i\in\mathcal{V}_k}\left\{\left(\frac{Z_i}{\pi}\left\{Y_i-h_1^*(X_i;\boldsymbol\epsilon^*)\right\}+h_1^*(X_i;\boldsymbol\epsilon^*)-\mu_1\right)\right\}+o_{P_n}(1).
\end{align*}

\subsection{Minimial prediction error: Empirical Efficiency Maximization}\label{app:minPredError}
We consider the different data-adaptive covariate adjusted estimators described in Section 3, which all belong to a broad class of augmented inverse probability weighting estimators. Cross-fitting disregarded, they can all be rewritten in the form
\begin{align*}
\frac{1}{n}\sum_{i=1}^n\left(\frac{Z_i}{\hat{\pi}}\left\{Y_i-\hat h_{1}(X_i)\right\}+\hat h_{1}(X_i)-\left[\frac{1-Z_i}{1-\hat{\pi}}\left\{Y_i-\hat h_{0}(X_i)\right\}+\hat h_{0}(X_i)\right]\right),
\end{align*}
where we consider the non-parametric predictions $\hat h_1(X_i)$ and $\hat h_0(X_i)$. The first prediction can be replaced by $h_1(X_i; \hat{\boldsymbol\gamma})$, $h_1(X_i; \tilde{\boldsymbol\gamma})$ or $\hat h^{(u)}_1(X_i; \hat\epsilon)$ (i.e., the updated prediction under $Z=1$ for the TMLE estimator), and the second by $h_0(X_i; \hat{\boldsymbol\eta})$, $h_0(X_i; \tilde{\boldsymbol\eta})$ or $\hat h^{(u)}_0(X_i; \hat\epsilon)$ (i.e., the updated prediction under $Z=0$ for the TMLE estimator). This is a consequence of using a generalized linear model with canonical link and intercept, as this guarantees that
\[\frac{1}{n}\sum_{i=1}^n\left(\frac{Z_i}{\hat{\pi}}\left\{Y_i-\hat h_{1}(X_i)\right\}\right)=0,\]
and similar for $\hat h_0(X_i)$. In addition, we have seen that the variance of these estimators is calculated as $1/n$ times the sample variance of
\[\frac{Z_i}{\hat{\pi}}\left\{Y_i-\hat h_{1}(X_i)\right\}+\hat h_{1}(X_i)-\left[\frac{1-Z_i}{1-\hat{\pi}}\left\{Y_i-\hat h_{0}(X_i)\right\}+\hat h_{0}(X_i)\right].\]

In what follows, we show that minimizing this variance is equivalent to minimizing the mean squared error of the outcome prediction (see also \citep{rubin2008empirical} and \citep{cao2009improving}):
\begin{align*}
    Var\left[\frac{Z}{\pi}\left\{Y-h_{1}(X)\right\}+h_{1}(X)-\mu_1\right]&=Var\left[\frac{Z}{\pi}\left\{Y-h_{1}(X)\right\}+h_{1}(X)\right]\\
    &=Var\left[\left(\frac{Z}{\pi}-1+1\right)\left\{Y(1)-h_{1}(X)\right\}+h_{1}(X)\right]\\
    &=Var\left[\left(\frac{Z}{\pi}-1\right)\left\{Y(1)-h_{1}(X)\right\}+Y(1)\right]\\
    &= Var\left(E\left[\left(\frac{Z}{\pi}-1\right)\left\{Y(1)-h_{1}(X)\right\}+Y(1)\biggr\rvert Y(1), X\right]\right)\\
    &+ E\left(Var\left[\left(\frac{Z}{\pi}-1\right)\left\{Y(1)-h_{1}(X)\right\}+Y(1)\biggr\rvert Y(1), X\right]\right)\\
    &=Var\{Y(1)\}+E\left[\left\{Y(1)-h_{1}(X)\right\}^2Var\left(\frac{Z}{\pi}\right)\right]\\
    &=Var\{Y(1)\}+E\left[\left\{Y(1)-h_{1}(X)\right\}^2\frac{\pi(1-\pi)}{\pi^2}\right]\\
    &=Var\{Y(1)\}+E\left[\left\{Y(1)-h_{1}(X)\right\}^2\frac{\pi(1-\pi)}{\pi^2}\right]\\
    &=Var\{Y(1)\}+\frac{\pi(1-\pi)}{\pi}E\left[\left\{Y(1)-h_{1}(X)\right\}^2\right]\\
    &=Var\{Y(1)\}+\frac{(1-\pi)}{\pi^2}E\left[Z\left\{Y-h_{1}(X)\right\}^2\right]
\end{align*}
where $Y(1)$ is the counterfactual outcome under $Z=1$. Thus, minimizing the variance of the imputaton estimator $\hat\mu_{1,DA}-\hat\mu_{0,DA}$ in the beginning of Section 3 (i.e., without cross-fitting), is equivalent to minimizing the mean squared error of the outcome predictions $h_1(X_i; \tilde{\boldsymbol\gamma})$ and $h_0(X_i; \tilde{\boldsymbol\eta})$. This, for example, applies for any flexible regression algorithm using least-squares estimation, even those that do not perform selection of variables (e.g., Lasso for linear models). Specifically, Lasso for linear models is exactly based on minimizing the MSE (subject to the sum of the absolute values of the non-intercept coefficients being less than a constraint).

However, to attain this efficiency guarantee, logistic regression models must be fitted via least squares instead of maximum likelihood estimation. For binary endpoints, using least squares (instead of maximum likelihood estimation) also has the consequence that the prediction unbiasedness condition (see Equation (1) in the main paper) no longer holds for regular standardization estimator in Section 2.3 (i.e., $\hat\theta$) and the data-adaptive approach in the beginning of Section 3 (i.e., $\hat\theta_{DA}$). In that case, to still obtain the robustness against model misspecification, these estimators need to be caclulated as
\begin{align*}
\hat\theta = \frac{1}{n}\sum_{i=1}^n\left(\frac{Z_i}{\hat{\pi}}\left\{Y_i-h_1(X_i; \hat{\boldsymbol\gamma})\right\}+\hat h_1(X_i; \hat{\boldsymbol\gamma})-\left[\frac{1-Z_i}{1-\hat{\pi}}\left\{Y_i-h_0(X_i; \hat{\boldsymbol\eta})\right\}+h_0(X_i; \hat{\boldsymbol\eta})(X_i)\right]\right),
\end{align*}
and
\begin{align*}
\hat\theta_{DA} = \frac{1}{n}\sum_{i=1}^n\left(\frac{Z_i}{\hat{\pi}}\left\{Y_i-h_1(X_i; \tilde{\boldsymbol\gamma})\right\}+\hat h_1(X_i; \tilde{\boldsymbol\gamma})-\left[\frac{1-Z_i}{1-\hat{\pi}}\left\{Y_i-h_0(X_i; \tilde{\boldsymbol\eta})\right\}+h_0(X_i; \tilde{\boldsymbol\eta})(X_i)\right]\right).
\end{align*}

Interestingly, the TMLE estimator in Section 3.2 does not need an adaptation. This is a consequence of using a canonical link function to update the initial predictions $h^{(1)}_1(X_i; \tilde{\tilde{\boldsymbol{\gamma}}})$ and $h^{(1)}_0(X_i; \tilde{\tilde{\boldsymbol{\eta}}})$.

This is also relevant for the estimators using cross-fitting (see Section 3.3). Choosing the prediction $\hat h_{1,k}(X)$ as the minimizer of $E\left[Z\left\{Y-h_1(X)\right\}^2\right]$ minimizes the variance of the main component of Equation \eqref{eq:influenceFunction1}. In addition, this also ensures that the remainder term in Equation \eqref{eq:influenceFunction1}, $E\left[\frac{Z}{\pi^2}\left\{Y-h_1(X)\right\}\right]$, converges to zero and is therefore negligible. This is appealing as cross-fitting allows the use of machine learning algorithms, many of which allow to choose the MSE as the loss function of interest. It can moreover be chosen as the loss function in the SuperLearner algorithm, which is also used to obtain the original predictions for the cross-validated TMLE (CV-TMLE; \cite{zheng2011cross}) approach.

Although this simple approach can be used when the randomization probability $\pi$ is estimated as a sample proportion (i.e., $\hat\pi$), \citet{cao2009improving} show that the approach becomes slightly more complicated when the propensity score is parametrically estimated (e.g., using a logistic regression model). 

\newpage
\section{Additional Simulation Results}\label{app:simulation}

\begin{table}[h!]
	\caption{\label{tab:MISTIE_Type1_Power_NoCorrection}Results for Type I error and Power. No small sample correction was used for the covariate adjustment approaches. $\#$ parameters depicts the (maximum) number of parameters (not including the intercept) in the outcome working models. Results based on 100,000 simulations.
	}
\centering
\begin{tabular}{ l  c ccc c ccc c ccc}
  \hline
  &&\multicolumn{3}{c}{\multirow{2}{*}{General}}&&\multicolumn{3}{c}{Data-adaptive}&&\multicolumn{3}{c}{Data-adaptive}\\
  &&\multicolumn{3}{c}{}&&\multicolumn{3}{c}{No cross-fitting}&&\multicolumn{3}{c}{Cross-fitting}\\
  \cline{3-5} \cline{7-9} \cline{11-13}\\
Setting & $\#$ parameters & 200 & 500 & 1000 & & 200 & 500 & 1000 & & 200 & 500 & 1000\\
  \hline  \\
  \multicolumn{13}{c}{\textbf{Type I error}}\\\\
  Unadjusted &0&2.6\%&2.5\%&2.4\%&&&&&&&&\\
  Cov. adj. (1) &1&2.7\% & 2.5\% & 2.4\%&&&&&&&&\\
  Cov. adj. (2) &1&2.7\% & 2.6\% & 2.4\%&&&&&&&&\\
  Cov. adj. (3) &1&2.7\% & 2.5\% & 2.4\%&&&&&&&&\\
  Cov. adj. (4) &$\leq 3$&2.8\%&2.5\%&2.5\%&&2.8\%&2.5\%&2.5\%&&2.9\%&2.5\%&2.5\%\\
  Cov. adj. (5) &$\leq 7$&3.1\%&2.6\%&2.6\%&&3.0\%&2.6\%&2.6\%&&2.8\%&2.5\%&2.5\%\\
  Cov. adj. (6) &$\leq 16$&3.9\%&3.0\%&2.8\%&&3.3\%&2.7\%&2.6\%&&2.8\%&2.6\%&2.5\%\\
  Cov. adj. (7) &$\leq 3$&2.8\%&2.5\%&2.5\%&&2.8\%&2.5\%&2.5\%&&2.9\%&2.5\%&2.5\%\\
  Cov. adj. (8) &$\leq 7$&3.1\%&2.6\%&2.6\%&&2.9\%&2.6\%&2.5\%&&2.8\%&2.5\%&2.5\%\\
  Cov. adj. (9) &$\leq 16$&3.8\%&2.9\%&2.7\%&&3.4\%&2.7\%&2.6\%&&2.8\%&2.6\%&2.5\%\\\\
  \multicolumn{13}{c}{\textbf{Power}}\\\\
  Unadjusted &0&11\%&22\%&38\%&&&&&&&&\\
  Cov. adj. (1) &1&12\% & 23\% & 41\%&&&&&&&&\\
  Cov. adj. (2) &1&12\% & 23\% & 40\%&&&&&&&&\\
  Cov. adj. (3) &1&12\% & 23\% & 40\%&&&&&&&&\\
  Cov. adj. (4) &$\leq 3$&15\%&28\%&49\%&&15\%&28\%&49\%&&14\%&28\%&49\%\\
  Cov. adj. (5) &$\leq 7$&15\%&29\%&50\%&&15\%&29\%&50\%&&14\%&28\%&49\%\\
  Cov. adj. (6) &$\leq 16$&17\%&30\%&50\%&&16\%&29\%&50\%&&14\%&28\%&49\%\\
  Cov. adj. (7) &$\leq 3$&15\%&28\%&49\%&&15\%&28\%&49\%&&14\%&28\%&49\%\\
  Cov. adj. (8) &$\leq 7$&15\%&29\%&49\%&&15\%&28\%&49\%&&14\%&28\%&49\%\\
  Cov. adj. (9) &$\leq 16$&17\%&29\%&50\%&&16\%&29\%&49\%&&14\%&28\%&49\%\\\\
   \hline
\end{tabular}\\

	 {\raggedright  \par}
\end{table}

  \begin{table}[h!]
\caption{\label{tab:MISTIE_Type1_Power_strongcorrection}Results for Type I error and Power. A small sample correction was used for the covariate adjustment approaches, with $p_j$ the number of parameters in the upper model (i.e., largest possible model allowed in the selection procedure). $\#$ parameters depicts the (maximum) number of parameters (not including the intercept) in the outcome working models. Results based on 100,000 simulations.
	}
 \centering
\begin{tabular}{ l  c ccc c ccc c ccc}
  \hline
  &&\multicolumn{3}{c}{\multirow{2}{*}{General}}&&\multicolumn{3}{c}{Data-adaptive}&&\multicolumn{3}{c}{Data-adaptive}\\
  &&\multicolumn{3}{c}{}&&\multicolumn{3}{c}{No cross-fitting}&&\multicolumn{3}{c}{Cross-fitting}\\
  \cline{3-5} \cline{7-9} \cline{11-13}\\
Setting & $\#$ parameters & 200 & 500 & 1000 & & 200 & 500 & 1000 & & 200 & 500 & 1000\\
  \hline  \\
  \multicolumn{13}{c}{\textbf{Type I error}}\\\\
  Unadjusted &0&2.6\%&2.5\%&2.4\%&&&&&&&&\\
  Cov. adj. (1) &1&2.6\% & 2.5\% & 2.4\%&&&&&&&&\\
  Cov. adj. (2) &1&2.6\% & 2.6\% & 2.4\%&&&&&&&&\\
  Cov. adj. (3) &1&2.6\% & 2.5\% & 2.4\%&&&&&&&&\\
  Cov. adj. (4) &$\leq 3$&2.6\%&2.5\%&2.5\%&&2.6\%&2.5\%&2.5\%&&2.7\%&2.5\%&2.5\%\\
  Cov. adj. (5) &$\leq 7$&2.7\%&2.4\%&2.5\%&&2.5\%&2.4\%&2.5\%&&2.4\%&2.4\%&2.4\%\\
  Cov. adj. (6) &$\leq 16$&2.7\%&2.5\%&2.5\%&&2.3\%&2.3\%&2.4\%&&1.9\%&2.2\%&2.3\%\\
  Cov. adj. (7) &$\leq 3$&2.6\%&2.5\%&2.5\%&&2.6\%&2.5\%&2.5\%&&2.7\%&2.5\%&2.4\%\\
  Cov. adj. (8) &$\leq 7$&2.6\%&2.4\%&2.5\%&&2.5\%&2.4\%&2.4\%&&2.4\%&2.4\%&2.4\%\\
  Cov. adj. (9) &$\leq 16$&2.6\%&2.5\%&2.5\%&&2.2\%&2.3\%&2.4\%&&1.8\%&2.2\%&2.3\%\\\\
  \multicolumn{13}{c}{\textbf{Power}}\\\\
  Unadjusted &0&11\%&22\%&38\%&&&&&&&&\\
  Cov. adj. (1) &1&12\% & 23\% & 41\%&&&&&&&&\\
  Cov. adj. (2) &1&12\% & 23\% & 40\%&&&&&&&&\\
  Cov. adj. (3) &1&12\% & 23\% & 40\%&&&&&&&&\\
  Cov. adj. (4) &$\leq 3$&14\%&28\%&49\%&&14\%&28\%&49\%&&14\%&28\%&49\%\\
  Cov. adj. (5) &$\leq 7$&14\%&28\%&49\%&&13\%&28\%&49\%&&13\%&27\%&49\%\\
  Cov. adj. (6) &$\leq 16$&13\%&27\%&49\%&&12\%&27\%&48\%&&10\%&26\%&48\%\\
  Cov. adj. (7) &$\leq 3$&14\%&28\%&49\%&&14\%&28\%&49\%&&14\%&28\%&49\%\\
  Cov. adj. (8) &$\leq 7$&14\%&28\%&49\%&&13\%&27\%&49\%&&13\%&27\%&48\%\\
  Cov. adj. (9) &$\leq 16$&13\%&27\%&48\%&&12\%&27\%&48\%&&10\%&25\%&47\%\\\\
   \hline
\end{tabular}\\
	 {\raggedright  \par}
\end{table}

\begin{table}[h!]
	\caption{\label{tab:MISTIE_bias}Results for (finite sample) bias. $\#$ parameters depicts the (maximum) number of parameters (not including the intercept) in the outcome working models. $\pi$: true propensity score is used. Results based on 100,000 simulations.
	}
\centering
{\small
\begin{adjustbox}{angle=90}
\begin{tabular}{ l  c ccc c ccc c ccc c ccc}
  \hline
  &&\multicolumn{3}{c}{\multirow{2}{*}{General}}&&\multicolumn{3}{c}{Data-adaptive}&&\multicolumn{3}{c}{Data-adaptive}&&\multicolumn{3}{c}{Data-adaptive}\\
  &&\multicolumn{3}{c}{}&&\multicolumn{3}{c}{No cross-fitting}&&\multicolumn{3}{c}{Cross-fitting}&&\multicolumn{3}{c}{Cross-fitting \& $\pi$}\\
  \cline{3-5} \cline{7-9} \cline{11-13} \cline{15-17}\\
Setting & $\#$ parameters & 200 & 500 & 1000 & & 200 & 500 & 1000 & & 200 & 500 & 1000 & & 200 & 500 & 1000\\
  \hline  \\
  \multicolumn{17}{c}{\textbf{Bias under weak null}}\\\\
  Unadjusted &0&-0.0002&-0.0001&-0.0002&&&&&&&&&&&&\\
  Cov. adj. (1) &1&-0.0002 & -0.0001 & -0.0002&&&&&&&&& &  &  & \\
  Cov. adj. (2) &1&-0.0002 & -0.0001 & -0.0002&&&&&&&&& &  &  & \\
  Cov. adj. (3) &1&-0.0002 & -0.0002 & -0.0002&&&&&&&&& &  &  & \\
  Cov. adj. (4) &$\leq 3$&-0.0001&-0.0001&-0.0002&&-0.0001&-0.0001&-0.0002&&-0.0002&-0.0001&-0.0002& & -0.0002 & -0.0001 & -0.0002\\
  Cov. adj. (5) &$\leq 7$&-0.0002&-0.0001&-0.0002&&-0.0001&-0.0001&-0.0002&&-0.0002&-0.0001&-0.0002& & -0.0002 & -0.0001 & -0.0002\\
  Cov. adj. (6) &$\leq 16$&-0.0003&-0.0002&-0.0002&&-0.0002&-0.0002&-0.0002&&-0.0001&-0.0001&-0.0002& & -0.0001 & -0.0001 & -0.0002\\
  Cov. adj. (7) &$\leq 3$&-0.0002&-0.0001&-0.0002&&-0.0001&-0.0001&-0.0002&&-0.0001&-0.0001&-0.0002& & -0.0001 & -0.0001 & -0.0002\\
  Cov. adj. (8) &$\leq 7$&-0.0003&-0.0001&-0.0002&&-0.0002&-0.0001&-0.0002&&-0.0002&-0.0001&-0.0002& & -0.0002 & -0.0001 & -0.0002\\
  Cov. adj. (9) &$\leq 16$&-0.0004&-0.0002&-0.0002&&-0.0003&-0.0002&-0.0002&&-0.0001&-0.0001&-0.0002& & -0.0002 & -0.0001 & -0.0002\\\\
  \multicolumn{17}{c}{\textbf{Bias under alternative}}\\\\
  Unadjusted &0&-0.0003&-0.0001&-0.0001&&&&&&&&& &  &  & \\
  Cov. adj. (1) &1&-0.0002 & -0.0001 & -0.0001&&&&&&&&& &  &  & \\
  Cov. adj. (2) &1&-0.0002& -0.0001 & -0.0002&&&&&&&&& &  &  & \\
  Cov. adj. (3) &1&-0.0003 & -0.0002 & -0.0001&&&&&&&&& &  &  & \\
  Cov. adj. (4) &$\leq 3$&-0.0002&-0.0001&-0.0002&&-0.0001&-0.0001&-0.0002&&-0.0002&-0.0001&-0.0002& & -0.0002 & -0.0001 & -0.0002\\
  Cov. adj. (5) &$\leq 7$&-0.0002&-0.0001&-0.0002&&-0.0001&-0.0001&-0.0002&&-0.0002&-0.0001&-0.0002& & -0.0002 & -0.0001 & -0.0002\\
  Cov. adj. (6) &$\leq 16$&-0.0004&-0.0002&-0.0002&&-0.0003&-0.0001&-0.0002&&-0.0002&-0.0001&-0.0002& & -0.0002 & -0.0001 & -0.0002\\
  Cov. adj. (7) &$\leq 3$&-0.0002&-0.0001&-0.0002&&-0.0002&-0.0001&-0.0002&&-0.0002&-0.0001&-0.0002& & -0.0002 & -0.0001 & -0.0002\\
  Cov. adj. (8) &$\leq 7$&-0.0003&-0.0001&-0.0002&&-0.0002&-0.0001&-0.0002&&-0.0002&-0.0001&-0.0002& & -0.0002 & -0.0001 & -0.0002\\
  Cov. adj. (9) &$\leq 16$&-0.0004&-0.0001&-0.0002&&-0.0003&-0.0001&-0.0002&&-0.0002&-0.0001&-0.0002& & -0.0002 & -0.0001 & -0.0002\\\\
   \hline
\end{tabular}
	 {\raggedright  \par}
  \end{adjustbox}
  }
\end{table}

\begin{table}[h!]
	\caption{\label{tab:MISTIE_conitional}Results for Type I error and Power based on the covariate adjustment approach and conditional effect estimates trough logistic regression. No small sample correction was used for the covariate adjustment approaches (i.e., `General covariate adjustment' and `Data-adaptive covariate adjustment'). Models were fitted in both treatment arms, and included an intercept, a treatment indicator and the variables as described in Section 4.1. `Logistic regression' includes the same variables as the corresponding `General covariate adjustment' approach, while `Data-adaptive logistic regression' includes the same variables as the corresponding `Data-adaptive covariate adjustment' approach. All simulations are without cross-fitting. $\#$ parameters depicts the (maximum) number of parameters (not including the intercept and coefficient corresponding with treatment indicator) in the outcome working models. Results based on 100,000 simulations.
	}
\centering
\begin{adjustbox}{angle=90}
\begin{tabular}{ l  c ccc c ccc c ccc c ccc}
  \hline
  &&\multicolumn{3}{c}{General}&&\multicolumn{3}{c}{Logistic}&&\multicolumn{3}{c}{Data-adaptive}&&\multicolumn{3}{c}{Data-adaptive}\\
  &&\multicolumn{3}{c}{cov. adj.}&&\multicolumn{3}{c}{regression}&&\multicolumn{3}{c}{cov. adj.}&&\multicolumn{3}{c}{Logistic regr.}\\
  \cline{3-5} \cline{7-9} \cline{11-13} \cline{15-17}\\
Setting & $\#$ parameters & 200 & 500 & 1000 & & 200 & 500 & 1000 & & 200 & 500 & 1000 & & 200 & 500 & 1000\\
  \hline  \\
  \multicolumn{17}{c}{\textbf{Type I error}}\\\\
  Setting (1) &1&2.7\% & 2.5\% & 2.4\%&&2.5\%&2.4\%&2.4\%&&&&&&&&\\
  Setting (2) &1&2.7\% & 2.6\% & 2.4\%&&2.5\%&2.5\%&2.4\%&&&&&&&&\\
  Setting (3) &1&2.7\% & 2.5\% & 2.4\%&&2.4\%&2.4\%&2.4\%&&&&&&&&\\
  Setting (4) &$\leq 3$&2.8\%&2.5\%&2.5\%&&2.5\%&2.4\%&2.5\%&&2.8\%&2.5\%&2.5\%&&2.5\%&2.4\%&2.5\%\\
  Setting (5) &$\leq 7$&3.1\%&2.6\%&2.6\%&&2.6\%&2.4\%&2.5\%&&3.0\%&2.6\%&2.6\%&&2.6\%&2.4\%&2.5\%\\
  Setting (6) &$\leq 16$&3.8\%&2.9\%&2.7\%&&2.9\%&2.6\%&2.6\%&&3.4\%&2.7\%&2.6\%&&2.9\%&2.6\%&2.5\%\\
  Setting (7) &$\leq 3$&2.8\%&2.5\%&2.5\%&&2.5\%&2.4\%&2.5\%&&2.8\%&2.5\%&2.5\%&&2.5\%&2.4\%&2.5\%\\
  Setting (8) &$\leq 7$&3.1\%&2.6\%&2.6\%&&2.6\%&2.4\%&2.5\%&&3.0\%&2.6\%&2.5\%&&2.6\%&2.4\%&2.5\%\\
  Setting (9) &$\leq 16$&3.7\%&2.9\%&2.7\%&&2.9\%&2.6\%&2.5\%&&3.4\%&2.6\%&2.5\%&&2.9\%&2.6\%&2.5\%\\\\
  \multicolumn{17}{c}{\textbf{Power}}\\\\
  Setting (1) &1&12\% & 23\% & 41\%&&12\%&23\%&41\%&&&&&&&&\\
  Setting (2) &1&12\% & 23\% & 40\%&&11\%&23\%&40\%&&&&&&&&\\
  Setting (3) &1&12\% & 23\% & 40\%&&11\%&23\%&40\%&&&&&&&&\\
  Setting (4) &$\leq 3$&15\%&29\%&49\%&&14\%&28\%&49\%&&15\%&28\%&49\%&&13\%&28\%&49\%\\
  Setting (5) &$\leq 7$&15\%&29\%&50\%&&14\%&28\%&49\%&&15\%&29\%&49\%&&14\%&28\%&49\%\\
  Setting (6) &$\leq 16$&17\%&30\%&50\%&&14\%&28\%&49\%&&16\%&29\%&50\%&&15\%&28\%&49\%\\
  Setting (7) &$\leq 3$&15\%&28\%&49\%&&14\%&28\%&49\%&&15\%&28\%&49\%&&13\%&27\%&49\%\\
  Setting (8) &$\leq 7$&15\%&29\%&50\%&&14\%&28\%&49\%&&15\%&28\%&49\%&&14\%&28\%&49\%\\
  Setting (9) &$\leq 16$&17\%&30\%&50\%&&14\%&28\%&49\%&&16\%&29\%&49\%&&15\%&28\%&49\%\\\\
   \hline
\end{tabular}
	 {\raggedright  \par}
  \end{adjustbox}
\end{table}

\begin{table}[h!]
	\caption{\label{tab:MISTIE_conitional_correction}Results for Type I error and Power based on the covariate adjustment approach and conditional effect estimates. A small sample correction ($(n-1)/(n-p-1)$, with $p$ the (maximum) number of parameters corresponding with the baseline covariates) was used for the covariate adjustment approaches targeting a marginal estimand (i.e., `General covariate adjustment' and `Data-adaptive covariate adjustment'). Models were fitted in both treatment arms, and included an intercept, a treatment indicator and the variables as described in Section 4.1. `Logistic regression' includes the same variables as the corresponding `General covariate adjustment' approach, while `Data-adaptive logistic regression' includes the same variables as the corresponding `Data-adaptive covariate adjustment' approach. All simulations are without cross-fitting. $\#$ parameters depicts the (maximum) number of parameters (not including the intercept and coefficient corresponding with treatment indicator) in the outcome working model. Results based on 100,000 simulations.
	}
\centering
\begin{adjustbox}{angle=90}
\begin{tabular}{ l  c ccc c ccc c ccc c ccc}
  \hline
  &&\multicolumn{3}{c}{General}&&\multicolumn{3}{c}{Logistic}&&\multicolumn{3}{c}{Data-adaptive}&&\multicolumn{3}{c}{Data-adaptive}\\
  &&\multicolumn{3}{c}{cov. adj.}&&\multicolumn{3}{c}{regression}&&\multicolumn{3}{c}{cov. adj.}&&\multicolumn{3}{c}{Logistic regr.}\\
  \cline{3-5} \cline{7-9} \cline{11-13} \cline{15-17}\\
Setting & $\#$ parameters & 200 & 500 & 1000 & & 200 & 500 & 1000 & & 200 & 500 & 1000 & & 200 & 500 & 1000\\
  \hline  \\
  \multicolumn{17}{c}{\textbf{Type I error}}\\\\
  Setting (1) &1&2.7\% & 2.5\% & 2.4\%&&2.5\%&2.4\%&2.4\%&&&&&&&&\\
  Setting (2) &1&2.7\% & 2.6\% & 2.4\%&&2.5\%&2.5\%&2.4\%&&&&&&&&\\
  Setting (3) &1&2.6\% & 2.5\% & 2.4\%&&2.4\%&2.4\%&2.4\%&&&&&&&&\\
  Setting (4) &$\leq 3$&2.6\%&2.5\%&2.5\%&&2.5\%&2.4\%&2.5\%&&2.7\%&2.5\%&2.5\%&&2.5\%&2.4\%&2.5\%\\
  Setting (5) &$\leq 7$&2.9\%&2.5\%&2.6\%&&2.6\%&2.4\%&2.5\%&&2.7\%&2.5\%&2.5\%&&2.6\%&2.4\%&2.5\%\\
  Setting (6) &$\leq 16$&3.3\%&2.7\%&2.6\%&&2.9\%&2.6\%&2.6\%&&2.8\%&2.5\%&2.5\%&&2.9\%&2.6\%&2.5\%\\
  Setting (7) &$\leq 3$&2.7\%&2.5\%&2.5\%&&2.5\%&2.4\%&2.5\%&&2.7\%&2.5\%&2.5\%&&2.5\%&2.4\%&2.5\%\\
  Setting (8) &$\leq 7$&2.9\%&2.5\%&2.5\%&&2.6\%&2.4\%&2.5\%&&2.7\%&2.5\%&2.5\%&&2.6\%&2.4\%&2.5\%\\
  Setting (9) &$\leq 16$&3.2\%&2.7\%&2.6\%&&2.9\%&2.6\%&2.5\%&&2.8\%&2.5\%&2.5\%&&2.9\%&2.6\%&2.5\%\\\\
  \multicolumn{17}{c}{\textbf{Power}}\\\\
  Setting (1) &1&12\% & 23\% & 41\%&&12\%&23\%&41\%&&&&&&&&\\
  Setting (2) &1&12\% & 23\% & 40\%&&11\%&23\%&40\%&&&&&&&&\\
  Setting (3) &1&12\% & 23\% & 40\%&&11\%&23\%&40\%&&&&&&&&\\
  Setting (4) &$\leq 3$&14\%&28\%&49\%&&14\%&28\%&49\%&&14\%&28\%&49\%&&13\%&28\%&49\%\\
  Setting (5) &$\leq 7$&15\%&28\%&49\%&&14\%&28\%&49\%&&14\%&28\%&49\%&&14\%&28\%&49\%\\
  Setting (6) &$\leq 16$&15\%&28\%&49\%&&14\%&28\%&49\%&&14\%&28\%&49\%&&15\%&28\%&49\%\\
  Setting (7) &$\leq 3$&14\%&28\%&49\%&&14\%&28\%&49\%&&14\%&28\%&49\%&&13\%&27\%&49\%\\
  Setting (8) &$\leq 7$&15\%&28\%&49\%&&14\%&28\%&49\%&&14\%&28\%&49\%&&14\%&28\%&49\%\\
  Setting (9) &$\leq 16$&15\%&28\%&49\%&&14\%&28\%&49\%&&14\%&28\%&49\%&&15\%&28\%&49\%\\\\
   \hline
\end{tabular}
	 {\raggedright  \par}
  \end{adjustbox}
\end{table}

\newpage
\bibliographystyle{Chicago}
\bibliography{sample}

\end{document}